\documentclass[aps,prb,twocolumn,floatfix,showpacs,citeautoscript,superscriptaddress,longbibliography,hyperlinks]{revtex4-2}
\usepackage{amsmath}
\usepackage{amssymb}
\usepackage{amsfonts}
\usepackage[colorlinks=true,citecolor=blue]{hyperref}
\usepackage{graphicx}
\usepackage{enumitem}
\setlist[enumerate]{leftmargin=6mm}
\usepackage[caption = false]{subfig}
\usepackage{braket}
\usepackage{siunitx}
\usepackage{tikz}
\usepackage{bbm}
\usepackage{circuitikz}
\usepackage[capitalise]{cleveref}
\usepackage{booktabs}
\usepackage[normalem]{ulem}

\usepackage{glossaries}
\glsdisablehyper    
\newacronym{mbs}{MBS}{Majorana bound state}
\newacronym{abs}{ABS}{Andreev bound state}
\newacronym{eqo}{EQO}{electron quantum optics}

\newcommand{\up}{\uparrow}
\newcommand{\dw}{\downarrow}
\newcommand{\e}{\mathrm{e}}

\newcommand{\sii}{\hat{\sigma}_{0}}

\newcommand{\ssx}{\hat{s}_{x}}

\newcommand{\ti}{\hat{\tau}_{0}}
\newcommand{\tx}{\hat{\tau}_{x}}

\newcommand{\tz}{\hat{\tau}_{z}}

\newcommand{\mean}[1]{\langle #1 \rangle}

\newcommand{\mbf}[1]{\mathbf{ #1 }}

\DeclareMathOperator{\sgn}{sgn}
\DeclareMathOperator{\tr}{Tr}

\newcommand{\md}{\mathrm{d}}
\newcommand{\mi}{\mathrm{i}}
\newcommand{\me}{\mathrm{e}}

\definecolor{crimson}{RGB}{255,102,255}
\definecolor{mahogany}{RGB}{192,64,0}

\usepackage{sidecap,tikz}
\definecolor{lime}{HTML}{A6CE39}
\DeclareRobustCommand{\orcidicon}{\hspace{-1mm}
	\begin{tikzpicture}
		\draw[lime, fill=lime] (0,0) 
		circle [radius=0.16] 
		node[white] {{\fontfamily{qag}\selectfont \tiny \,ID}};
		\draw[white, fill=white] (-0.0525,0.095) 
		circle [radius=0.007];
	\end{tikzpicture}
	\hspace{-3mm}
}
\foreach \x in {A, ..., Z}{\expandafter\xdef\csname orcid\x\endcsname{\noexpand\href{https://orcid.org/\csname orcidauthor\x\endcsname}
		{\noexpand\orcidicon}}
}

\begin{document}

\title{Floquet-Nambu theory of electron quantum optics with superconductors} 

\author{Pablo Burset\orcidA{}}
\affiliation{Department of Theoretical Condensed Matter Physics\char`,~Universidad Aut\'onoma de Madrid, 28049 Madrid, Spain}
\affiliation{Condensed Matter Physics Center (IFIMAC), Universidad Aut\'onoma de Madrid, 28049 Madrid, Spain}
\affiliation{Instituto Nicol\'as Cabrera, Universidad Aut\'onoma de Madrid, 28049 Madrid, Spain}

\author{Benjamin Roussel\orcidB{}}
\affiliation{Department of Applied Physics,
	Aalto University, 00076 Aalto, Finland}

\author{Michael Moskalets\orcidC{}}
\affiliation{Institute for Cross-Disciplinary Physics and Complex Systems 
	IFISC (UIB-CSIC), 07122 Palma de Mallorca, Spain}
\affiliation{Department of Metal and Semiconductor Physics\char`,~NTU “Kharkiv Polytechnic Institute”, 61002 Kharkiv, Ukraine}

\author{Christian Flindt\orcidD{}}
\affiliation{Department of Applied Physics,
	Aalto University, 00076 Aalto, Finland}
\affiliation{RIKEN Center for Quantum Computing, Wakoshi, Saitama 351-0198, Japan}

\date{\today}

\begin{abstract}
We present a comprehensive Floquet-Nambu theory to describe the time-dependent quantum transport in mesoscopic circuits involving superconductors. The central object of our framework is the first-order correlation function, which accounts for the excitations that are generated by a time-dependent voltage and their coherent scattering off the interface with a superconductor. 
We analyze the time-dependent current generated by periodic voltage pulses and how it depends on the excitation energies of the voltage drive compared to the gap of the superconductor. Our general formalism allows us to identify the conditions for the excitations that are scattered off the superconductor to become coherent electron-hole superpositions. To this end, we consider the purity of the outgoing states, which characterizes their ability to carry quantum information. 
To illustrate our formalism, we apply it to a system composed of chiral quantum Hall edge states connected to a superconductor, and we calculate the current in the outgoing lead and the purity of the outgoing states for Lorentzian and harmonic voltage drives. 
Our framework paves the way for systematic investigations of time-dependent scattering problems involving superconductivity, and it may help interpret future experiments in electron quantum optics with superconductors.
\end{abstract}

\maketitle

\section{Introduction \label{sec:intro}}

\begin{figure*}
\includegraphics[width=1.0\textwidth]{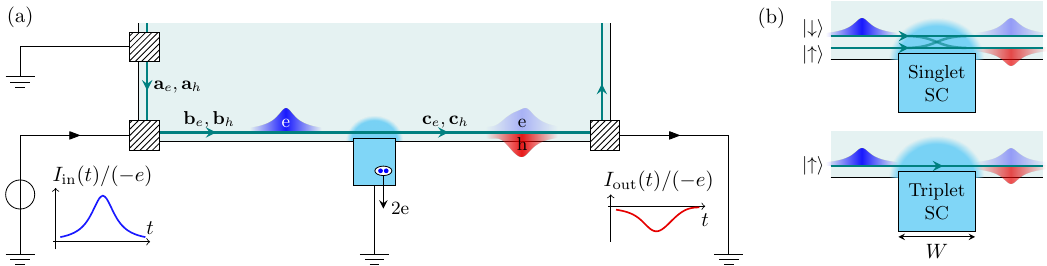}
	\caption{Quantum Hall sample with chiral edge states coupled to a superconductor. 
    (a) Through an external source (circle), voltage pulses are applied to the contact (square box) of a quantum Hall sample with chiral edge states (arrows) that are connected to a superconductor (blue box). 
    The injected electric current ($I_\text{in}(t)$ in blue) is measured in the outputs ($I_\text{out}(t)$ in red) after the incoming charges have been either normal or Andreev scattered at the interface with the superconductor. 
    The operators $\boldsymbol{a}_{e,h}$ describe electrons ($e$) and holes ($h$) before the voltage pulses are applied. The operators $\boldsymbol{b}_{e,h}$ describe particles that have been excited by the pulses, while $\boldsymbol{c}_{e,h}$ correspond to the particles that have reached the outputs. 
    (b) For superconductors of width $W$ and singlet pairing, the spin of an Andreev converted particle is flipped during the scattering process and the particle appears in the other edge channel. For triplet superconductors, the spin is preserved in the scattering process and both normal and Andreev scattered particles remain in the same edge channel. 
}\label{fig:setup} 
\end{figure*}

Electron quantum optics is an emerging field of condensed matter physics which brings ideas and concepts from the quantum theory of light into the domain of high-frequency quantum transport~\cite{Erwann_AdP,Janine_PSS,Waintal_RPP,Weinbub_2022}. Many experiments are carried out with mesoscopic conductors at subkelvin temperatures using the chiral edge states of a quantum Hall sample, which play the role of electronic wave guides for gigahertz charge pulses~\cite{Feve_2007,Bocquillon_2013,Dubois_2013,Jullien:2014,Assouline:2023,Chakraborti2025}. Electrons and holes can be made to interfere in a controllable manner at quantum point contacts~\cite{Dubois_2013,Jullien:2014,Freulon_2015,Assouline:2023,Wang:2023,Fletcher:2023,Ubbelohde:2023,Chakraborti2025}, serving as electronic beam splitters for incoming charges~\cite{Jones_1988,Foxon_1988}. Moreover, several quantum point contacts can be combined to realize fermionic interferometers such as Mach-Zehnder~\cite{Ji_2003,Mailly_2008,Strunk_2008,West_2009,Tewari_2016,Roulleau_2021,Chakraborti2025} and Hanbury Brown-Twiss~\cite{Henny1999,Oliver1999} setups. Based on these ideas, the Hong-Ou-Mandel effect has been observed by emitting single electrons onto each side of a quantum point contact and measuring the noise in the outputs~\cite{Bocquillon_2013,Dubois_2013,Jullien:2014,Freulon_2015,Assouline:2023,Chakraborti2025}. Experiments have also demonstrated wavefunction tomography~\cite{Grenier_2011,Jullien_2014,Fletcher2019,Bisognin_2019,Chakraborti2025} and time-of-flight measurements~\cite{Kataoka_2013,Kataoka_2015,Kataoka_2016,Bauerle_2018}. 

So far, experiments in electron quantum optics have been carried out with normal-state conductors. However, the inclusion of superconductors would unlock a wide range of physical phenomena that cannot be observed in  experiments with photons. In particular, electrons may be transformed into holes through the Andreev scattering on a superconductor, which is a conversion process of a particle into an antiparticle that has no counterpart in quantum optics. 
This electron-hole degree of freedom can for example be exploited for generating quantum entanglement~\cite{Beenakker_2003,Trauzettel_2005,Dasenbrook_2015,Chen2022,Martin_2023,Chen2025}. 
Experimentally, quantum Hall edge states have recently been connected to superconductors~\cite{Zulicke_2005,Ustinov_2007,Schonenberger_2012,Rokhinson_2015,Calado2015,BenShalom2016,Amet_2016,Lee_2017,Ren_2017,Das_2018,Xu_2019,Finkelstein_2020,Das_2021,Kim2022,Shabani_2022,Finkelstein_2023,Vignaud2023,Uday2023}, and supercurrents have been mediated over micrometer distances~\cite{Amet_2016}. Quantum-coherent coupling between a chiral edge state and a conventional $s$-wave superconductor has also been observed~\cite{Finkelstein_2020,Vignaud2023}. Moreover, transport measurements with static voltages have revealed signatures of electron-hole conversion at low filling factors~\cite{Lee_2017,Kim2022,Shabani_2022,Uday2023}. For these reasons, experiments in electron quantum optics with superconductors seem feasible. 
Theoretically, there have been advances in describing the interface between chiral edge states and superconductors~\cite{Beenakker_2011,Beenakker_2014,Clarke_2014,Nazarov_2017,Nazarov_2019,Trauzettel_2019,Idrisov_2020,Akhmerov_2022,Klinovaja_2022,Kurilovich2023,Oreg_2023,Schmidt_2023,Houzet_2023,Balseiro_2023,Baba2025}.  Dynamic quantum transport involving superconductors has also been considered~\cite{Tien_1963,Vanevic_2015,Martin_2019,Sassetti_2020,Vasenko_2020,Martin_2023c,Ronetti2024}.
As such, it appears to be the right moment to expand the field of electron quantum optics by incorporating superconductivity, both in theory and experiments.

In this article, we develop a Floquet-Nambu formalism to evaluate the time-dependent electric current in an edge state coupled to a superconductor. To this end, we extend the dynamic scattering theory of mesoscopic transport with time-dependent voltages by incorporating superconducting correlations. We find the Floquet-Nambu scattering matrix, which describes the excitations generated by voltage pulses as well as their scattering off a superconductor. The central object of our formalism is the associated excess correlation function, which allows us to determine the time-dependent electric current as well as the purity of the outgoing state and, consequently, its usefulness for carrying quantum information. In a companion paper, we make use of this formalism to propose and analyze a tunable electron-hole converter based on superconductors under realistic conditions~\cite{Burset_short}. 

Here, we use our methodology to calculate the time-dependent current that is generated by periodic voltage pulses as illustrated in \cref{fig:setup}. Lorentzian voltage pulses are of particular interest as their amplitude can be tuned so that only an integer number of electrons are excited out of the Fermi sea without any electron-hole excitations. These clean single-particle excitations are known as levitons~\cite{Levitov_2006,Dubois_2013}. We also consider a harmonic drive, which, by contrast, creates electronic states that are accompanied by a cloud of electron-hole excitations. We analyze the time-dependent excess correlation function of the outgoing pulses and demonstrate that levitons with energies below the gap can become coherent superpositions of single electrons and holes after being reflected on the superconductor. On the other hand, with energies above the gap, parts of the charge pulses may be transmitted into the superconductor. We extend these results by exploring multi-charge pulses as well as singlet and triplet pairings in the superconductor. 

We also quantify the usefulness of the outgoing states for quantum information purposes. Specifically, we consider the single-particle purity of the excitations that reach the outputs and how it may be reduced by transmissions into the superconductor above the gap. We also introduce a measure of the purity of states above the Fermi level, which can be used to characterize the single-particle content of the emitted states. We analyze both purities for Lorentzian and harmonic voltage drives and find that the Lorentzian pulses generate highly pure states for a wide range of realistic operating conditions. Our formalism can be applied to a variety of time-dependent scattering problems involving superconductors, and it may be useful for describing future experiments in electron quantum optics with superconductors. 

The rest of the paper is organized as follows. In \cref{sec:FNform}, we describe the scattering setup consisting of chiral edge states coupled to a superconductor as illustrated in \cref{fig:setup}. 
We then develop a Floquet-Nambu formalism that combines dynamic scattering theory with the Nambu description of superconductor interfaces. In \cref{sec:results-1}, we use our methodology to calculate the time-dependent electric current in the outgoing edge states for superconductors that feature either singlet or triplet pairing. In \cref{sec:purity}, we introduce the single-particle purities of the outgoing states, which we illustrate with several examples. In \cref{sec:results-2}, we calculate the purites for Lorentzian and harmonic voltage drives to characterize their usefulness for quantum information purposes. Finally, in \cref{sec:conc}, we present our conclusions together with an outlook on possible developments for the future. Several technical details are presented in the appendices.

\begin{table*}[t]\centering
\setlength{\tabcolsep}{8pt}
\begin{tabular}{@{}ccc@{}} 
 \toprule
\bf Section  & \bf Title & \bf Content and important concepts\\ 
    \midrule
  \ref{sec:overview} & Overview of section & Brief overview of \cref{sec:FNform}.  \\ 
 \ref{subsec:setup} & Scattering setup & Physical setup based on a quantum Hall sample, operators.\\ 
 \ref{subsec:current} & First-order correlation function & Nambu formalism, excess correlation function, and electric current.\\ 
\ref{sec:scattsup} & Scattering off superconductor &  Scattering matrix of the superconductor given by \cref{eq:NS_SM}. \\ 
 \ref{subsec:pulses} & Voltage pulses &  Floquet scattering matrix of voltage pulses.\\ 
 \ref{subsec:FNscat} & Floquet-Nambu scattering matrix & Scattering matrix of superconductor and voltage pulses.\\ 
 \ref{sec:curr}
 & Outgoing current & 
 Excess correlation function, \cref{eq:GF-final}, and time-dependent current, \cref{eq:current-ac-right}.
 \\
\bottomrule
\end{tabular}
\caption{Overview of \cref{sec:FNform} on our Floquet-Nambu scattering formalism and its applications.} 
\label{table:overview}
\end{table*}

\section{Floquet-Nambu formalism}
\label{sec:FNform}

\subsection{Overview of section } 
\label{sec:overview}

In this section, we consider the time-dependent electric current that is generated as incoming charges, which are excited by a time-dependent voltage, are scattered off the interface with a superconductor. \cref{table:overview} provides an overview of the section together with a short description of each subsection. 
These theoretical developments ultimately lead us to the Floquet-Nambu excess correlation function in \cref{eq:GF-final}, which is the key result of this section. From the excess correlation function, we can define observables like the time-dependent current in \cref{eq:current-ac-right}. 

\subsection{Scattering setup} \label{subsec:setup}

We consider the scattering setup in \cref{fig:setup}(a) consisting of two chiral edge states of a quantum Hall sample which are coupled to a superconductor. The edge states act as waveguides for charge pulses that are injected into the circuit by applying voltage pulses to an Ohmic contact. 
Our formalism can readily be extended to describe more external reservoirs as well as more complex scatterers, but here we concentrate on the setup in \cref{fig:setup}(a). 
The chirality of the edge channels simplifies the analysis of the scattering processes, however, it would also be possible to treat non-chiral transport channels within our formalism. 

We denote creation and annihilation operators for the equilibrium particles by $\boldsymbol{a}^{\dagger}_{\sigma}(E)$ and $\boldsymbol{a}_{\sigma}(E)$, with $\sigma=\uparrow,\downarrow$ labelling the spin. Particles in the normal contacts have energy $E$ measured with respect to the chemical potential of the superconductor, $\mu_S$, which is externally fixed and we set it to zero from now on~\cite{Anantram-Datta}. 
To describe superconducting correlations, we introduce hole states (in addition to electron states) and label their fermionic operators by $\boldsymbol{a}_{\alpha\sigma}(E)$, with $\alpha=e$ for electrons and $\alpha=h$ for holes. 
If superconducting correlations couple electrons with opposite spins, as for singlet $s$-wave superconductors, we identify the electron states as $\boldsymbol{a}_{e\sigma}(E)= \boldsymbol{a}_{\sigma}(E)$ and the hole states as $\boldsymbol{a}_{h\sigma}(E)= \boldsymbol{a}^\dagger_{\sigma}(-E)$. 
The same identifications hold for unconventional superconductors, where correlations couple electrons with the same spin into triplet Cooper pairs as shown in \cref{fig:setup}(b). 

The time-dependent voltages are applied to the normal leads far from the superconductor. Moreover, close to the Fermi level, we can linearize the dispersion relation as 
\begin{equation}\label{eq:linear-disp}
	E\simeq \hbar v_F(k-k_F),
\end{equation} 
where $k_F$ is the Fermi wave vector, implying that all relevant excitations propagate with the Fermi velocity~$v_F$. 
The effective Hamiltonian for superconducting correlations in one dimension then reads~\cite{BTK,Anantram-Datta}
\begin{equation}\label{eq:h-BCS}
    \hat{H}_\Delta = \int \! \frac{\md k}{2\pi} 
        \Delta_{\sigma\sigma'}(k) \boldsymbol{a}^{\dagger}_{\sigma}(k)\boldsymbol{a}^{\dagger}_{\sigma'}(-k) + \text{H.c.} 
        , 
\end{equation}
with $\Delta_{\sigma\sigma'}(k) $ the pairing potential. Note that \cref{eq:h-BCS} only considers even-frequency correlations~\cite{Cayao2021Mar,Ahmed2025Jan}. 

As shown in \cref{fig:setup}(a), we denote the creation and annihilation operators for particles that have been excited by the voltage pulses by $\boldsymbol{b}^\dagger_{\alpha\sigma}(E)$ and $\boldsymbol{b}_{\alpha\sigma}(E)$, while the operators $\boldsymbol{c}^\dagger_{\alpha\sigma}(E)$ and $\boldsymbol{c}_{\alpha\sigma}(E)$ correspond to particles that arrive in the outgoing lead. Below, we derive a Floquet-Nambu scattering matrix that connects the operators in the outgoing lead to the equilibrium ones by accounting for the excitations by the voltage drive and the scattering off the superconductor. All averages are then taken with respect to the equilibrium operators, and we have 
\begin{equation}\label{eq:op-averages}
\mean{\boldsymbol{a}^\dagger_{\alpha'\sigma'}(E') \boldsymbol{a}_{\alpha\sigma}(E)} = \delta_{\alpha\alpha'}\delta_{\sigma\sigma'} \delta(E'-E) f_{\alpha}(E) ,
\end{equation}
for $f_{e,h}(E)=f_{\pm V}(E)$, where
\begin{equation}\label{eq:Fermi-eq}
	f_{V}(E)=\frac{1}{\e^{(E- eV)/k_BT}+1}
\end{equation}	
is the Fermi distribution at temperature $T$ with $k_B$ being the Boltzmann constant. Here, we have included a constant voltage offset, $V$, while the time-dependent part will be treated separately in the following sections. We have also used that the annihilation and creation operators fulfill the anticommutation relations
\begin{equation}\label{eq:comm-rels-a1}
		\{\boldsymbol{a}^\dagger_{\alpha\sigma}(E), \boldsymbol{a}_{\alpha'\sigma'}(E') \} = \delta_{\alpha\alpha'}\delta_{\sigma\sigma'}\delta(E'-E)  
\end{equation}
and
\begin{equation}\label{eq:comm-rels-a2}
	\{ \boldsymbol{a}_{\alpha\sigma}(E), \boldsymbol{a}_{\alpha'\sigma'}(E') \} = 
     \delta_{\bar{\alpha}\alpha'} 
    \delta_{\sigma\sigma'}\delta(E'+E),
\end{equation}
with $\bar{\alpha}$ denoting the opposite of $\alpha$. The second anticommutation relation is a consequence of the particle-hole symmetry of the Bogoliubov quasiparticles~\cite{Beenakker_2014}.  

\begin{table*}[t]
\centering
\setlength{\tabcolsep}{10pt}
\begin{tabular}{ccc} 
\toprule
\addlinespace[7pt] 
Symbol & Name & Definition
\vspace{2pt} \\ 
\midrule
\addlinespace[3pt] 
$\hat{\mathcal{G}}(x,t;x't')$  & First-order correlation function & \cref{eq:coherence} \vspace{2pt} \\
$\hat{\mathcal{G}}_\text{off}(x,t;x't')$  & Equilibrium first-order correlation function & \cref{eq:coherence,eq:excess,eq:GF-equilibrium} \vspace{2pt} \\
$\hat{\mathcal{G}}_\text{on}(x,t;x't')$  & First-order correlation function with voltage drive & \cref{eq:coherence,eq:excess,eq:G-on_final} \vspace{2pt} \\
$\hat{G}(x,t;x't')$  & Excess correlation function & \cref{eq:excess,eq:GF-final} \vspace{2pt} \\
$\hat{G}_\text{in,out}(x,t;x't')$  & Excess correlation function on the incoming or outgoing leads & \cref{eq:excess,eq:GF-final} \vspace{2pt} \\
$\hat{G}_l(\omega)$ & Excess correlation function in frequency domain & \cref{eq:G-freq} \vspace{2pt} \\
$Q$ & Transmitted charge per period & \cref{eq:charge} \vspace{2pt} \\
$\boldsymbol{\mathcal{\hat{G}}}$ & First-order correlation function in operator form & \cref{eq:G-operator} \vspace{2pt} \\
$\boldsymbol{\hat{G}}$ & Excess correlation function in operator form & \cref{eq:excess-operator} \vspace{2pt} \\
$\boldsymbol{\hat{G}}_{\epsilon\epsilon'}$ & Projection onto energies above and/or below the Fermi level ($\epsilon,\epsilon'=\pm$)  & \cref{eq:G-projected} \vspace{2pt} \\
$\gamma$ & Single-particle purity of the outgoing state & \cref{eq:purity_1} \vspace{2pt} \\
$\gamma_+$ & Single-particle purity above the Fermi level & \cref{eq:purity_2} \vspace{2pt} \\
\bottomrule
\end{tabular}
\caption{Notation for the Floquet-Nambu formalism. The hat symbol in $\hat{\mathcal{G}}$ or $\hat{G}$ indicates a Nambu (electron-hole) degree of freedom. Bold font, like for $\boldsymbol{\hat{\mathcal{G}}}$ or $\boldsymbol{\hat{G}}$, is used for operators without a specific representation.}
\label{table:notation}
\end{table*}

\subsection{First-order correlation function} \label{subsec:current}

Our goal is to evaluate the time-dependent electric current in the outgoing lead. To this end, we follow the conventional scattering formalism by defining the field operator in the outgoing edge channels as~\cite{Anantram-Datta}
\begin{equation}\label{eq:field-op_1}
	\hat{\Psi} (x,t) = \int \frac{\mathrm{d}E}{\sqrt{h v_F}}
	\e^{-\mi Et/\hbar} \hat{\phi}_E (t,x) \hat{\boldsymbol{c}} (E),
\end{equation}
where $\hat{\boldsymbol{c}} (E)$ generally has the full spin-Nambu structure
\begin{equation}
	\label{eq:fullNambu}
\hat{\boldsymbol{c}}(E)=[\boldsymbol{c}_{e\uparrow}(E), \boldsymbol{c}_{e\downarrow}(E),\boldsymbol{c}_{h\uparrow}(E), \boldsymbol{c}_{h\downarrow}(E)]^T.
\end{equation}
The hat symbol in $\hat{\Psi}$, $\hat{\phi}_E$, or $\hat{\boldsymbol{c}}$ above indicates that operators and matrices have a Nambu (electron-hole) degree of freedom. 
The outgoing eigenstates of electrons and holes are contained in the matrix $\hat{\phi}_E(t,x)$, which we specify below. 
The Nambu degree of freedom introduces a double-counting of fermionic states that we take into account when defining observables like the electric current. 
Indeed, all relevant observables can be obtained from the excess correlation function. To find it, we need the Floquet-Nambu first-order correlation function, which is defined as the expectation value of the tensor product of the field operator in \cref{eq:field-op_1} with itself, 
\begin{equation}\label{eq:coherence}
	\hat{\mathcal{G}} (x,t;x',t') = \mean{ \hat{\Psi}^{\dagger}(x',t') \otimes \hat{\Psi}(x,t)} .
\end{equation}
The excess correlation function then reads~\cite{Grenier_2011,Moskalets_2020,Kotilahti_2021}
\begin{equation}\label{eq:excess}
	\hat{G} (x,t;x',t') = \hat{\mathcal{G}}_\text{on}(x,t;x',t') - \hat{\mathcal{G}}_\text{off}(x,t;x',t') ,
\end{equation}
where the equilibrium contribution $\hat{\mathcal{G}}_\text{off}$ (with the voltage drive turned off) has been subtracted from the correlation function that includes the voltage drive $\hat{\mathcal{G}}_\text{on}$. 
\Cref{eq:excess} contains all the relevant information about the emitted state and is the central object of our Nambu-Floquet formalism. 
For example, observables like the electric current in the outgoing lead can be obtained from the excess correlation function as 
\begin{equation}\label{eq:current-full}
	I(t) = -ev_F \mathrm{Tr} \left\{\hat{G} (x,t;x,t) \tz \sii \right\}/2 ,
\end{equation}
where the Pauli matrices $\hat{\sigma}_{i}$ and $\hat{\tau}_{i}$ act on the spin and Nambu degrees of freedom, respectively, and $\sii$ and $\ti$ are identities. 
The excess correlation function at $x'=x$ and $ t'=t$ provides the density of electrons and holes that propagate with the Fermi velocity $v_F$ at a particular time and place. The Pauli matrix $\hat{\tau}_{z}$ inside the trace accounts for the opposite charge of electrons and holes, while the factor of one half compensates for the double-counting because of the Nambu structure. Due to the linear dispersion, \cref{eq:linear-disp}, we are free to consider the current at any position in the outgoing lead. 

In the following, we consider superconductors with either singlet or triplet pairings as shown in \cref{fig:setup}(b). Both the singlet and the non-polarized triplet pairings couple electrons and holes with opposite spins, which effectively decouples \cref{eq:fullNambu} into two separate spin channels that can be described by two reduced Nambu spinors reading
\begin{equation}\label{eq:spin-channels}
	\begin{split}
	\hat{\boldsymbol{c}}_\uparrow (E)=[\boldsymbol{c}_{e\uparrow}(E), \boldsymbol{c}_{h \downarrow }(E)]^T,\\
	\hat{\boldsymbol{c}}_\downarrow (E)=[\boldsymbol{c}_{e\downarrow}(E), \boldsymbol{c}_{h \uparrow }(E)]^T.
	\end{split}
\end{equation} 
In this case, the non-zero components of the pairing potential in \cref{eq:h-BCS} are $\Delta_{\up\dw}$ and $\Delta_{\dw\up}$. 
Similarly, the polarized triplet pairings with $\Delta_{\sigma\sigma}\neq0$ and $\Delta_{\up\dw}=\Delta_{\dw\up}=0$ couple electrons and holes with the same spin, and the reduced Nambu spinors can be modified accordingly. Still, in the absence of any effects that break spin-rotational symmetry resulting in triplet pairings that mix polarized and non-polarized states, we may work with two uncoupled spin channels. Thus, to simplify the notation, we replace the operator in \cref{eq:fullNambu} by the~reduced~Nambu~spinor
\begin{equation}
	\label{eq:redNambu}
	\hat{\boldsymbol{c}}(E)=[\boldsymbol{c}_{e}(E), \boldsymbol{c}_{h}(E)]^T  
\end{equation}
for each decoupled spin channel, where the explicit spin index has been omitted. 
For the reduced notation, the tensor product in \cref{eq:coherence} is defined as 
\begin{equation}
	\label{eq:tensorprod}
    \hat{\boldsymbol{c}}^\dagger (E') \otimes \hat{\boldsymbol{c}} (E) = \begin{pmatrix}
        \boldsymbol{c}_{e}^\dagger (E') \boldsymbol{c}_{e} (E) & 
        \boldsymbol{c}_{h}^\dagger (E') \boldsymbol{c}_{e} (E) \\
        \boldsymbol{c}_{e}^\dagger (E') \boldsymbol{c}_{h} (E) & 
        \boldsymbol{c}_{h}^\dagger (E') \boldsymbol{c}_{h} (E)
    \end{pmatrix} .
\end{equation}
Correspondingly, the matrix in \cref{eq:field-op_1} reduces to
\begin{equation}
	\hat{\phi}_E(t,x) = [ \hat{\phi}_{e}(E;t,x), \hat{\phi}_{h}(E;t,x)],
    \label{Nm}
\end{equation}
which contains the outgoing eigenstates for electrons and holes as its columns. Specifically, we have
\begin{equation}
    \hat{\phi}_{e}(E;t,x)= \e^{-\mi \mu t/\hbar} \begin{pmatrix} 1 \\ 0 \end{pmatrix} \e^{\mi k_{e}(E)x}, 
    \label{eq:normal-eigenstates1}
\end{equation}
and
\begin{equation}
\hat{\phi}_{h}(E;t,x)= \e^{\mi \mu t/\hbar} \begin{pmatrix} 0 \\ 1 \end{pmatrix} \e^{-\mi k_{h}(E)x} ,
\label{eq:normal-eigenstates2}
\end{equation}
where $\mu$ is the chemical potential in the outgoing leads, shown for the sake of clarity. However, in what follows, we only consider cases where the chemical potential in the outgoing channels is zero. 
We have also assumed that the chiral propagation is in the positive direction of the $x$-axis. 
The reduced Nambu matrix then reads
\begin{equation}
	\label{eq:NambuMnormal}
\hat{\phi}_E (x) = \begin{pmatrix} \e^{\mi k_{e}(E)x} & 0 \\ 0 & \e^{-\mi k_{h}(E)x} \end{pmatrix},
\end{equation} 
where the electrons and holes have the wavenumbers
\begin{equation}\label{eq:wavenumber-N}
k_{e,h}(E) = k_F \pm E/(\hbar v_F).
\end{equation} 

In the reduced notation, the electric current in the outgoing lead becomes 
\begin{equation}\label{eq:current}
	I(t) = -g_s ev_F \mathrm{Tr}_\text{N} \left\{\hat{G} (x,t;x,t) \tz \right\}/2 ,
\end{equation}
where $\mathrm{Tr}_\text{N}$ is the trace in the reduced Nambu space, and $g_s=1,2$ is the spin degeneracy factor for triplet and singlet pairings. 
Finally, we can associate a spin-singlet pairing in the superconductor with the quantum Hall filling factor of 2, where the two spin orientations are possible~\cite{Schmidt_2023,Houzet_2023,Balseiro_2023,Baba2025}. By contrast, the chiral state is spin-polarized at filling factor 1, which we describe as a polarized spin-triplet pairing according to \cref{fig:setup}(b). 

\subsection{Scattering off superconductor}
\label{sec:scattsup}

Next, we relate the creation and annihilation operators in the outgoing lead to those that describe particles before the superconductor. Eventually, they will be connected to the operators that describe particles in equilibrium before the voltage pulses are applied. Since these processes conserve energy, we seek a relation of the form
\begin{equation}
	\label{eq:sc_scat}
\hat{\boldsymbol{c}}(E)	= \hat{S}(E)\hat{\boldsymbol{b}}(E)  ,
\end{equation}
which describes the scattering of charges on the superconductor, encoded in the Nambu scattering matrix
\begin{equation}
	\label{eq:NSM}
\hat{S}(E)= \begin{pmatrix}
S_{ee} & S_{eh} \\ 
S_{he} & S_{hh} 
	\end{pmatrix}.
\end{equation}
The electron-hole transformation imposes the following constraint on operators in the full Nambu space, \cref{eq:fullNambu}, 
\begin{equation}\label{eq:phs}
	\hat{\boldsymbol{c}}(E)= \tx \sii [ \hat{\boldsymbol{c}}^\dagger(-E) ]^T,
\end{equation}
where the Pauli matrices $\hat{\sigma}_{i}$ and $\hat{\tau}_{i}$ respectively act on the spin and the Nambu spaces. 

\begin{figure}[t!]
	\includegraphics[width=1.0\columnwidth]{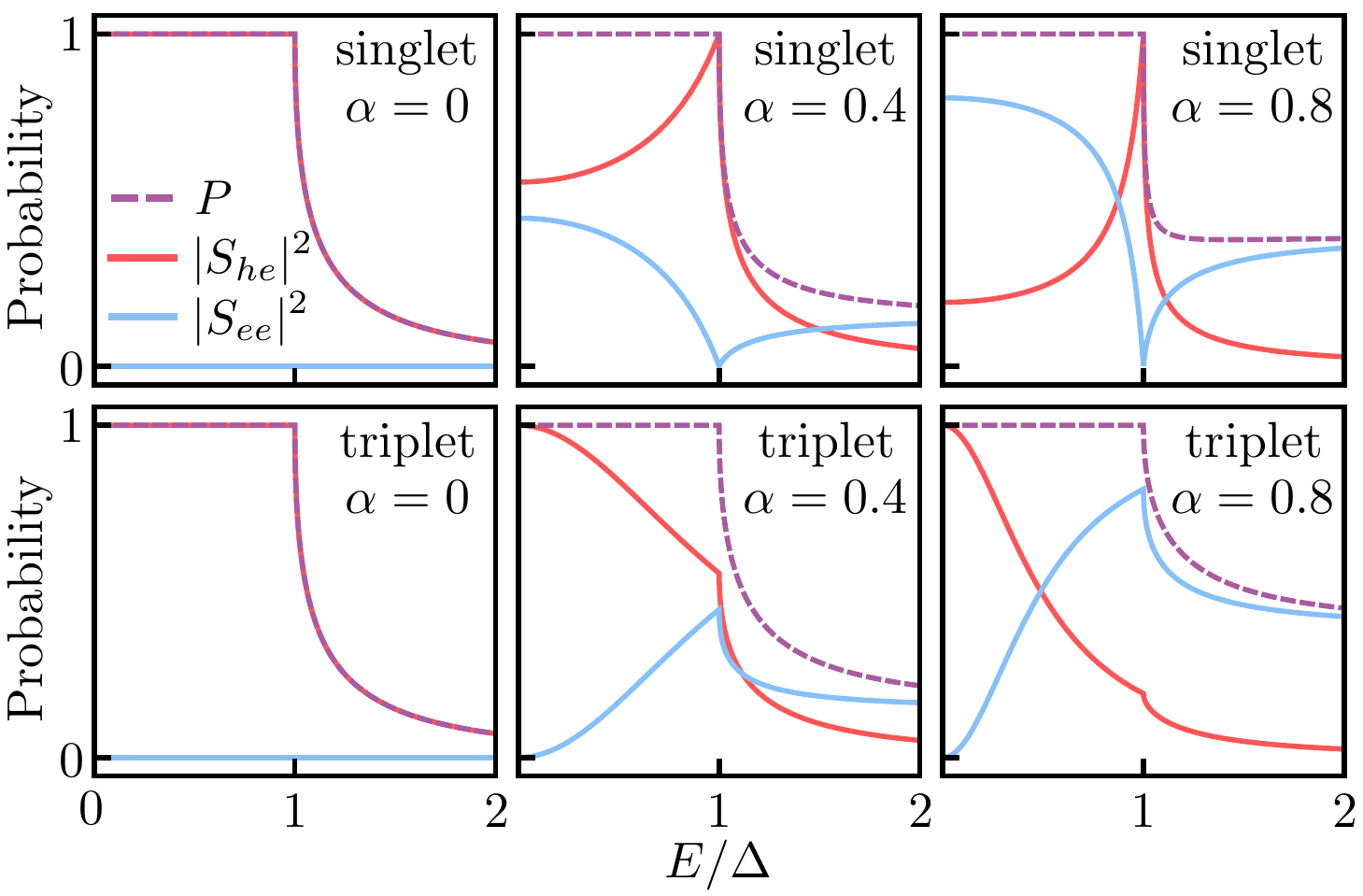}
	\caption{\label{fig:support-1}
		Scattering off the superconductor. Probabilities for Andreev conversion, $|S_{he}|^2$, and normal transmission, $|S_{ee}|^2$, in \cref{eq:NS_SM} and their sum, $P=|S_{he}|^2+|S_{ee}|^2$, as a function of the energy normalized to the gap of the superconductor. The top and bottom rows show results for singlet and triplet pairings, respectively, with different degrees of conversion $\alpha$.
	}
\end{figure}

The symmetry operation for spinless triplet pairings adopts the same form in the space spanned by \cref{eq:redNambu},  
\begin{equation}\label{eq:phs_t}
	\hat{\boldsymbol{c}}(E)= \tx [\hat{\boldsymbol{c}}^\dagger(-E)]^T,
\end{equation}
with a similar transformation for $\hat{\boldsymbol{b}}(E)$. Consequently, the scattering matrix for spinless triplet pairings fulfills the electron-hole symmetry 
\begin{equation}\label{eq:phs-t_SM}
    \hat{S}(E) = \tx \hat{S}^*(-E) \tx .
\end{equation}

For singlet pairings, the electron-hole transformation mixes the decoupled spin channels in \cref{eq:spin-channels} and reads
\begin{equation}\label{eq:phs_s-red}
	\hat{\boldsymbol{c}}_{\sigma}(E)= 
    \ssx
    [\hat{\boldsymbol{c}}_{\bar\sigma}^\dagger(-E)]^T ,
\end{equation}
with $\bar\sigma=\dw,\up$ for $\sigma=\up,\dw$ and the Pauli matrices $\hat{s}_{i}$ acting on the reduced Nambu space. The symmetry of the scattering matrix also mixes channels as
\begin{equation}\label{eq:phs-s_SM}
    \hat{S}_\sigma(E) = \ssx \hat{S}_{\bar\sigma}^*(-E) \ssx ,
\end{equation}
where $\hat{S}_\sigma$ is the scattering matrix in \cref{eq:NSM} for spin channel $\hat{\boldsymbol{c}}_\sigma$ according to \cref{eq:spin-channels}. 
In the following, we use the reduced notation in \cref{eq:redNambu} and take into account the electron-hole transformations for each spin channel. 

To find the scattering matrix of the superconductor, we match the solutions at the interface~\cite{BTK,Anantram-Datta,Beenakker_2011}. 
The eigenstates for quasiparticles in the superconductor are
\begin{equation}
\hat{\phi}^S_{e} (E,x)= \frac{ \e^{-\mi k^{S}_{e}(E)x} }{ \sqrt{\mathcal{N}(E)} } 
\begin{pmatrix} 1 \\ \Lambda(E) \end{pmatrix}, 
\label{eq:super-eigenstates1}
\end{equation}
and
\begin{equation}
\hat{\phi}^S_{h} (E,x)= \frac{\e^{\mi k^{S}_{h}(E)x}}{\sqrt{\mathcal{N}(E)}}  
\begin{pmatrix} \eta \Lambda(E) \\ 1 \end{pmatrix},
\label{eq:super-eigenstates2}
\end{equation}
where $\eta=\pm 1$ distinguishes singlet ($\eta=+1$) and triplet ($\eta=-1$) superconductors. We have also defined 
\begin{equation}\label{eq:lambda}
	\Lambda(E) = \frac{|\Delta|}{E+\zeta(E)} 
\end{equation}
and
\begin{equation}\label{eq:N(E)}
\mathcal{N}(E)= 1 + |\Lambda(E)|^2.
\end{equation}
We note that a singlet superconductor requires that we change $\Lambda(E)$ to $s_\sigma\Lambda(E)$ in \cref{eq:super-eigenstates1,eq:super-eigenstates2}, with $s_\sigma=+1,-1$ for the spin channels $\sigma=\up,\dw$ according to \cref{eq:spin-channels}. Since we only consider spin-degenerate cases for singlet pairing, this sign is not important in the following. Consequently, to keep the discussion simple, we only consider the case $s_\up=+1$. We have also expressed the wave numbers for electrons and holes as
\begin{equation}\label{eq:wavenumber-S}
	k^S_{e,h} = k_F \pm \zeta(E)/(\hbar v_F),
\end{equation}
where the function
\begin{equation}\label{eq:zeta}
	\zeta(E)= \left\{ \begin{array}{cr}
		\sgn\left(E\right)\sqrt{E^2-|\Delta|^2}, & |E|>|\Delta| \\ 
		\mi\sqrt{|\Delta|^2-E^2}, & |E|\leq|\Delta|
	\end{array}\right. 
\end{equation}
is real outside the gap and imaginary inside the gap, where it describes the decay of the wavefunction. For a vanishing gap, $\Delta=0$, we recover the eigenvectors in \cref{eq:normal-eigenstates1,eq:normal-eigenstates2} since $\Lambda=0$ and $\mathcal{N}=1$.

\begin{table*}[t]
\centering
\setlength{\tabcolsep}{10pt}
\begin{tabular}{ccc} 
\toprule
\addlinespace[7pt] 
Voltage  & $eV_\text{ac}(t)$  & Floquet scattering amplitude, $J_n$ 
\vspace{5pt}\\ 
\midrule
\addlinespace[10pt] 
Harmonic & $eV_0\cos(\Omega t)$  &  $\displaystyle \mathcal{J}_n (z)=\frac{1}{2\pi}\int_{-\pi}^{\pi}d\tau\me^{\mi(n\tau- z\sin\tau)},\,\, z=eV_0/\hbar \Omega$ \vspace{10pt}\\
  \parbox[c]{3cm}{Leviton train \\($q\in \mathbb{R}$)} & 
    $ \displaystyle
    -\sum_{m=-\infty}^{\infty}\frac{ 2q\hbar\tau_0}{(t-m\mathcal{T})^2+\tau_0^2}+q \hbar\Omega 
    $
    &  
        $
        \displaystyle
        \begin{array}{cr} \displaystyle
	    \sum\limits_{k=0}^{\infty}  \frac{(-1)^{q+k} \me^{-(2k+n) \Omega\tau_0}\Gamma(q+n+k)}{\Gamma(k+1)\Gamma(q+1-k)\Gamma(k+n+1)} , & n\geq 0 \vspace{5pt} \\
     \displaystyle
            \sum\limits_{k=0}^{\infty} \frac{(-1)^{q+k+n} \me^{-(2k+|n|) \Omega\tau_0} \Gamma(q+k)}{\Gamma(k+1)\Gamma(q+1-k-|n|)\Gamma(k+|n|+1)} , & n<0
	\end{array}
        $ \vspace{10pt}\\
        \vspace{10pt}
    \parbox[c]{3cm}{Leviton train \\($q=1$)} & 
    $\displaystyle -\sum_{m=-\infty}^{\infty}\frac{ 2\hbar\tau_0}{(t-m\mathcal{T})^2+\tau_0^2} $ &
     $
        \displaystyle
        \begin{array}{cr} 
	    -2 \sinh (\Omega\tau_0) \me^{-n\Omega\tau_0} ,  & n>0 \\
            \me^{-\Omega\tau_0}, &  n=0 \\
                0, & n<0 
	\end{array}
        $
   \\
   \vspace{5pt}
   \parbox[c]{3cm}{Single leviton\\ \mbox{($q=1, \Omega\tau_0\ll 1$)}}
     & $\displaystyle -\frac{2\hbar\tau_0} {t^2+\tau_0^2}$ & 
    $
        \displaystyle
        \begin{array}{cr} 
	    -2\Omega\tau_0 \me^{-n\Omega\tau_0},  & n>0 \\
            1-\Omega\tau_0, &  n=0 \\
                0, & n<0 
	\end{array}
        $ \\
\bottomrule
\end{tabular}
\caption{Time-dependent voltages and their Floquet scattering amplitudes. The Gamma function is denoted by $\Gamma(x)$. }
\label{table:voltages}
\end{table*}

Following Ref.~\cite{BTK}, the scattering matrix of the contact between the chiral edge channel and the superconductor can now be written as
\begin{equation}\label{eq:NS_SM}
	\hat{S}(E) = \frac{\begin{pmatrix}
			\left( 1 - \eta \me^{-2\mi \gamma} \right) \alpha & 
			\left(1-\alpha^2\right) \me^{-\mi\gamma} \me^{\mi\phi} \\ 
			\eta\left(1-\alpha^2\right) \me^{-\mi\gamma} \me^{-\mi\phi} & 
			\left( 1 -\eta \me^{-2\mi\gamma} \right) \alpha
	\end{pmatrix}}{ 1- \alpha^2 \eta \me^{-2\mi\gamma} } ,
\end{equation}
where $\phi$ is the superconducting phase, and the energy dependence is contained in the angle $\gamma = \cos^{-1}(E/\Delta)$ for $|E|\leq\Delta$, and $\gamma = -\mi \cosh^{-1}(E/\Delta)$ for $E>\Delta$. 
The parameter $\alpha$ controls the degree of electron-hole mixing at the interface, with $\alpha = 0$ for perfect Andreev conversion and $\alpha = 1$ for full normal transmission as seen in \cref{fig:support-1}. 
The electron-hole mixing is governed by the ratio of the width $W$ of the contact region over the coherence length $\xi=\hbar v_\text{F}/\Delta$ [\cref{fig:setup}(b)], assuming that the magnetic penetration length is shorter than $\xi$, which is also smaller than the magnetic length of the quantum Hall edge state~\cite{Beenakker_2011,Beenakker_2014,Schmidt_2023}. 
We note that the model of Ref.~\cite{BTK} has been used to explain several experiments by assuming that the parameter $\alpha$ is related to the electron-hole mixing taking place along the interface between the superconductor and the edge states~\cite{Lee_2017}. 

For energies well inside the gap, we get $\me^{\mi\gamma}\simeq \mi$, and we can write the scattering matrix as
\begin{equation}
\label{eq:NS_SM-expand}
	\hat{S}(E) \simeq \frac{\begin{pmatrix}
			\left( 1+\eta \right) \alpha & 
			\mi\left(\alpha^2-1\right) e^{\mi\phi} \\ 
			\mi\eta\left(\alpha^2-1\right) e^{-\mi\phi} & 
			\left( 1+\eta \right) \alpha
	\end{pmatrix}}{1+\alpha^2 \eta}, \,\,
 |E|\ll|\Delta| .
\end{equation}
For  the triplet case ($\eta=-1$), we then find
\begin{equation}\label{eq:NS_SM-expand-trip}
	\hat{S}(E) \simeq  
	\begin{pmatrix}
			0 & 
		    -\mi\me^{\mi\phi} \\ 
		    \mi\me^{-\mi\phi} & 
			0
	\end{pmatrix},\,\,|E|\ll|\Delta|,
\end{equation}
showing that only Andreev conversions occur. For singlet pairings ($\eta=1$), that only happens for $\alpha=0$. 

\subsection{Voltage pulses} 
\label{subsec:pulses}

We are now ready to describe the injection of charge pulses from the source electrode. To this end, we consider periodic voltage pulses of the form
\begin{equation}
V(t)= V + V_\text{ac}(t),
\label{eq:vol_decomp}
\end{equation}
where $V$ is a constant offset, while
\begin{equation}
	 V_\text{ac}(t)=V_\text{ac}(t+\mathcal{T}) 
\end{equation}
is time-dependent with period $\mathcal{T}$. The offset $V$ can be included in the Fermi functions, and we now focus on the time-dependent part $ V_\text{ac}(t)$. As we discuss below, there is some freedom in how the voltage pulses are split into an offset and a time-dependent part, and the offset may in some cases be included in the time-dependent part.

The drive can be accounted for by Floquet theory, which describes how particles change their energy~as
\begin{equation}
	E\rightarrow E_{n}=E+n\hbar\Omega,
\end{equation}
where $\Omega=2\pi/\mathcal{T}$ is the frequency of the drive, and the integer $n$ can be both positive and negative~\cite{Moskalets_book,Kotilahti_2021}. 
The periodic voltage introduces the phase factor
\begin{equation}\label{eq:Fourier-trans-Vac}
	J(t) = \e^{ \mi  \int_{-\infty}^{t}\mathrm{d}t' eV_{\mathrm{ac}}(t')/\hbar} 
	=  \sum\limits_{n}J_n \e^{-\mi n\Omega t},
\end{equation}
which can be expanded in  its Fourier components 
\begin{equation}\label{eq:Fourier-coeffs}
	J_n= \frac{1}{\mathcal{T}}\int_{0}^{\mathcal{T}}  \mathrm{d}t J(t)\me^{\mi n\Omega t}.
\end{equation}
Those are the probability amplitudes for a particle to absorb $n$~energy quanta of size~$\hbar\Omega$. 
Since the pulses are applied to the input lead, far from the superconductor, the excitations at the drive are independent from the scattering at the superconductor. Specifically, the operators for particles that propagate towards the superconductor can be written in terms of the equilibrium ones as
\begin{equation}\label{eq:ferm-tdep}
		\boldsymbol{b}_{e}(E) = \sum\limits_{n=-\infty}^{\infty} J_n \boldsymbol{a}_{e}(E_{-n}) 
\end{equation}
and
\begin{equation}
				\boldsymbol{b}_{h}(E) = \sum\limits_{n=-\infty}^{\infty} J^*_{-n} \boldsymbol{a}_{h}(E_{-n})  , 
\end{equation}
where we have used that an electron with energy $-E$ corresponds to a hole with energy $+E$.

In the Nambu notation, we can relate the annihilation operators before and after the drive as 
\begin{equation}\label{eq:ops-after-drive}
	\hat{\boldsymbol{b}} (E)= \sum\limits_{n=-\infty}^{\infty} \hat{J}_{n} \hat{\boldsymbol{a}}(E_{-n}) ,
\end{equation}
where the matrix
\begin{equation}
\hat{J}_{n} = 
\begin{pmatrix}
	J_{n} & 
	0 \\ 
	0 & 
	J^*_{-n}
\end{pmatrix}
\end{equation}
contains the  amplitudes for electron and hole excitations.

We now consider a series of Lorentzian voltage pulses 
\begin{equation}
\label{eq:volt-lor}
\begin{split}
eV(t)=&-\sum_{m=-\infty}^{\infty}\frac{ 2q\hbar\tau_0}{(t-m\mathcal{T})^2+\tau_0^2}\\
=& -q\hbar\Omega\frac{\sinh(\Omega\tau_0)}{\cosh(\Omega\tau_0)-\cos(\Omega t)},
    \end{split}
\end{equation}
where $\tau_0$ denotes the half-width and $q$ is the average charge per pulse. To proceed, we decompose the voltage as in \cref{eq:vol_decomp} with the constant offset $eV= -q \hbar \Omega$ if $q$ is not an integer. For integer values, the constant voltage can be included in the time-dependent part. For example, for Lorentzian pulses with one  charge per period, the Floquet scattering amplitudes for the total voltage can be related to those without the constant offset by changing the index as $n\rightarrow n-1$~\cite{hofer:2014}. The  Floquet scattering amplitudes~\cite{Dubois_2013b} are listed in \cref{table:voltages} together with those for a harmonic drive given by Bessel functions. 

The Lorentzian voltage pulses with integer charges are special, since $J_{n}=0$ for $n<0$, which implies that no particles below the Fermi sea are excited. These pulses generate clean few-electron excitations that are emitted on top of the Fermi sea. By contrast, for non-integer values of the average charge, both electrons and holes are emitted from the electrode. For a pulse with exactly one charge per period, the Floquet scattering amplitudes become particularly simple as shown in \cref{table:voltages}. Moreover, to describe the emission of a single electron, we can take the period to be much longer than any other time scale in the system, including $\Omega\tau_0\ll 1$, and we then obtain the expression in \cref{table:voltages}.

\begin{figure}	\includegraphics[width=1.0\columnwidth]{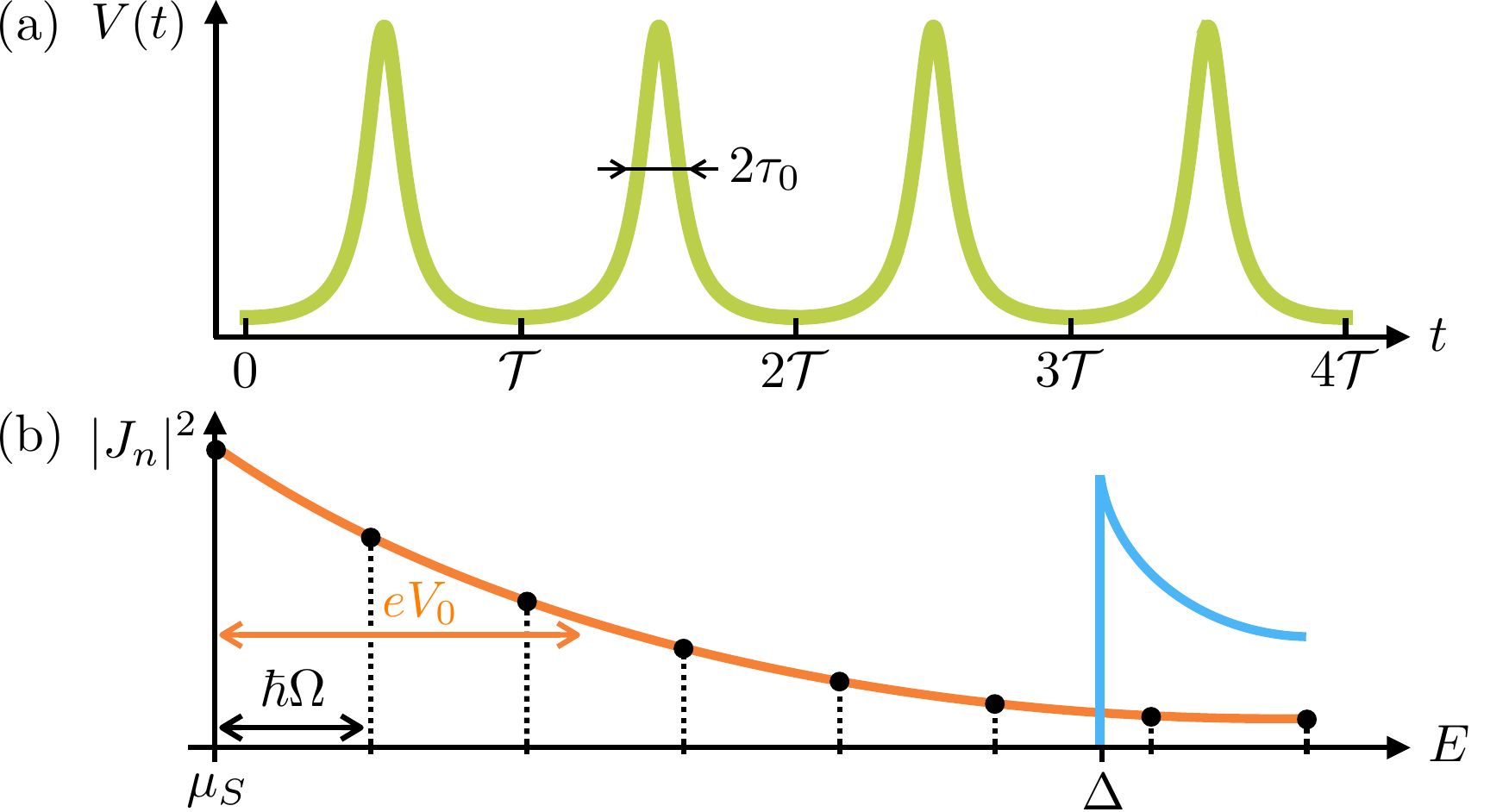}
\caption{\label{fig:support-2} 
     Voltage pulses. (a) Lorentzian voltage pulses with period $\mathcal{T}$, half-width $\tau_0$, and amplitude $eV_0=2q\hbar/\tau_0$. 
     (b) Excitation energies compared with the gap, $\Delta$, and the density of states of the superconductor (in blue). 
    }
\end{figure}

\Cref{fig:support-2}(a) shows the periodic Lorentzian voltages pulses given by \cref{eq:volt-lor}. The specific shape of the pulses is determined by their period, half-width, and amplitude. 
As shown in \cref{fig:support-2}(b), the driving frequency determines the possible excitation energies, $E_n=E+n\hbar\Omega$, while the width and the amplitude control the probability amplitudes in \cref{eq:Fourier-coeffs}. The excitation energies should be compared to the superconducting gap, since that will determine if quasiparticles are transmitted into the superconductor. 
Because the sum in \cref{eq:ops-after-drive} extends to infinity, there is a finite probability of exciting particles over the gap for any value of~$\Delta$. In practice, however, the drive amplitude determines how fast the excitation probabilities~$|J_n|^2$ decay with the excitation number $n$ as illustrated in \cref{fig:support-2}(b). 
Technically, we typically need to include between 10 and 100 probability amplitudes to ensure convergence of our calculations. Experiments in electron quantum optics operate in the gigahertz regime using small amplitudes so that only a few particles are excited~\cite{Aluffi2023}. 
Consequently, in the next sections, we set $eV_0\lesssim\Delta$ so that the ratio $\hbar\Omega/\Delta$ determines if transport is mostly subgap ($\hbar\Omega\ll\Delta$), or if there is a finite probability of exciting particles over the gap~($\hbar\Omega\gtrsim\Delta$). 

\subsection{Floquet-Nambu scattering matrix} \label{subsec:FNscat}

We now assemble the Floquet-Nambu scattering matrix that relates the operators in the outgoing lead to the equilibrium operators before the voltage drive as seen in \cref{fig:setup}(a). Thus, we combine \cref{eq:sc_scat,eq:ferm-tdep} and find
\begin{equation}
	\label{eq:fn_scat}
	\begin{split}
	\hat{\boldsymbol{c}}(E)	&=  \sum\limits_{n=-\infty}^{\infty} \hat{S}(E)\hat{J}_{n} \hat{\boldsymbol{a}}(E_{-n})\\
	& =\sum\limits_{n=-\infty}^{\infty} \hat{S}(E)\hat{J}_{-n} \hat{\boldsymbol{a}}(E_{n})\\
	&=\sum\limits_{n=-\infty}^{\infty} \hat{S}_F(E,E_{n}) \hat{\boldsymbol{a}}(E_{n}),
		\end{split}
\end{equation}
having introduced the Floquet-Nambu scattering matrix
\begin{equation}\label{eq:scat-mat-simp1}
	\hat{S}_F(E,E_n)= \hat{S}(E) \hat{J}_{-n},
\end{equation}
which can also be written as
\begin{equation}\label{eq:scat-mat-simp2}
	\hat{S}_F (E_n,E) = \hat{S}(E_n) \hat{J}_n.
\end{equation}
The Floquet-Nambu scattering matrix describes the generation and scattering of charge pulses on the superconductor, and it is the essential building block for the calculations of the time-dependent current. 

To find the correlation function in \cref{eq:coherence} we need to evaluate averages of the scattered particles as follows
\begin{equation}
\label{eq:op-avg}   
\begin{split}    
\mean{\boldsymbol{c}^\dagger_{\alpha}(E')\boldsymbol{c}_{\beta}(E)} = \sum\limits_{n,m} \delta(E'_m -E_n) \sum\limits_{\nu=e,h} f_{\nu}(E_n) \\
     \times  [S_F^*(E_{n-m},E_n)]_{\alpha\nu} [S_F(E,E_n)]_{\beta\nu},
\end{split}
\end{equation}
where we have used \cref{eq:op-averages} for the expectation value in equilibrium. The Floquet-Nambu scattering matrix must be unitary so that the operators in the outgoing leads, which are out of equilibrium, comply with the anticommutation rules in \cref{eq:comm-rels-a1,eq:comm-rels-a2}~\cite{Moskalets_2002}. 
Therefore, the Floquet-Nambu scattering matrix must fulfill that
\begin{equation}
\label{eq:unitarity_2}
    \begin{split}
        \delta_{\alpha\beta} \delta_{jk} = & 
	\sum\limits_{n} \sum\limits_{\nu} [S_F(E_{j},E_n)]_{\alpha\nu} [S_F^*(E_{k},E_n)]_{\beta\nu} \\
	= & \sum\limits_{n} \sum\limits_{\nu} [S_F^*(E_n,E_{j})]_{\nu\beta} [S_F(E_n,E_{k})]_{\nu\alpha} .
    \end{split}
\end{equation}

\subsection{Outgoing current}
\label{sec:curr}

The final step is to define the field operators to compute the Floquet-Nambu first-order correlation function and thereby find the time-dependent electric current. By substituting \cref{eq:fn_scat} into \cref{eq:field-op_1}, we obtain
\begin{equation}\label{eq:field-op_2}
    \begin{split}
		\hat{\Psi} (x,t)=& \int \frac{\mathrm{d}E}{\sqrt{h v_F}} \e^{-\mi Et/\hbar}\hat{\phi}_E(x) \\ & \times\sum\limits_{n=-\infty}^{\infty} \hat{S}_F(E,E_{n}) \hat{\boldsymbol{a}}(E_{n}).
   \end{split}
\end{equation}
We can then write the correlation function as
\begin{widetext}
\begin{equation}\label{eq:correlator-on}
		\hat{\mathcal{G}}_\text{on} (x,t;x',t') = 
		\iint\frac{\mathrm{d}E\mathrm{d}E'}{h v_F}  \e^{\mi(E't'-Et)/\hbar} 
	\sum_{n,m} \langle [\hat{\phi}_{E'}(x')\hat{S}_F(E',E'_{m})\boldsymbol{\hat a}(E'_m)]^\dagger\otimes [\hat{\phi}_E(x)\hat{S}_F(E,E_{n})\boldsymbol{\hat a}(E_n)]\rangle. 
\end{equation}
Using \cref{eq:scat-mat-simp1} for the Floquet-Nambu scattering matrix, we find 
\begin{equation}
	\hat{\mathcal{G}}_\text{on} (x,t;x',t')  = 
\iint\frac{\mathrm{d}E\mathrm{d}E'}{h v_F}  \e^{\mi(E't'-Et)/\hbar} 
\sum_{n,m} \sum_{\mu=e,h} \langle \boldsymbol{\hat a}^\dagger_\mu(E'_m) \boldsymbol{\hat a}_\mu(E_n) \rangle [\hat J_{-m}^\dagger]_{\mu\mu} [\hat J_{-n}]_{\mu\mu} \hat{M}_{\mu\mu}(E',E;x',x), 
\end{equation}
where we have defined the Nambu matrices 
\begin{equation}\label{eq:M-matrix}
	\hat{M}_{\mu\nu}(E',E;x',x) = 
		\begin{pmatrix}
			\hat{\phi}_e^*(E',x') [\hat{S}]^*_{e\mu}(E') & \hat{\phi}_h^*(E',x') [\hat{S}]^*_{h\mu}(E')
		\end{pmatrix} 
		\otimes
		\begin{pmatrix}
			\hat{\phi}_e(E,x) [\hat{S}]_{e\nu}(E) \\ \hat{\phi}_h(E,x) [\hat{S}]_{h\nu}(E)
		\end{pmatrix} .
\end{equation} 
Evaluating the equilibrium averages according to \cref{eq:op-averages} together with a change of variables, we then obtain 
\begin{equation}\label{eq:G-on_final}
	\begin{split}	
		\hat{\mathcal{G}}_\text{on} (x,t;x',t')  
		 = &
		\int\frac{\mathrm{d}E}{h v_F}  \e^{-\mi E(t_x-t'_{x'})/\hbar} \sum_{n,m} 
		\e^{-\mi n\Omega  t_x} \me^{\mi \Omega m t'_{x'}} \\
		& \times \left\{ 
		      f_{V}(E)  J_{n} J^*_{m} \hat{M}_{ee}(E_{m},E_{n};x',x) +
		      f_{-V}(E)  J_{-n}^* J_{-m} \hat{M}_{hh}(E_{m},E_{n};x',x) 
		 \right\}, 
	\end{split}
\end{equation}
To find the correlation function without an applied voltage, $\hat{\mathcal{G}}_\text{off} (t_x;t'_{x'})$, we set $J_n=\delta_{n0}$ and $V=0$ in the expression above. The first-order excess correlation function is then given by their difference, \cref{eq:excess}, becoming
\begin{equation}
	\label{eq:GF-final}
	\begin{split}
		\hat{G} (x,t;x',t') = &
		\int\frac{\mathrm{d}E}{h v_F}  \sum_{n,m} \left\{ 
		\left[ f_{V}(E)  - f_{0}(E_n) \right] J_{n} \e^{-\mi E_n t_x/\hbar} J^*_{m} \e^{\mi E_m t'_{x'}/\hbar} \hat{M}_{ee}(E_{m},E_{n};x',x) \right. 
		\\
		&
        \left. +
		\left[ f_{-V}(E) - f_{0}(E_{-n}) \right] J^*_{n} \e^{-\mi E_{-n} t_x/\hbar} J_{m} \e^{\mi E_{-m} t'_{x'}/\hbar} \hat{M}_{hh}(E_{-m},E_{-n};x',x) 
		\right\}. 
	\end{split}
\end{equation}
Here, we have made use of the unitarity of the Floquet-Nambu scattering matrix as expressed by \cref{eq:unitarity_2}, which in our case becomes an identity for the Fourier coefficients of the phase factor $J(t)$, see \cref{eq:Fourier-trans-Vac}, i.e., $\sum_{n} J_n J^*_{m+n} = \delta_{m0}$. 
The expression for the first-order correlation function is the main result of this section and it allows us to evaluate the time-dependent current in the output in response to the time-dependent voltage. Indeed, the current in the output is obtained from the correlation function according to \cref{eq:current}, and we then find
\begin{equation}\label{eq:current-ac-right}
	I(t) = 
	\frac{g_s e}{h} 
	\sum\limits_{n,m} 
	\mathrm{Re} \{
	J_n J^*_m \e^{\mi(m-n)\Omega t} \int_{-\infty}^{\infty}  \mathrm{d}E \left[ f_{V}(E)  - f_{0}(E_n) \right]  
	\left[ S^*_{ee}(E_{m}) S_{ee}(E_{n})  - S^*_{he}(E_{m}) S_{he}(E_{n}) \right] \},
\end{equation}
where we have made use of the electron-hole symmetry in \cref{eq:phs}. 

It is instructive to consider the current in a few specific cases. For a constant voltage, we have $J_n=\delta_{n0}$, and we then find the constant current~\cite{Anantram-Datta,Lambert-Raimondi}
\begin{equation}\label{eq:current-dc}
	I = 
	\frac{g_s e}{h} 
	 \int_{-\infty}^{\infty}  \mathrm{d}E \left[ f_{V}(E)  - f_{0}(E) \right]  
	\left[ |S_{ee}(E)|^2  - |S_{he}(E)|^2 \right],
\end{equation}
where $|S_{ee}(E)|^2$ is the probability for an electron to be normal-transmitted at the interface, while $|S_{he}(E)|^2$ is the probability for an Andreev conversion. Also, we can write the general time-dependent current in \cref{eq:current-ac-right} as $I(t)=I_{\mathrm{dc}}+I_{\mathrm{ac}}(t)$ and evaluate the constant term by setting $n=m$. We then find
\begin{equation}\label{eq:tien-gordon}
	I_{\mathrm{dc}}= 
	\frac{g_s e}{h} 
	\sum\limits_{n} 
	|J_n|^2 \int_{-\infty}^{\infty} \mathrm{d}E \left[ f_{V}(E) - f_{0}(E_n) \right]  
	\left[|S_{ee}(E_n)|^2  - |S_{he}(E_n)|^2 \right],
\end{equation}
which is similar to the expression found by Tien and Gordon for a related setup~\cite{Tien_1963}.
\end{widetext}

The formalism presented in this section, leading to the definition of the excess correlation function in \cref{eq:GF-final} and the corresponding time-dependent current in \cref{eq:current-ac-right}, is general and not restricted to the quantum Hall setup shown in \cref{fig:setup}. \Cref{eq:GF-final} is defined for a two-terminal configuration without backscattering, but it can be generalized to a multi-terminal setup~\cite{Anantram-Datta}. The excess correlation function could also describe more complicated scatterers by simply inserting their static scattering matrix into \cref{eq:M-matrix}. Other observables, like the current noise, can also be computed from \cref{eq:GF-final} by extending previous results~\cite{Moskalets_2016b}. 

\begin{table*}[t]\centering
\setlength{\tabcolsep}{8pt}
\begin{tabular}{@{}ccc@{}} 
 \toprule
\bf Section  & \bf Title & \bf Content and important results\\ 
    \midrule
  \ref{sec:results-1-overview} & Overview of results & Brief overview of \cref{sec:results-1}.  \\ 
  \ref{sec:results-1-large-gap} & Excitations inside the gap & Time-dependent current for excitations inside the superconducting gap.  \\ 
  \ref{sec:results-1-leak} & Quasiparticle leakage & Time-dependent current for excitations above the superconducting gap.  \\ 
  \ref{sec:results-1-multi} & Multi-particle interference & Voltage drives with more than one charge per pulse.  \\ 
  \ref{sec:results-1-charge} & Transmitted charge & Analysis of transmitted charge per period.  \\ 
  \ref{sec:results-1-summary} & Summary of transport results & Summary of main results.  \\ 
 \bottomrule
\end{tabular}
\caption{Overview of \cref{sec:results-1} on the time-dependent electric current.}
\label{table:overview2}
\end{table*}

\section{Time-dependent current\label{sec:results-1}}

\subsection{Overview of results}\label{sec:results-1-overview}

In this section, we apply our Floquet-Nambu formalism to analyze the electric current that is generated by the scattering of incoming charge pulses on the interface with a superconductor. First, we consider the case where the excitations are inside the superconducting gap, and the probability that quasiparticles are transmitted into the superconductor is negligible. We then examine larger excitation energies where, in addition to subgap transport, quasiparticles may be transmitted into the superconductor above the gap. \Cref{table:overview2} provides an overview of this section together with a summary of each subsection. 

\subsection{Excitations inside the gap\label{sec:results-1-large-gap}}

We start by considering excitations with energies inside the gap of the superconductor. For the sake of simplicity, we only discuss spin-singlet superconductors here. We consider the Lorentzian pulses in \cref{eq:volt-lor} with one electron emitted per period and the harmonic drive
\begin{equation}\label{eq:volt-sin}
V(t)=V+V_0\cos(\Omega t),
\end{equation}
where $V$ is a constant offset and $V_0$ controls the amplitude of the drive. The mean number of emitted charges per period of the drive is then given as $q =eV/\hbar\Omega$, while the amplitude of the time-dependent part is controlled by the ratio $z=eV_0/\hbar\Omega$ as shown in \cref{table:voltages}. 

\begin{figure*}[t]
	\includegraphics[width=1.0\textwidth]{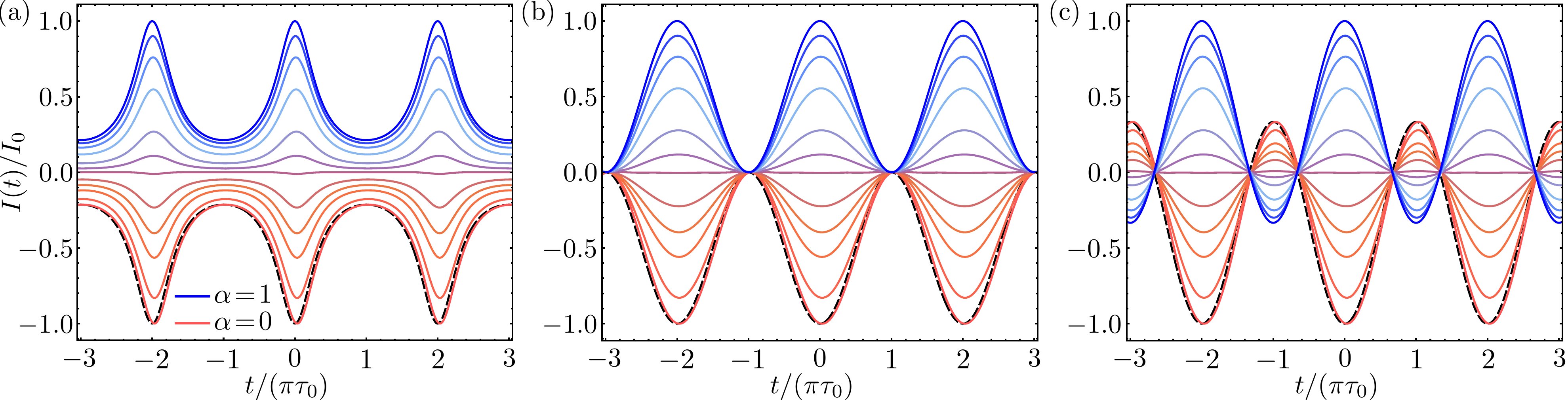}
	\caption{\label{fig:current-large-D} 
		Electric current. (a) Outgoing current for levitons with charge $q=1$ and width $\tau_0=\hbar/\Delta$ and driving frequency below the gap of the superconductor, $\hbar\Omega=0.1\Delta$. We show results for different values of the electron-hole conversion parameter $\alpha=0$ (full conversion), 0.15, 0.25, 0.3, 0.35, 0.42, 0.45, 0.5, 0.6, 0.7, 0.8, and 1 (normal transmission). 
		(b,c) Similar results for a  harmonic drive with $z=eV_0/(\hbar\Omega)=1$ and $q=eV/(\hbar\Omega)=1$ in panel (b) and $q=1/2$ in panel (c). The current is normalized with respect to the maximum of the injected current, $I_0$, which is plotted (with the sign changed) with a black dashed line. 
	}
\end{figure*}

In \cref{fig:current-large-D}, we show results for the time-dependent current based on \cref{eq:current-ac-right} for the Lorentzian pulses and for the harmonic drive with two different offsets. The sign of the outgoing current depends on the degree of Andreev conversion at the superconductor, and we show results for a range of values between the two limiting cases of full Andreev conversion ($\alpha=0$) and normal-transmission only ($\alpha=1$) according to \cref{eq:NS_SM}. For these limiting cases, levitons are either fully converted into anti-levitons or are completely normal-transmitted at the interface with the superconductor as seen in \cref{fig:current-large-D}(a). The injected current is shown with a black dashed line. We find similar results for the harmonic drive, which are shown in \cref{fig:current-large-D}(b) and \cref{fig:current-large-D}(c). For intermediate degrees of Andreev conversion, the injected pulses are partially Andreev converted and partially normal-transmitted. Thus, for a particular degree of Andreev conversion, the outgoing current vanishes as shown with purple lines in \cref{fig:current-large-D}. In that case, the outgoing pulses are charge-neutral but they can still be detected through noise measurements. 

To corroborate the results in \cref{fig:current-large-D}, we can derive analytic expressions for the current in the regime, where the excitations are well inside the superconducting gap. In this limit, the scattering matrix can be simplified as in \cref{eq:NS_SM-expand} and the time-dependent current becomes 
\begin{equation}
\label{eq:current_Large-Delta}
	I(t)= g_s(e^2/h)(2P_A - 1 ) V(t), 
\end{equation}
where the probability for Andreev conversion reads 
\begin{equation}\label{eq:Andreev-prob_Large-Delta}
	P_A=\left|S_{he}(E)\right|^2\simeq \left[(1-\alpha^2)/(1+\alpha^2)\right]^2, E\ll\Delta ,
\end{equation} 
as detailed in \cref{sec:app_large-D}. 
\Cref{eq:current_Large-Delta} accurately reproduces the currents in \cref{fig:current-large-D} (not shown here), and the current vanishes for $\alpha=\sqrt{2}-1\simeq0.41$, since $P_A=1/2$. 

We can also analyze the particle content of the current. To this end, we consider the first-order excess correlation function in \cref{eq:GF-final}, which can be written as 
\begin{equation}\label{eq:excess-GF_general}
	\hat{G}_\text{out}(x,t;x',t') = \hat{M}_{ee}^{(0)} G_\text{in}(t,t') + \hat{M}_{hh}^{(0)} G^*_\text{in}(t,t'),
\end{equation}
by decoupling the energy dependence of the scattering matrix from the time-dependence of the drive. Here, the correlation function for the incoming current reads
\begin{equation}
\begin{split}
    G_\text{in}(t,t')=  \int_{-\infty}^{\infty}  &\frac{\mathrm{d}E}{h v_F} \e^{\mi E(t' - t)/\hbar} \\
    \times & \left[ f_V(E) J(t)J^*(t') - f_0(E) \right] ,
\end{split}
\end{equation}
and we have defined
\begin{equation}
 \hat{M}_{\mu\mu}^{(0)}= \hat{M}_{\mu\mu}(E,E'\ll\Delta;x,x')   
\end{equation}
following \cref{eq:M-matrix}. 
The time-dependent voltage is encoded in the diagonal elements of $G_\text{in}(t,t')$ as
\begin{equation}
    h v_F G_\text{in}(t,t)= e V(t).
\end{equation} 
As such, the correlation function and the outgoing  current consist of contributions from positive and negative voltages weighted by the scattering probabilities in~$\hat{M}_{\mu\mu}^{(0)}$. 

\begin{figure*}
	\includegraphics[width=1.0\textwidth]{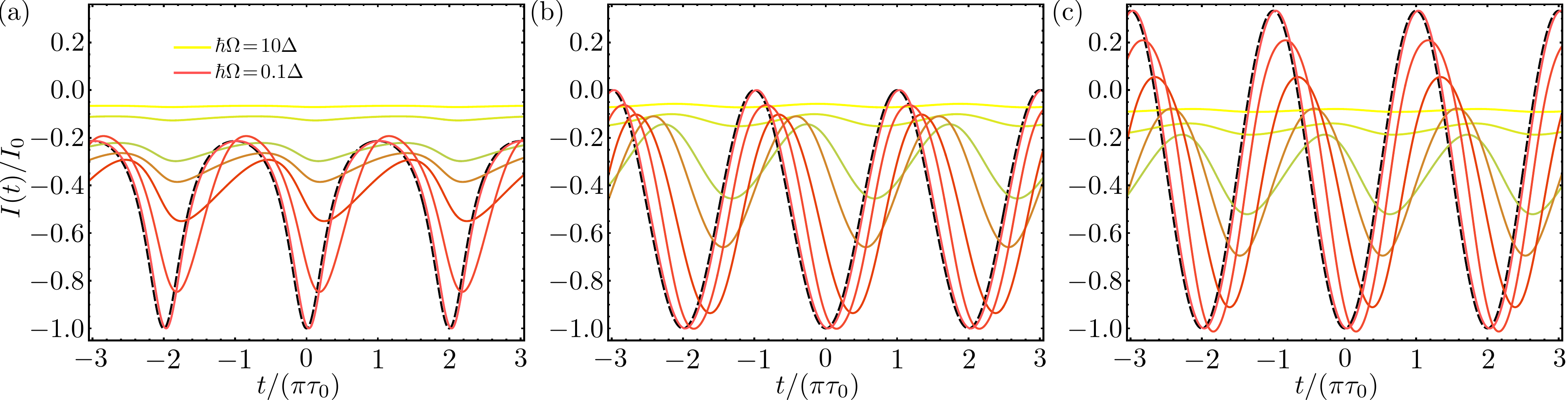}
    \caption{\label{fig:current-small-D} 
		Influence of driving frequency on the current. (a) Lorentzian drive with $q=1$ and $\tau_0=\hbar/\Delta$ and a superconductor with perfect Andreev conversion, $\alpha=0$. The driving frequencies are $\hbar\Omega/\Delta= 0.1, 0.2, 0.5, 0.66, 1, 2, 10$. (b,c) Similar results for a harmonic drive with $z=eV_0/(\hbar\Omega)=1$ and $q=eV/(\hbar\Omega)=1$ in panel (b) and $q=1/2$ in panel (c). 
		The black dashed line represents the injected current (with the sign changed), and its maximum is $I_0$. }
\end{figure*}

For a single Lorentzian pulse, we have 
\begin{equation}
    G_\text{in}(t,t')=G_{+}(t,t')
\end{equation}
and 
\begin{equation}
    G^*_\text{in}(t,t')=G_{-}(t,t'), 
\end{equation} 
with $G_{\pm}(t,t')$ being the excess correlation function for a leviton ($+$) or an anti-leviton ($-$). For a Lorentzian pulse, we can express the correlation function as
\begin{equation}\label{eq:excess-GF_lev}
	G_{\pm}(t,t')= \pm \frac{1}{v_F} \Psi_{\mp}^*(t') \Psi_{\mp}(t) ,
\end{equation}
where the wave function reads
\begin{equation}\label{eq:wf_lev}
	\Psi_{\mp}(t) = \sqrt{\frac{\tau_0}{\pi}} \frac{1}{t \mp \mi \tau_0},
\end{equation}
and $\tau_0$ is the half-width of the Lorentzian pulse in \cref{eq:volt-lor}. Hence, the current is composed of contributions from levitons and anti-levitons as 
\begin{equation}\label{eq:current_Large-Delta_2}
	I(t)= -g_s e v_F \left[ P_N G_{+}(t,t) + P_A G_{-}(t,t) \right] ,
\end{equation}
which is consistent with \cref{eq:current_Large-Delta} using that $P_N=1-P_A$. 

The information carried by the transmitted particles is encoded in their wavefunctions, which may be accessed through noise measurements~\cite{Bisognin_2019}. For the Lorentzian pulses, the excess correlation function determines the current according to \cref{eq:excess-GF_lev,eq:current_Large-Delta_2}, and it is given by the wavefunction of the leviton in \cref{eq:wf_lev}. As a result, all the quantum information that is carried by the leviton is available through tomographic measurements of the time-dependent current~\cite{Bisognin_2019,Roussel_2023}. In \cref{sec:results-2}, we analyze in detail the purity of the outgoing state after the scattering off the interface with the superconductor.

Besides giving access to observables such as the current, the excess correlation function contains information about the single-particle wavefunctions in the system~\cite{Roussel_2021}, which can be probed experimentally using tomography protocols~\cite{Bisognin_2019}.
For a single leviton, the excess correlation function depends only on the wave function of the leviton in \cref{eq:wf_lev}. Furthermore, the excess correlation function determines the current according to \cref{eq:excess-GF_lev,eq:current_Large-Delta_2}.
In \cref{sec:results-2}, we analyze in detail the purity of the outgoing state after the scattering off the interface with the superconductor.

\begin{figure*}[t]
	\includegraphics[width=1.0\textwidth]{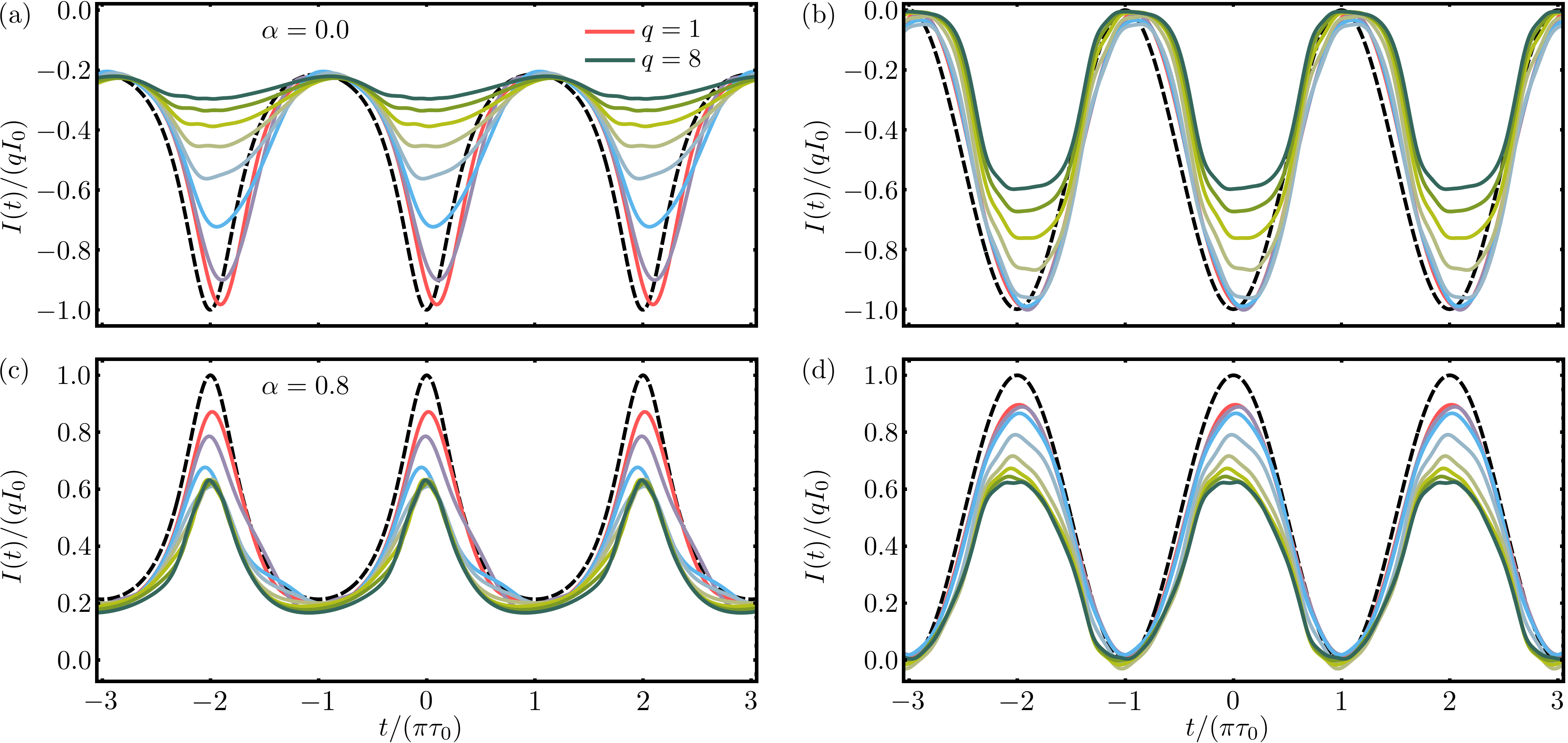}
	\caption{\label{fig:current-Q-change} 
		Time-dependent current for pulses with multiple charges. (a) Current for a Lorentzian drive of  width $\tau_0=\hbar/\Delta$ and driving frequency below the gap, $\hbar\Omega=0.25\Delta$, for a superconductor with perfect Andreev conversion, $\alpha=0$. The number of injected charges varies from $q=1$ to $q=8$. (b) Similar results for a harmonic drive with $z=1$ and $q=1$. (c,d) Same conditions as in panels (a) and (b) but for a superconductor with $\alpha=0.8$. In all panels, the current is normalized by its maximum,~$q I_0$. The injected current is plotted using black dashed lines (with the sign changed in the top panels). 
	}
\end{figure*}

\subsection{Quasiparticle leakage}
\label{sec:results-1-leak}

We now consider excitation energies that exceed the gap of the superconductor such that quasiparticles may be transmitted into the superconductor. \Cref{fig:current-small-D} shows the electric current for Lorentzian and harmonic drives with different excitation energies. 
We consider here a superconducting junction that fully Andreev converts incoming charges. Thus, for excitations inside the gap, the sign of the current is reversed, and its shape is given by the applied voltage, shown by a dashed line.

\Cref{fig:current-small-D} shows how the outgoing pulses experience a delay (a shift towards larger times) as the driving frequency becomes comparable to the superconductor gap. A careful analysis of the response function of the superconducting lead shows that this time delay is always present~\cite{Roussel_2023}, and it is related to the characteristic time scale of an Andreev reflection~\cite{Vasenko_2020}. For larger excitation energies, the pulses start to deform and their amplitude is reduced. The reduction of the amplitude occurs because of the photo-assisted tunneling of quasiparticles into the superconductor outside the gap. 
Specifically, when the drive can excite quasiparticles into Floquet states with energies outside the superconducting gap, quasiparticles can leak into the superconductor together with the quantum information that they carry. This loss of quantum information is further analyzed in \cref{sec:results-2}, where we consider the purity of the states. We note that similar results are obtained for triplet pairings, which are also dominated by the Andreev reflections. 

\subsection{Multi-particle interference}
\label{sec:results-1-multi}

So far, we have focused on  voltage pulses that carry one electron per period. We now consider the scattering of multi-particle pulses that carry more than one charge. We consider again Lorentzian and harmonic drives, and compare the case of perfect Andreev conversion to a situation where normal transmission dominates. The time-dependent currents for these different cases are shown in \cref{fig:current-Q-change}. The first row shows results for perfect Andreev conversions, while the second row corresponds to the situation where normal transmission dominates. 

In \cref{fig:current-Q-change}(a) and \cref{fig:current-Q-change}(b), we see that the multi-particle pulses also experience a time delay.
Moreover, the excitation energies increase with the number of charges, which causes quasiparticle tunneling into the superconductor. This effect can also be seen in \cref{fig:current-Q-change}(c) and \cref{fig:current-Q-change}(d), which show how the amplitude of the current is reduced, and the deformation of the pulses becomes stronger as the number of charges is increased. The deformation of the multi-particle pulses is different from the one for single-electron pulses. The multi-particle pulses exhibit small oscillations in the tails, which resemble multi-particle interference in normal-state interferometers~\cite{Kotilahti_2021,Shimizu2024,Iyoda2024}.

\begin{figure*}[t]  
	\includegraphics[width=1.0\textwidth]{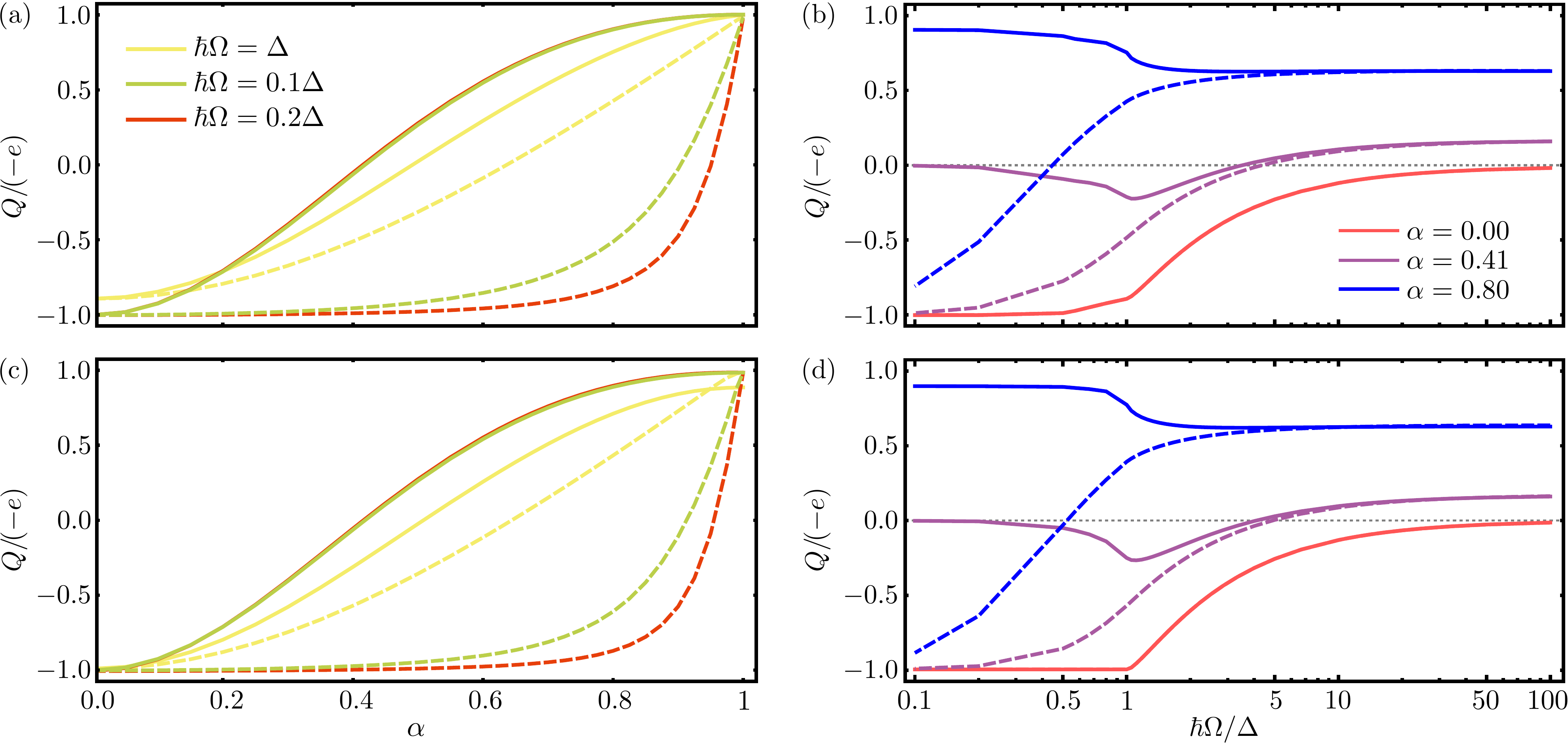}
	\caption{\label{fig:alp-D-evol}	
		Average transmitted charge. (a,b) Average charge per period of the drive for Lorentzian voltage pulses with charge $q=1$ and width $\tau_0=\hbar/\Delta$. Results are shown for different driving frequencies and degrees of conversion. Full lines correspond to  singlet pairings and dashed lines to triplet pairings. (c,d) Similar results for a harmonic drive with $q=1$ and $z=1$. 
	}
\end{figure*}

\subsection{Transmitted charge}
\label{sec:results-1-charge}

We now explore the difference between singlet and triplet pairings by considering the transmitted charge 
\begin{equation}\label{eq:charge}
	Q=\int_0^\mathcal{T} \mathrm{d}t I(t),
\end{equation}
where the electric current is given by \cref{eq:current-ac-right}. The scattering off the superconductor for singlet and triplet  pairings is very different inside the gap. In this regime, the degree of Andreev conversion for singlet pairings is only maximal for $\alpha=0$, where $|S_{he}|^2=1$ inside the gap according to \cref{eq:NS_SM-expand}, and it is otherwise reduced below one. For triplet pairings, by contrast, the degree of Andreev conversion is maximal at low energies with $|S_{he}|^2=1$ for all values of $\alpha$. As such, the transmitted charge at low energies is highly sensitive to the pairing potential inside the superconductor. 

In \cref{fig:alp-D-evol}, we show the transmitted charge for Lorentzian and harmonic drives and compare results for singlet and triplet pairings. In \cref{fig:alp-D-evol}(a) and~\cref{fig:alp-D-evol}(c), we show results for a Lorentzian drive and a harmonic drive, respectively, as a function of the degree of Andreev conversion. We show the singlet case with full lines and the triplet case with dashed lines. The transmitted charge evolves monotonically from a perfectly converted pulse for a fully Andreev-reflecting interface until it is completely normal-transmitted. For the singlet pairing, the dependence is essentially the same as long as the excitation energies are below the superconducting gap. For triplet pairings, the charge pulses are fully Andreev-converted for a much larger range of the degree of mixing, $\alpha$. However, as the excitation energies are increased, quasiparticles are transmitted into the superconductor, and the average amount of transmitted charge is reduced. Again, the behavior is qualitatively the same for Lorentzian and harmonic pulses. 

\Cref{fig:alp-D-evol}(b) and \cref{fig:alp-D-evol}(d) show the transmitted charge as a function of the driving frequency over the superconducting gap. With a large driving frequency compared to $\Delta$, most of the charge is transmitted into the superconductor if the degree of Andreev conversion is large. For lower driving frequencies, the transmitted charge approaches the behavior for subgap driving, in particular for the harmonic drive which was also considered in \cref{fig:current-large-D}. In this regime, the transmitted charge is determined by the probability for Andreev conversion according to \cref{eq:current_Large-Delta,eq:Andreev-prob_Large-Delta}. Singlet and triplet pairings yield very different results because of their different subgap scattering properties as illustrated in \cref{fig:support-1}. 

\subsection{Summary of transport results}\label{sec:results-1-summary}

In this section, we have seen how the degree of Andreev conversion determines the time-dependent current that runs in a chiral quantum Hall edge connected to a superconductor. If the excitation energies due to the voltage pulses are smaller than the superconducting gap, the outgoing current can be expressed in terms of contributions from normal-transmitted pulses and Andreev-converted pulses. Moreover, if single-electron pulses are injected into the edge state, the current corresponds to a coherent superposition of single-electron states and single-hole states according to \cref{eq:current_Large-Delta_2}.  
We have also identified other features of the time-dependent current, such as the time delay of Andreev-converted pulses, oscillations arising from multi-particle interference, and the tunneling of quasiparticles into the superconductor above the gap. 

\begin{table*}[t]
\centering
\setlength{\tabcolsep}{8pt}
\begin{tabular}{@{}ccc@{}} 
\toprule
 \bf Section  & \bf Title & \bf Content and important concepts\\ \midrule
  \ref{sec:overview2} & Overview of section & Brief overview of \cref{sec:purity}.  \\ 
 \ref{sec:def_pur} & Single-particle purity & Discussion of many-body and single-particle purity.\\ 
 \ref{sec:ope_nota} & Operator notation & Convenient operator notation and projectors. \\ 
\ref{sec:purity_cond} & Measure of single-particle purity & Definition of single-particle purity in terms of correlation functions. \\ 
 \ref{sec:purity_emit} & Purity above the Fermi level  & Single-particle purity of states above the Fermi level. \\ 
 \ref{sec:eh_sup} & Examples &  Examples of coherent superpositions and their single-particle purity. \\ 
 \bottomrule
\end{tabular}
\caption{Overview of \cref{sec:purity} on the purity of excitations generated by voltage pulses. }
\label{table:purity}
\end{table*}

\section{Measures of purity}
\label{sec:purity}
\subsection{Overview of section} 
\label{sec:overview2}

In this section, we focus on characterizing the states that are scattered off the superconductor. In particular, we introduce the purity of the states, which can be used to analyze their usefulness for quantum information processing~\cite{Roussel_2021}. 
Without the superconductor, quantum tomography protocols have been developed for extracting information about the electronic single-particle states and their wave functions~\cite{Grenier_2011,Bisognin_2019}. 
Here, we generalize this approach to include the electron-hole degree of freedom, which is relevant when superconductors are involved. An overview of this section is provided in \cref{table:purity}. 

\subsection{Single-particle purity}
\label{sec:def_pur}

In general, a pure many-body state $|\Psi\rangle$ can be represented by the  density matrix $\rho=|\Psi\rangle\!\langle \Psi|$, which is a projector, since $\rho^2=\rho$, and whose purity is equal to one, $\tr \{ \rho^2\}=1$. Mixed states, by contrast, have a purity below one. The purity is conserved under unitary transformations, including coherent time-evolution. 

Generally, since this notion of purity requires access to the full many-body state, it is hard to evaluate. As such, one may be interested in a more restrictive definition which reflects the information that is available in practice. In our case, we assume that the first-order coherence is known, and we can then introduce a corresponding notion of purity following Refs.~\cite{Grenier_2011,Bisognin_2019}. 
In particular, because the first-order coherence contains the complete information about the single-particle content of a given quantum state, we can introduce a notion of purity at the single-particle level. 
As we will see, if a quantum state is composed of independent particles (electrons, holes, or Bogoliubov quasiparticles), it is pure both at the single-particle level and at the quantum many-body level. However, because this notion of purity does not describe many-body correlations, quantum states that are pure at the many-body level may not be pure at the single-particle level, for example, states that describe entanglement between two or more particles. 

In addition to the single-particle purity, we introduce a notion of purity that only considers energies above the Fermi level. For normal metals, this notion of purity characterizes the ability of a single-particle source to generate clean single-particle excitations. If superconductivity is involved, it quantifies the ability to generate clean Bogoliubov excitations. Those are quantum superpositions of electrons and holes, and each Bogoliubov quasiparticle can thus be understood as a mobile qubit. 

To define the single-particle purity, we note that the correlation function, just as the density matrix, is a Hermitian and positive operator with eigenvalues between zero and one. Moreover, if the correlation function is a projector, the corresponding quantum state is pure at the single-particle level. For example, the Fermi sea at zero temperature is represented by the correlation function 
\begin{equation}\label{eq:GF-equilibrium}
    \hat{\mathcal{G}}_{\text{off}}(\omega',\omega)=
    \Theta(-\omega) \delta(\omega-\omega')
    \begin{pmatrix}
    	1 & 0\\
    	0 & 1
    \end{pmatrix},
\end{equation}
obtained from \cref{eq:op-averages}, which is a projector since
\begin{equation}\label{eq:purity-off}
		\int \mathrm{d}\omega \hat{\mathcal{G}}_{\text{off}} ( \omega_1, \omega)\hat{\mathcal{G}}_{\text{off}} (\omega, \omega_2) = \hat{\mathcal{G}}_{\text{off}} ( \omega_1, \omega_2).
\end{equation}
Thus, the Fermi sea at zero temperature is pure at the single-particle level, which is consistent with the fact that the many-body state can be expressed as a product of single-particle operators applied to the vacuum state~\cite{Roussel_2021}.

In the following, it will be convenient to use a compact notation for expressions like \cref{eq:purity-off}, which we write as 
\begin{equation}
\hat{\mathcal{G}}_{\text{off}} \circ \hat{\mathcal{G}}_{\text{off}}=\hat{\mathcal{G}}_{\text{off}}. 
\label{eq:pureFermi}
\end{equation}
By writing the correlation function in \cref{eq:correlator-on} as
\begin{equation}
	\hat{\mathcal{G}}_{\text{on}} = \hat{S}_F \hat{\mathcal{G}}_{\text{off}}  \hat{S}_F^\dagger, 
\end{equation}
we then have 
\begin{equation}\label{eq:purity-coherence}
\begin{split}
   \hat{\mathcal{G}}_{\text{on}}  \circ  \hat{\mathcal{G}}_{\text{on}}  ={}& \hat{S}_F \hat{\mathcal{G}}_{\text{off}}  \hat{S}_F^\dagger\circ \hat{S}_F \hat{\mathcal{G}}_{\text{off}}  \hat{S}_F^\dagger\\
   ={}& \hat{S}_F \hat{\mathcal{G}}_{\text{off}}  \hat{S}_F^\dagger=\hat{\mathcal{G}}_{\text{on}},
\end{split}
\end{equation}
having assumed that the scattering is unitary, such that $\hat{S}_F^\dagger\hat{S}_F=\hat{1}$, and  the initial state is pure. In that case, the quantum state after the scattering processes is also pure at the single-particle level. 

\subsection{Operator notation}
\label{sec:ope_nota}
For the following discussion, we introduce a convenient operator notation. First, we define the  operator~\cite{Roussel_2021}
\begin{equation}\label{eq:G-operator} 
    \boldsymbol{\mathcal{\hat{G}}} = \int \mathrm{d} r\mathrm{d} r' \hat{\mathcal{G}}(r,r') \ket{r} \! \bra{r'},
\end{equation}
which is expressed here in an arbitrary single-particle basis $\{\ket{r}\}$ ($r$ can describe either the time or the energy, for example), with the matrix elements reading
\begin{equation}
\hat{\mathcal{G}}(r,r')= \bra{r} \boldsymbol{\mathcal{\hat{G}}} \ket{r'}.
\end{equation}
To simplify the discussion, we assume that $\boldsymbol{\mathcal{\hat{G}}}$ acts on the whole spin-Nambu space, \cref{eq:fullNambu}. 
In this representation, convolutions as in \cref{eq:purity-off,eq:pureFermi} take the form 
\begin{equation}\label{eq:G2-operator}
	\mean{r_1| \boldsymbol{\mathcal{\hat{G}}}_1
	\boldsymbol{\mathcal{\hat{G}}}_2 | r_2} = \int \mathrm{d}r
	\hat{\mathcal{G}}_1 (r_1,r) \hat{\mathcal{G}}_2 (r,r_2)
	=(\hat{\mathcal{G}}_{1} \circ \hat{\mathcal{G}}_{2})(r_1, r_2),
\end{equation}
where we have used the completeness relation, $\int \mathrm{d}r \ket{r}\!\bra{r} =1$. If we use a basis of localized states $\{\ket{s}\}$, with $\mean{s|s'}=\delta(s-s')$, we recover the coordinate representation of the correlation function, $\hat{\mathcal{G}}(s,s')= \bra{s} \boldsymbol{\mathcal{\hat{G}}} \ket{s'}$. 

Below, we use a frequency representation since the operator for the Fermi sea then takes a simple form. Also, we define operators that project onto single-particle states with energies above or below the Fermi level~\cite{Roussel_2021},
\begin{equation}
\mbf{\Pi}_{\pm}= \mbf{\Pi}_{\pm}^2,
\end{equation}
with the  properties $\mbf{\Pi}_{+}+\mbf{\Pi}_{-}= \mbf{1}$ and $\mbf{\Pi}_{\pm}\mbf{\Pi}_{\mp}= \mbf 0$. 
We note that the Fermi sea at zero temperature is simply 
\begin{equation}
\boldsymbol{\hat{\mathcal{G}}}_\text{off} = \mbf{\Pi}_{-}.
\end{equation}

\subsection{Measure of single-particle purity}
\label{sec:purity_cond}

To define a measure of the purity  at the single-particle level, we consider the excess correlation function 
\begin{equation}\label{eq:excess-operator}
	\mbf{\hat{G}} = \boldsymbol{\hat{\mathcal{G}}}_\text{on} - \boldsymbol{\hat{\mathcal{G}}}_\text{off}.
\end{equation}
Since we consider here the full Nambu space, not all the matrix elements are independent. In particular, due to the particle-hole symmetry, \cref{eq:phs}, we have,
\begin{equation}
	\hat{G}(E,E')
	=
	(\tx \sii) \hat{G}^*(-E,-E') (\tx \sii),
\end{equation}
with $\tx$ a Pauli matrix acting on the Nambu space and $\sii$ the identity in spin space.

We now assume that the states with and without the drive are both pure at the single-particle level, implying that $\boldsymbol{\hat{\mathcal{G}}}_\text{on}^2=\boldsymbol{\hat{\mathcal{G}}}_\text{on}$ and $\boldsymbol{\hat{\mathcal{G}}}_\text{off}^2=\boldsymbol{\hat{\mathcal{G}}}_\text{off}$. For the state without the drive, we take the Fermi sea at zero temperature, $\boldsymbol{\hat{\mathcal{G}}}_\text{off}= \mbf{\Pi}_{-}$, such that $\boldsymbol{\hat{\mathcal{G}}}_\text{on} = \mbf{\hat{G}} + \mbf{\Pi}_{-}$ and
\begin{equation}
	( \mbf{\hat{G}} + \mbf{\Pi}_{-} )^2 = \mbf{\hat{G}} + \mbf{\Pi}_{-}.
\end{equation}
After a bit of algebra, we then find
\begin{equation}
	\mbf{\hat{G}}^2 = \mbf{\hat{G}} - \mbf{\Pi}_{-} \mbf{\hat{G}} - \mbf{\hat{G}} \mbf{\Pi}_{-}.
\end{equation}
Moreover, by inserting identities of the form $\mbf{\Pi}_{+}+\mbf{\Pi}_{-}= \mbf{1}$, we arrive at the expression 
\begin{equation}
	\mbf{\hat{G}}^2 =\mbf{\Pi}_{+} \mbf{\hat{G}} \mbf{\Pi}_{+} - \mbf{\Pi}_{-} \mbf{\hat{G}} \mbf{\Pi}_{-} =\mbf{\hat{G}}_{++} - \mbf{\hat{G}}_{--} ,
 \label{eq:purity-cond-1}
\end{equation}
where the projections of the excess correlation function onto energies above and below the Fermi level are
\begin{equation}\label{eq:G-projected}
    \mbf{\hat{G}}_{\epsilon\epsilon'}=\mbf{\Pi}_{\epsilon} \mbf{\hat{G}} \mbf{\Pi}_{\epsilon'} ,
\end{equation}
with $\epsilon,\epsilon'=+,-$. 
Thus, for any drive of the Fermi sea at zero temperature with unitary scattering, the excess correlation function must fulfill \cref{eq:purity-cond-1}. Therefore, we define the single-particle purity of the outgoing state as 
\begin{equation}\label{eq:purity_1}
	\gamma = \frac{ \tr \{  \mbf{\hat{G}}^2 \}}{2 \tr \{  \mbf{\hat{G}}_{++} \} }, 
\end{equation}
which is a number between zero and one. The purity is normalized with respect to the number of electrons and holes, $ \tr \{  \mbf{\hat{G}}_{++}\} = - \tr \{ \mbf{\hat{G}}_{--}  \} = N_e + N_h$. Pure states at the single-particle level can be expressed as products of single-particle states, including Bogoliubov excitations. 

\subsection{Purity above the Fermi level}
\label{sec:purity_emit}

For many applications, it would be useful to have a single-particle emitter that generates pure excitations above the Fermi sea. However, the single-particle purity defined above does not distinguish between a voltage drive that excites electron-hole pairs in the Fermi sea, like the harmonic voltage, and the Lorentzian voltage pulses that produce clean excitations above the Fermi level while leaving the Fermi sea unaltered. 
Therefore, we introduce a measure of single-particle purity that is restricted to energies above the Fermi level. Specifically, we consider a state to be pure above the Fermi level if $\mbf{\hat{G}}_{++}$ is a projector, such that $
	\mbf{\hat{G}}_{++}^2 = \mbf{\hat{G}}_{++}$.
We then define the corresponding single-particle purity as
\begin{equation}\label{eq:purity_2}
	\gamma_+ = \frac{\tr \{\mbf{\hat{G}}_{++}^2 \}}{ \tr \{\mbf{\hat{G}}_{++}\}
	}.
\end{equation}
Since $\tr \{\mbf{\hat{G}}_{++}\}$ is the number of excited electrons and holes, the purity is well defined, except for the Fermi sea at zero temperature for which both the numerator and the denominator vanish. 
With the definition above, we remove the electron-hole coherences that appear if the drive excites the Fermi sea, while keeping information about both electrons and holes that are generated by the drive. In particular, we have that 
\begin{equation}\label{eq:usp-pep-connection}
\gamma = \gamma_+ + \frac{ \tr \{ \mbf{\hat{G}}^\dagger_{-+} \mbf{\hat{G}}_{-+} \}} 
{\tr \{ \mbf{\hat{G}}_{++} \} } , 
\end{equation}
where $\mbf{\hat{G}}_{\pm\mp}$ is defined in \cref{eq:G-projected}. 
Since the last term is positive, the purity above the Fermi level is never larger than the single-particle purity 
\begin{equation}
\gamma_+ \le \gamma.
\end{equation}
Furthermore, the equality only holds if the coherence function is block diagonal, $\mbf{\hat{G}} = \mbf{\hat{G}}_{++} + \mbf{\hat{G}}_{--}$. 
In that case, the many-body state can be represented by a product of Bogoliubov excitations on top of the Fermi sea,
\begin{equation}
	\label{eq:pep:manybody}
	\ket{\Psi}
	=
	\prod_i \left(\alpha_i \psi^\dagger[\varphi_i^{(e)}]
	+
	\beta_i \psi[\varphi_i^{(h)}]\right) \ket{F},
\end{equation}
where the operator $\psi^\dagger[\varphi]$ creates electrons in the wavefunction $\varphi$. The wavefunctions $\varphi^{(e)}_i$ constitute a set of orthogonal electronic excitations containing only positive frequencies and, conversely, the hole states $\varphi^{(h)}_i$ comprise a set of orthogonal hole excitations with only negative frequencies. 
This distinction between positive and negative frequencies makes the purity of the states above the Fermi sea more restrictive than the single-particle purity.

Importantly, because of the Nambu structure in \cref{eq:fullNambu}, the correlation function $\mbf{\hat{G}}_{++}$ contains information about the electron and hole excitations as well as the superconducting correlations. Consequently, only the coherences $\mbf{\hat{G}}_{+-}$ and $\mbf{\hat{G}}_{-+}$ are projected away. As such, \cref{eq:purity_1,eq:purity_2} generalize earlier definitions of the purity for electronic excitations only~\cite{Bisognin_2019,Roussel_2021}. 

\subsection{Examples}
\label{sec:eh_sup}

To better understand the purity measures from above, we illustrate them with a few examples. To begin with, we consider an exciton-like electron-hole pair
\begin{equation}\label{eq:exciton_wf}
	\ket{\Psi_\text{exc}} = \left(\alpha +\beta\Psi_{h}\Psi_{e}^\dagger\right) \ket{F},
\end{equation}
where $\alpha$ and $\beta$ are complex numbers with $|\alpha|^2+|\beta|^2=1$, and the Fermi sea at zero temperature is denoted by~$\ket{F}$. The operators $\Psi^\dagger_e$ and $\Psi_h$ create electrons and holes in the states~$\varphi_e$ and $\varphi_h$ above and below the Fermi level, respectively. 
Note that $\Psi_e$ and $\Psi_h$ are fermionic operators, instead of Nambu quasiparticle operators, so that the creation of a hole below the Fermi level is represented by $\Psi_h \ket{F} \neq 0$. 
Being a combination of the Fermi sea and an electron-hole pair, the state is charge neutral (with respect to the charges in the Fermi sea), with no superconducting correlations since $\bra{\Psi_{\text{exc}}}\Psi_e \Psi_h\ket{\Psi_{\text{exc}}}=0$. 

Consequently, the excess correlation function in Nambu space takes the form
\begin{equation}
	\hat{\mathbf{G}} =
	\begin{pmatrix}
		\mathbf{G}^{(e)} & 0 \\
		0 & \mathbf{G}^{(h)}
	\end{pmatrix}.
	\label{eq:diagonal_coherence}
\end{equation}
Moreover, in the frequency representation, we have
\begin{equation}
	\hat{\mathbf{G}} =
\int \mathrm{d} \omega\mathrm{d} \omega' \hat G (\omega,\omega')
 \ket{\omega} \! \bra{\omega'},
\end{equation}
where the matrix elements read
\begin{equation}
\begin{split}
	G^{(e)}(\omega, \omega') 
&=
	|\beta|^2 [\varphi_e^*(\omega') \varphi_e(\omega) -
	\varphi_h^*(\omega') \varphi_h(\omega)]\\
	&+ \alpha^* \beta \varphi_h^*(\omega') \varphi_e(\omega)
	+ \alpha \beta^* \varphi_e^*(\omega') \varphi_h(\omega) ,
	\label{eq:coherence:elec:crossterms}
\end{split}
\end{equation}
with $\langle\omega|\omega'\rangle=\delta(\omega-\omega')$. We also have 
\begin{equation}
G^{(h)}(\omega, \omega') = -G^{(e)}(-\omega, -\omega'), 
\end{equation}
which follows from the anticommutation relations. 

From these expressions, we find that
\begin{equation}
G^{(e)} \circ G^{(e)} =
	|\beta|^2 [\varphi_e^*(\omega') \varphi_e(\omega) +
	\varphi_h^*(\omega') \varphi_h(\omega)]\,.
\end{equation}
Equivalently, we have
\begin{equation}
[\mathbf{G}^{(e/h)}]^2 = \mathbf{G}^{(e/h)}_{++} -
\mathbf{G}^{(e/h)}_{--},
\end{equation}
implying that $\hat{\mathbf{G}}$ obeys \cref{eq:purity-cond-1}. Thus, the state is pure at the single-particle level, and it can be expressed as a product of single-electron operators. Indeed, by defining the single-electron operator $\Phi^\dagger=\alpha \Psi_{h}^\dagger-\beta\Psi_{e}^\dagger$, the state can be written as $\ket{\Psi_\text{exc}} = \Phi^\dagger\Psi_h \ket{F}$.

Despite the state being pure at the single-particle level, it may not describe pure single-particle states above the Fermi level. To see this, we  project the excess correlation function onto energies above the Fermi level and consider~$\hat{\mathbf{G}}_{++}$, whose matrix elements are
\begin{equation}
	\hat{G}_{++}(\omega, \omega') = |\beta|^2 \begin{pmatrix}
		\varphi_e^*(\omega') \varphi_e(\omega) & 0 \\
		0 &  \varphi_h(-\omega') \varphi_h^*(-\omega)
	\end{pmatrix}.
\end{equation}
We then find $\gamma_+ = |\beta|^2$, showing that the state above the Fermi level is only pure at the single-particle level if the electron-hole pair is generated with certainty, $|\beta|^2 = 1$. 

Another example is the entangled state
\begin{equation}\label{eq:entangled_wf}
	\ket{\Psi_\text{ent}}
	=
	\left(
		\alpha \Psi_{h1}^{\phantom{\dagger}} \Psi_{e1}^\dagger
		+
		\beta \Psi_{h2}^{\phantom{\dagger}}\Psi_{e2}^\dagger
	\right)
	\ket{F},
\end{equation}
where one electron and one hole are created in orthogonal states.
In this case, the correlation function reads
\begin{equation}
	\hat{\mathbf{G}} =
	|\alpha|^2 \hat{\mathbf{G}}_1
	+ |\beta|^2 \hat{\mathbf{G}}_2,
\end{equation}
where $\hat{\mathbf{G}}_{i=1,2}$ are the correlation functions for electrons and holes that are emitted in the wavefunctions $(\varphi_{e,i}, \varphi_{h,i})$ on top of the Fermi sea. We immediately see that the first-order correlator does not contain information about the entanglement, as it is given by a statistical mixture. Furthermore, since the wavefunctions are orthogonal, we have $\hat{\mathbf{G}}_1 \hat{\mathbf{G}}_2 = \hat{\mathbf{G}}_2 \hat{\mathbf{G}}_1 = 0$ and $\hat{\mathbf{G}}^2 = |\alpha|^4 \hat{\mathbf{G}}_1 + |\beta|^4 \hat{\mathbf{G}}_2$, such  that $\gamma = \gamma_+ = |\alpha|^4 + |\beta|^4$. 

Finally, we consider a superposition of a single electronic excitation and a single hole reading
\begin{equation}\label{eq:superp_wf}
	\ket{\Psi_\text{sup}} = \left(\alpha \Psi_{e}^\dagger + \beta
	\Psi_{h} \right) \ket{F} . 
\end{equation}
Such a state can be realized by the Andreev reflection of an electron on a superconductor. For this state, the correlation function is block diagonal, $\mbf{\hat{G}} = \mbf{\hat{G}}_{++} + \mbf{\hat{G}}_{--}$. In this case, the anomalous correlator $\bra{\Psi_{\text{sup}}}\Psi_e \Psi_h\ket{\Psi_{\text{sup}}} = - \alpha\beta^*$ is non-zero, and we have
\begin{multline}
	\label{eq:gpp:supra}
	\hat{G}_{++}(\omega, \omega') =\\\begin{pmatrix}
		|\alpha|^2 \varphi_e^*(\omega') \varphi_e(\omega) &
		\alpha \beta^* \varphi_h(-\omega') \varphi_e(\omega) \\
		\alpha^* \beta \varphi_e^*(\omega') \varphi_h^*(-\omega) &
		|\beta|^2 \varphi_h(-\omega') \varphi_h^*(-\omega)
	\end{pmatrix}.
\end{multline}
For the single-particle purity, we then find $\gamma = \gamma_+ = 1$, as one we would expect, since the state is a particular case of \cref{eq:pep:manybody}. Interestingly, the state $\ket{\Psi_\text{sup}}$ is pure for all $\alpha$ and $\beta$. As such, we can consider it as a charge qubit in the basis of the electron and the hole states. 

\begin{table*}[t]
\centering
\setlength{\tabcolsep}{8pt}
\begin{tabular}{@{}ccc@{}} 
\toprule
 \bf Section  & \bf Title & \bf Content and important concepts\\ \midrule
 \ref{sec:results-2-overview} & Overview of results & Brief description of \cref{sec:results-2}.  \\ 
 \ref{sec:results-2-purity} & Purity of the outgoing state & Relevant quantities in the frequency domain.  \\
 \ref{sec:results-2-small-drive} & Low driving frequencies & Purity of the outgoing states for slow drives.  \\ 
 \ref{sec:results-2-large-drive} & Large driving frequencies & Quasiparticle leakage and the purity of the outgoing states.  \\ 
 \ref{sec:results-2-comparable-drive} & Intermediate driving frequencies & Distinguishing between singlet and triplet pairings.  \\ 
 \ref{sec:results-2-exp} & Measuring the purity & Possible connections to experiments. \\ 
 \bottomrule
\end{tabular}
\caption{Overview of \cref{sec:results-2} with results for the purity of the single-particle excitations. }
\label{table:purity_results}
\end{table*}

These examples illustrate different situations that we will encounter in the following section. 
The state in \cref{eq:exciton_wf} is characteristic for a voltage drive that emits pure states at the single-particle level which, however, are not pure for energies above the Fermi level. That is the case for most voltage drives, including the harmonic drive in \cref{eq:volt-sin}. The entangled state in \cref{eq:entangled_wf} illustrates the loss of information that can occur, for example, if quasiparticles are transmitted into the superconductor. Finally, the state in \cref{eq:superp_wf} represents the superposition of single electron and a hole, representing a mobile charge qubit, which is one of our main focuses.

\section{Purity of outgoing excitations}
\label{sec:results-2}

\subsection{Overview of results}
\label{sec:results-2-overview}

In this section, we present results for the purity of the few-particle excitations that are generated by a time-dependent voltage. We first derive some useful expressions in the frequency representation and then go on to analyze the purity for different driving frequencies compared to the superconducting gap. At the end, we provide a short perspective on future measurements of the purity. An overview of the section is provided in \cref{table:purity_results}.

\subsection{Purity of the outgoing state}
\label{sec:results-2-purity}

We now consider the purity of the outgoing state after the superconductor. It is convenient to work in the frequency representation of the correlation function in \cref{eq:GF-final}. To this end, we note that the correlation function is not periodic in both time arguments, however, we can  express it in terms of the two time variables
\begin{equation}
\bar t =(t+t')/2 \quad\mathrm{and}\quad \tau=t-t'. 
\end{equation}
The correlation function is then periodic in $\bar t$, but not in~$\tau$. As detailed in \cref{sec:app-freq}, we can perform a discrete Fourier transformation over $\bar t$ and a continuous Fourier transformation over $\tau$, such that we find
\begin{equation}\label{eq:G-freq}
	\hat{G}_l(\omega) =    \int_{-\infty}^{\infty} 
    \mathrm{d}\tau \int_0^\mathcal{T}  \frac{\mathrm{d}\bar{t}}{\mathcal{T}}\e^{\mi l \Omega \bar{t}}  \e^{\mi \omega \tau} \hat{G} (\bar{t}+\tau/2,\bar{t}-\tau/2).
\end{equation}
The trace of the correlation function then takes the form
\begin{equation}\label{eq:trace-freq}
    \tr \{ \mbf{\hat{G}} \} = \int^{\infty}_{-\infty} \frac{ \md \omega }{ 2\pi } \tr_\text{N} \{ \hat{G}_{l=0}(\omega) \}, 
\end{equation}
where $\tr_\text{N}$ denotes the trace in Nambu space. 

In the following, we analyze the single-particle purity of the outgoing state in \cref{eq:purity_1} as well as its purity above the Fermi level in \cref{eq:purity_2}. In particular, we compare a sequence of levitons to a harmonic drive. 
We focus on driving amplitudes that are comparable to the superconducting gap but consider the full range of driving frequencies, so that the number of particles with significant excitation probabilities varies from a few to many. 

\subsection{Low driving frequencies}
\label{sec:results-2-small-drive}

\begin{figure*}[t]
    \includegraphics[width=1.0\textwidth]{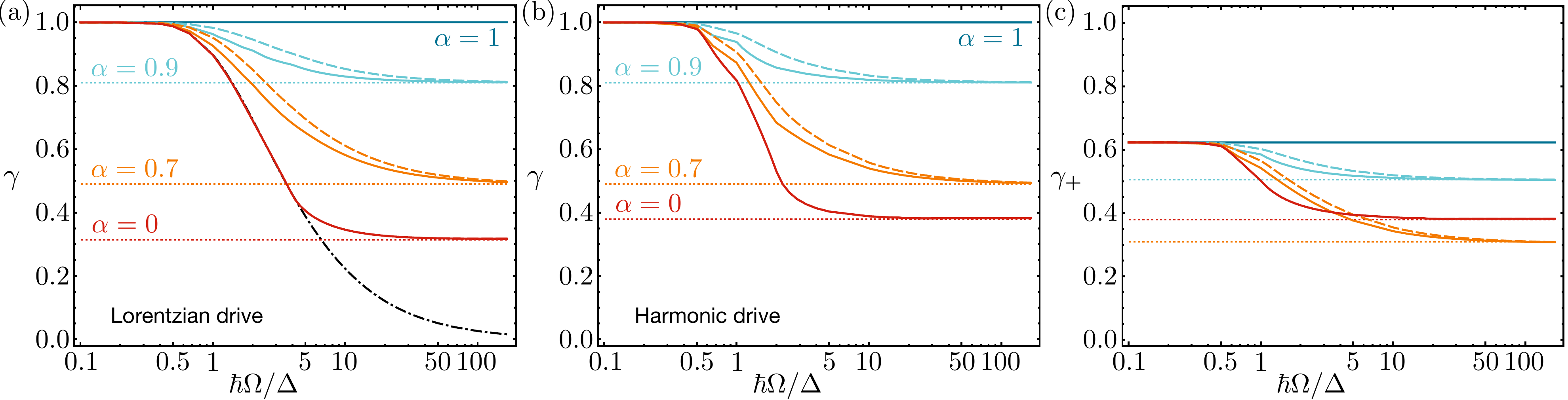}
	\caption{\label{fig:purity}
    Purity. (a) Purity for Lorentzian voltage pulses of width $\tau_0=0.25\hbar/\Delta$ and $q=1$ as a function of the driving frequency. The purity and the purity above the Fermi level are the same, $\gamma=\gamma_+$. Full lines correspond to singlet pairings and dashed lines to triplet pairings. The dot-dashed black line shows \cref{eq:purity-lev-intermediate}. 
    (b) Purity for a harmonic drive with $q=0$ and $z=2$. (c) Purity above the Fermi level for the harmonic drive. The limiting values for $\hbar\Omega\gg\Delta$ are shown by dotted lines.
	}
\end{figure*}

We first take the  driving frequency to be much smaller than the superconducting gap, so that no quasiparticles are transmitted into the superconductor. According to \cref{eq:current_Large-Delta_2}, the outgoing current is then a combination of positive and negative pulses weighted by the probabilities for Andreev and normal scattering. 
In contrast to the current in \cref{eq:current}, the trace in Nambu space in \cref{eq:trace-freq} becomes a sum of the two probabilities, which then yields $|S_{ee}|^2+|S_{he}|^2=1$ for all relevant energies. Thus, the states are pure at the single-particle level as seen in \cref{fig:purity}(a) and~\cref{fig:purity}(b) for low driving frequencies. 

Next, we consider the purity of the states above the Fermi level. The special property of Lorentzian pulses with integer charge is that they only excite single-particle states with energies above the Fermi level, making them useful for quantum information 
applications~\cite{Roussel_2021,Martin_2023b,Assouline:2023,Chakraborti2025}. By contrast, any other voltage drive excites a combination of states with positive and negative energies. 
Thus, for the Lorentzian pulses with integer charge, we have 
\begin{equation}
\mbf{\hat{G}} = \mbf{\hat{G}}_{++} \quad\mathrm{and}\quad \mbf{\hat{G}}_{--}=0,
\end{equation}
showing that the Lorentzian pulses that have scattered off the superconductor are pure, both at the single-particle level and above the Fermi level~\cite{Moskalets_2015,Roussel_2021}. 
In \cref{sec:app-purity} we show that the pulses, before they have scattered off the superconductor, are pure for all driving frequencies.

For the harmonic drive the situation is different. For small driving frequencies, the outgoing states are pure at the single-particle level as seen in \cref{fig:purity}(b), since no quasiparticles are transmitted into the superconductor. By contrast, the purity above the Fermi level is reduced below one, because electron-hole pairs are excited by the drive. If only a few particles are excited, the coherence function is almost diagonal, and the purity of states above the Fermi level can be approximated as
\begin{equation}
\label{eq:pep-large-D}
    \gamma_+\simeq \sum_{n>0} n|J_n(z)|^2, \quad\hbar\Omega\ll\Delta ,
\end{equation}
with $z=eV_0/\hbar\Omega$. 
For a harmonic drive with $V=0$ in \cref{eq:volt-sin}, the sum above can be carried out and, using properties of the Bessel functions in \cref{table:voltages}, one finds
\begin{equation}
    \gamma_+\simeq   
    z \left[ z \left(\mathcal J_0(z)^2 + \mathcal J_1(z)^2 \right) -\mathcal J_0(z) \mathcal J_1(z) \right]/2 .
\end{equation}
Taking $eV_0=2\hbar\Omega$, we get $\gamma_+\simeq 0.64$, which fits well for small frequencies in \cref{fig:purity}(c). 
 
\subsection{Large driving frequencies}
\label{sec:results-2-large-drive}

Next, we consider large driving frequencies such that quasiparticles may be transmitted into the superconductor. In this regime, the single-particle purity is well-approximated by the squared conversion probability, $\gamma\simeq\alpha^2$, as seen in \cref{fig:purity}(a) and \cref{fig:purity}(b). The transmitted pulses can essentially be understood as the emitted ones times the scattering probability, $\mathbf{G}^\text{out}\simeq\alpha^2\mathbf{G}^\text{in}$. 
Since the emitted pulses are pure at the single-particle level, we have $\mathrm{Tr}\{(\mathbf{G}^\text{in})^2\}\simeq \mathrm{Tr} \{\mathbf{G}^\text{in}\}$, and we immediately see that 
\begin{equation}\label{eq:usp-large-drive}
    \gamma \simeq  \alpha^2, \quad \hbar\Omega\gg\Delta. 
\end{equation}

At low driving frequencies, a similar argument can be made for the purity above the Fermi sea, which is rescaled by the factor in \cref{eq:pep-large-D} that accounts for the information that is lost by tracing out the states with negative energies. 
By contrast, for large driving frequencies, the purity above the Fermi level is well-approximated by  the purity of the injected state times the conversion probability, $\gamma_+\simeq \alpha^2\sum_{n>0}n|J_n|^2$.
The sum can be cut off from above, when $n$ reaches the ratio of the driving frequency over the amplitude, $\hbar\Omega/eV_0$, which is roughly the number of particles that are excited by the drive.

The case of perfect Andreev reflection ($\alpha = 0$)  is special and must be treated separately. For that, we find 
\begin{equation}
\label{eq:asymptotic-low-energy}
    \gamma
    \simeq
    \gamma_+
    \simeq \frac{4}{5} \sum_{n>0}|J_n|^2, \quad \alpha=0, \,\hbar\Omega\gg\Delta,
\end{equation}
where the prefactor of $4/5$ is a result of an integration over the frequency as detailed in \cref{sec:app-purity}. 

For periodic levitons, the sum can be cut off at a value that is determined by the overlap between the pulses, which is controlled by the half-width of the pulses compared to the driving frequency. For example, in \cref{fig:purity}(a), we consider well-separated levitons and only the first term in the sum is significant, so that $\gamma_+\simeq (4/5)\left|J_1\right|^2$. Interestingly, the purities coincide at large frequencies for the harmonic drive in \cref{fig:purity}(b) and~\cref{fig:purity}(c) and the value of $\gamma\simeq\gamma_+\simeq 0.38$ is reached with three terms in the sum. 

\subsection{Intermediate driving frequencies}
\label{sec:results-2-comparable-drive}

We now consider the intermediate regime, where the driving frequency is on the order of the superconducting gap.
For a sequence of levitons that are perfectly Andreev-converted by the superconductor ($\alpha=0$), we show in \cref{sec:app-purity} that the outgoing state can be described as either a single hole emitted with probability $P_+$ or the undisturbed Fermi sea with probability $1-P_+$. The purity above the Fermi level is then given by the probability, $\gamma_+\simeq P_+$, as
\begin{equation}\label{eq:purity-lev-intermediate}
    \gamma_+ \simeq 1 + \frac{1+2\tau_0\Delta/\hbar}{\left(\tau_0\Delta/\hbar\right)^2} \me^{-\tau_0\Delta/\hbar} - 2 \mathcal{K}_2\left( 2\tau_0\Delta/\hbar \right) ,
\end{equation}
where $\mathcal{K}_n(z)$ is the modified Bessel function of the second kind. 
In \cref{fig:purity}(a), we show this approximation and we see that it works well for driving frequencies that are not too large. This expression relies on well-separated levitons with $\tau_0\Omega\ll1$. Consequently, it improves as we reduce the half-width as seen in \cref{fig:purity-app} of \cref{sec:app-purity}. 

\Cref{fig:purity} shows the purities both for singlet and triplet pairings. For low and high frequencies, the purities are the same for the two pairings, since there is either no Andreev conversion, or it completely dominates the transport. We also find the same purities for perfect Andreev conversion ($\alpha=0$) and for complete normal transmission ($\alpha=1$). By contrast, the difference between singlet and triplet pairings is clear when the driving frequency is on the order of the superconducting gap, $\hbar\Omega\simeq\Delta$, and the contrast between them becomes more pronounced as the degree of conversion is reduced. The purity for the triplet pairing is usually higher than for the singlet pairing, because of its characteristic zero-energy resonance~\cite{Tanaka2011}. 

Additionally, as we approach complete normal transmission, the purity displays a series of nonanalytic features at $\hbar\Omega/\Delta=1/2,1,$ and 2, which are more remarked in \cref{fig:purity}(b). 
These features occur because of resonances when the amplitude of the drive matches the superconducting gap for its lowest modes, $n=1,2,4$. 
Indeed, at the gap edge, the total transmission probability, including normal transmission and Andreev conversion, $|S_{ee}|^2+|S_{he}|^2$, features a strong nonlinearity due to the onset of quasiparticle leakage into the superconductor. For energies below the gap, we have $|S_{ee}|^2+|S_{he}|^2=1$ since there is no leakage while, right above the gap, the probability is reduced. This transition is not smooth and it takes the form of a sharp dip, which is more pronounced for the singlet pairing and it is mostly due to the Andreev conversion processes being suppressed above the gap. This effect also explains why the purity for the triplet pairing is higher than for the singlet pairing. 

\subsection{Measuring the purity}
\label{sec:results-2-exp}

In this section, we have analyzed the purity of excitations emitted by a Lorentzian or a harmonic drive as they scatter off a superconductor. 
If the driving frequency is lower that the  superconducting gap, the purity is mostly determined by the drive and less by the scattering off the superconductor. As such, the scattering of a leviton leads to a coherent superposition of single-electron and single-hole excitations. In the opposite regime, where the driving frequency is large, quasiparticles are transmitted into the superconductor and the purity of the outgoing state is reduced. If the driving frequency is comparable to the gap, the purity becomes sensitive to the energy-dependent scattering at the superconductor and the outgoing excitations carry information about the pairing mechanism of the superconductor. 

The purity can be accessed through observables such as the current and the noise~\cite{Bisognin_2019,Roussel_2017}. The excess correlation function can be reconstructed from two-particle interferometry using advanced signal processing~\cite{Marguerite2017,Roussel_2017,Roussel_2021,Roussel_2023}. 
Analyzing the single-particle content in this way would be an important step forward towards future applications in quantum sensing~\cite{sequoia2022} and for realizing flying qubits~\cite{Edlbauer2022}. Quantum tomography protocols for detecting quantum signals have also been implemented using electronic interferometry of electrons with energies in the range of millielectronvolts~\cite{Grenier_2011,Jullien_2014,Fletcher2019}. It is thereby possible to extract the wavefunction of the emitted particles together with the emission probabilities and thus develop a complete description of the quantum many-body state~\cite{Bisognin_2019}. 

Finally, we comment on the possible influence of interactions. The considerations above are based on non-interacting particles, however, in a quantum Hall sample, interactions may lead to a spin-charge separation of incoming pulses~\cite{Ferraro_2014,Iyoda2024}. The spin and charge modes propagate with different velocities that we denote by $v_s$ and $v_c$. A pulse with a duration of $\tau_0$ then decoheres over a length scale given by $l_\text{dec}\simeq v_\text{int} \tau_0$, where $v_\text{int}^{-1}= v_s^{-1} - v_c^{-1}$ is the interaction-dependent velocity~\cite{Ferraro_2014,Cabart_2018}. 
The decoherence length in typical experiments is around $l_\text{dec} \simeq \SI{1}{\micro\meter}$~\cite{Freulon_2015,Marguerite_2016}, but it can be increased by reducing interchannel interactions~\cite{Altimiras_2010,Huynh_2012}. 
An outgoing electron-hole superposition will thus maintain its coherence over distances of $d< l_\text{dec}$, but the visibility is eventually reduced for $d\simeq l_\text{dec}$~\cite{Roussel_2017,Martin_2023b}. 
Still, it has been suggested that interactions may not be detrimental, and they may for example be used to encode the spin of an electron into spatially separated modes~\cite{Iyoda2024,Shimizu2024}. 

\section{Conclusions and outlook\label{sec:conc}}

We have presented a detailed Floquet-Nambu theory to describe the dynamic quantum transport in mesoscopic circuits involving superconductors. 
A central result is the Floquet-Nambu first-order correlation function in \cref{eq:GF-final}, from which one can express the outgoing current in \cref{eq:current-ac-right} in terms of the time-dependent voltage and the scattering matrix of the superconductor. 
Based on our formalism, we have analyzed the time-dependent current in a chiral edge state that is coupled to a superconductor. The degree of Andreev reflection at the superconductor determines the sign of the outgoing current, and it makes it possible to generate charge-neutral pulses that carry an equal amount of electrons and holes. For realistic parameters, we have found that an incoming single-electron state can be transformed into a coherent superposition of single-electron and single-hole states as it is reflected on the superconductor. To quantify the usefulness of the reflected charge pulses for quantum information purposes, we  considered their purity in the second part of our manuscript. Here, we identified the necessary operating conditions for generating pure charge qubits that can be expressed in terms of single-electron and single-hole states. 
Our formalism paves the way for future electron quantum optics applications that exploit the electron-hole degree of freedom. Thus, in a companion paper, we describe a tunable interferometer that makes it possible to generate and manipulate coherent superpositions of single-electron and single-hole states~\cite{Burset_short}. 

Finally, we conclude with an outlook on possible developments for the future, including potential experiments. First, we note that the magnetic fields required for the quantum Hall effect may constitute a technical challenge when combined with superconductivity. However, our formalism itself is general, and it can also be applied to scattering problems without edge states. Also experimentally, edge states are not necessary for implementing single-electron sources, although they are useful in the design of electronic interferometers. As such, experiments with single-electron emitters and superconductors may be realized with two-dimensional electron gases and no magnetic fields~\cite{Dubois_2013,Bauerle_2018}. Alternatively, electronic circuits could exploit other types of edge states, for example, those based on the topological quantum spin Hall effect~\cite{QSHI_2007,Wu2018}.
In terms of applications, there is a growing interest in developing flying qubits based on electrons as a way to circumvent the large hardware footprint of most solid-state architectures~\cite{Edlbauer2022}. Flying qubits are scalable and offer a universal set of gates for quantum computing~\cite{Furusawa2017}. Flying electronic qubits, combined with superconductors, could harness the spin entanglement of Cooper pairs for the design of novel devices~\cite{Burkard2000,Mazza2013}. 
As such, superconductivity appears to be a valuable addition to electron quantum optics, as it makes it possible to exploit the electron-hole degree of freedom. 

\acknowledgments
We thank A. A. Clerk, P. Dutta and J. Keeling for valuable discussions. 
The work was supported by Spanish CM ``Talento Program'' project No.~2019-T1/IND-14088 and No.~2023-5A/IND-28927, the Agencia Estatal de Investigaci\'on project No.~PID2020-117992GA-I00, No.~PID2024-157821NB-I00 and No.~CNS2022-135950 and through the ``María de Maeztu'' Programme for Units of Excellence in R\&D (CEX2023-001316-M), the CSIC/IUCRAN2022 under Grant No.~UCRAN20029, and Research Council of Finland through the Finnish Centre of Excellence in Quantum Technology (project~number~352925).

\appendix 

\section{Levitons below the gap} 
 \label{sec:app_large-D}

If the temperature and all relevant excitation energies are much smaller than the superconducting gap, $ k_BT,\hbar\Omega,\hbar/\tau_0 \ll \Delta$, we can evaluate the Nambu scattering matrix at $E=0$, which simplifies the further calculations. Specifically, \cref{eq:GF-final} becomes
\begin{equation}
\begin{split}
	 \hat{G} & (x,t;x',t') \simeq 
	\int_{-\infty}^{\infty} \frac{\mathrm{d}E}{h v_F} \e^{\mi E(t'-t)/\hbar} f_0(E) \\ 
 \times &\left\{
	\e^{\mi\mu_N(t'-t)/\hbar} [ J(t)J^*(t') - 1 ] \hat{M}_{e}(x';x)\right. \\
	 +& \left.\e^{-\mi\mu_N(t'-t)/\hbar} [ J^*(t)J(t') - 1 ] \hat{M}_{h}(x';x)
	\right\}, 
 \end{split}
\end{equation}
where $\mu_N$ is the chemical potential of the outgoing lead. In the main text we set it to zero, but we keep it here for the sake of clarity. We have also defined the matrices
\begin{equation}\label{eq:SM-large-D}
	\hat{M}_{\mu}(x';x) = 
	\begin{pmatrix}
		\e^{\mi k_F(x'-x)} \left|S_{e\mu}\right|^2 & \e^{\mi k_F(x'+x)} S_{e\mu}^*S_{h\mu} \\ \e^{-\mi k_F(x'+x)} S_{h\mu}^*S_{e\mu} & \e^{-\mi k_F(x'-x)} \left|S_{h\mu}\right|^2 
	\end{pmatrix} .
\end{equation}
The spatial and temporal parts of the excess correlation function are now decoupled. We can then focus on the time-dependent part to find the correlation functions for levitons and anti-levitons. Specifically, we have 
\begin{equation}
\begin{split}
G_{+}(t,t')=&\int_{-\infty}^{\infty} \! \frac{\mathrm{d}E}{hv_F} \e^{\mi E(t'-t)/\hbar} f_0(E) \left[ J(t)J^*(t') - 1 \right]\\ 
	=&
	\frac{\tau_0}{v_F\pi} \frac{1}{t'+ \mi \tau_0} \frac{1}{t - \mi \tau_0} = \frac{1}{v_F}\Psi^*_{-}(t')\Psi_{-}(t)  
\end{split}
\end{equation}
for levitons, while for anti-levitons we find
\begin{equation}
\begin{split}
G_{-}(t,t')	= &\int_{-\infty}^{\infty} \! \frac{\mathrm{d}E}{hv_F} \e^{\mi E(t'-t)/\hbar} f_0(E) \left[ J^*(t)J(t') - 1 \right] \\
      	=&
	-\frac{\tau_0}{v_F\pi} \frac{1}{t'- \mi \tau_0} \frac{1}{t+ \mi \tau_0} = -\frac{1}{v_F}\Psi^*_{+}(t')\Psi_{+}(t),
\end{split}
\end{equation}
where we have introduced the wave functions
\begin{equation}
\Psi_{\pm}(t) = \sqrt{\frac{\tau_0}{\pi}} \frac{1}{t\pm \mi \tau_0}.
\end{equation}
The excess-correlation functions are orthogonal, 
\begin{equation}
		v_F \int_{-\infty}^{\infty}\!\mathrm{d}t G_\pm(t_1,t) G_\mp(t,t_2) = 0,
\end{equation}
and they correspond to pure states since 
\begin{equation}
 \label{eq:orthonormal-largeD}
		v_F \int_{-\infty}^{\infty}\mathrm{d}t G_\pm(t_1,t) G_\pm(t,t_2) = \mp G_\pm(t_1,t_2) .
\end{equation}
For this reason, the outgoing correlation function reads
\begin{equation}\label{eq:app:excess}
\begin{split}
    	\hat{G} (x,t;x',t') &\simeq 
	\e^{\mi\mu_N(t'-t)/\hbar} G_{-}(t,t') \hat{M}_{e}(x';x) \\
	& + \e^{-\mi\mu_N(t'-t)/\hbar} G_{+}(t,t') \hat{M}_{h}(x';x).
\end{split}
\end{equation}
Inserting this expression into \cref{eq:current}, we find
\begin{equation}
	I(t)= \frac{e^2}{h} \left( \left|S_{ee}(0)\right|^2 - \left|S_{he}(0)\right|^2 \right) V(t),
\end{equation}
showing that the outgoing current is a sum of contributions from levitons and antilevitons, weighted by the probabilities for normal transmission and Andreev conversion, respectively. Also, since there are no transmissions into the superconductor above the gap, the scattering probabilities fulfill the unitarity condition $\left|S_{ee}\right|^2 + \left|S_{he}\right|^2=1$. We can thus define $P_A=\left|S_{he}\right|^2$ as the probability for Andreev conversion and write the current as 
\begin{equation}\label{eq:current_Large-Delta-app}
	I(t)= \frac{e^2}{h} \left( 1 - 2P_A \right) V(t). 
\end{equation}

\section{Frequency domain\label{sec:app-freq}}

Here, we evaluate the excess correlation function in the Fourier domain. The excess correlation function in the time domain is given by \cref{eq:GF-final} and reads
\begin{widetext}
\begin{equation}
	\begin{split}
		\hat{G} (t,t') = 
		 &\sum_{n,m} \left\{ 
		  J_{n}J^*_{m} \e^{-\mi n \Omega t}  \e^{\mi m \Omega t'} \int_{-n\Omega}^{0}\frac{\mathrm{d}E}{h v_F} \e^{-\mi E(t- t')/\hbar} \hat{M}_{ee}(E_{m},E_{n}) \right. 
		\\
		& \phantom{\sum_{n,m}}\left. +
	J_{n}^*  J_{m}\e^{\mi n \Omega t} \e^{-\mi m \Omega t'} \int_{n\Omega}^{0}\frac{\mathrm{d}E}{h v_F} \e^{-\mi E(t- t')/\hbar}\hat{M}_{hh}(E_{-m},E_{-n}) 
		\right\},
	\end{split} \label{eq:GexcT0}
\end{equation}
where we have omitted the spatial dependence and taken the temperature to be zero. Since the voltage drive is periodic, $V(t)=V(t+\mathcal{T})$, we in turn have that $\hat{G}(t+\mathcal{T},t'+\mathcal{T})=\hat{G}(t,t')$. Therefore, we define the time variables
\begin{equation}
\bar t =(t+t')/2 \quad\mathrm{and}\quad \tau=t-t' ,
\end{equation}
and write the correlation function as
\begin{equation}
	\begin{split}
		\hat{G} (\bar{t}+\tau/2,\bar{t}-\tau/2) = 
		\sum_{n,m} & \left\{ 
		J_{n} J^*_{m} \e^{-\mi (n-m)\Omega\bar{t}} \int_{-n\Omega}^{0}\frac{\mathrm{d}E}{h v_F} \e^{-\mi E \tau/\hbar}  \e^{-\mi (n+m)\Omega\tau/2} \hat{M}_{ee}(E_{m},E_{n}) \right. 
		\\
		+ & \left. 
		J_{n}^*J_{m} \e^{\mi (n-m)\Omega\bar{t}} \int_{n\Omega}^{0}\frac{\mathrm{d}E}{h v_F} \e^{-\mi E\tau/\hbar} \e^{\mi (n+m)\Omega\tau/2} \hat{M}_{hh}(E_{-m},E_{-n}) 
		\right\} .
	\end{split} \label{eq:GexcT02}
\end{equation}
We can then Fourier transform the correlation function. First, we perform a discrete Fourier transform,
\begin{equation}\label{eq:disFT}
	\hat{G}_l(\tau) =  \int_0^\mathcal{T} \frac{\mathrm{d}\bar{t} }{\mathcal{T}}\e^{\mi l \Omega \bar{t}} \hat{G} (\bar{t}+\tau/2,\bar{t}-\tau/2),
\end{equation}
since the function is periodic in $\bar{t}$. Next, we perform a continuous Fourier transform over the remaining time variable, 
\begin{equation}\label{eq:contFT}
	\hat{G}_l(\omega) =  \int_{-\infty}^{\infty} \mathrm{d}\tau \e^{\mi \omega \tau} \hat{G}_l(\tau) .
\end{equation}
We then insert \cref{eq:GexcT02} into \cref{eq:disFT} to find
\begin{equation}
	\begin{split}
		\hat{{G}}_l (\tau) = 
		\sum_{n,m} & \left\{ 
		J_{n} J^*_{m} \left(  \int_0^\mathcal{T} \frac{\mathrm{d}\bar{t} }{\mathcal{T}} \e^{\mi (l-n+m) \Omega \bar{t}} \right) \int_{-n\Omega}^{0}\frac{\mathrm{d}E}{h v_F} \e^{-\mi E \tau/\hbar}  \e^{-\mi (n+m)\Omega\tau/2} \hat{M}_{ee}(E_{m},E_{n}) \right. 
		\\
		 + & \left.
		J_{n}^*J_{m} \left(  \int_0^\mathcal{T} \frac{\mathrm{d}\bar{t} }{\mathcal{T}} \e^{\mi (l+n-m) \Omega \bar{t}} \right) \int_{n\Omega}^{0}\frac{\mathrm{d}E}{h v_F} \e^{-\mi E\tau/\hbar} \e^{\mi (n+m)\Omega\tau/2} \hat{M}_{hh}(E_{-m},E_{-n}) 
		\right\} .
	\end{split} \label{eq:GexcFT1}
\end{equation}
The integrals over time yield Kronecker deltas, and inserting the result into \cref{eq:contFT}, we find
\begin{equation}
	\begin{split}
		\hat{G}_l (\omega) = 
		\sum_{n} & \left\{ 
		J_{n} J^*_{n-l} \int_{-n\Omega}^{0}\frac{\mathrm{d}E}{h v_F} \hat{M}_{ee}(E_{n-l},E_{n}) \left(  \int_{0}^{\mathcal{T}} \frac{\mathrm{d}\bar{t} }{\mathcal{T}} \e^{\mi \omega \tau} \e^{-\mi E \tau/\hbar}  \e^{-\mi (2n-l)\Omega\tau/2} \right) \right. 
		\\
		+ & \left. 
		J_{n}^*J_{n+l}  \int_{n\Omega}^{0}\frac{\mathrm{d}E}{h v_F}  \hat{M}_{hh}(E_{-n-l},E_{-n})  \left(  \int_{0}^{\mathcal{T}} \frac{\mathrm{d}\bar{t} }{\mathcal{T}} \e^{\mi \omega \tau} \e^{-\mi E\tau/\hbar} \e^{\mi (2n+l)\Omega\tau/2} \right) 
		\right\} .
	\end{split} \label{eq:GexcFT2}
\end{equation}
Now, the integrals over time become Dirac delta functions and those related to electrons can be evaluated as
\begin{equation}
\int_{-n\Omega}^{0} \mathrm{d}E M(E) \delta(\omega_0-E) = \left\{\begin{array}{cr}
M(\omega_0) \Theta[\omega_0-n\Omega]\Theta[-\omega_0], & n\geq 0 \\
-M(\omega_0) \Theta[|n|\Omega-\omega_0]\Theta[\omega_0], & n< 0 
\end{array}\right. ,
\end{equation}
while for those related to  holes, we find 
\begin{equation}
\int_{n\Omega}^{0} \mathrm{d}E M(E) \delta(\omega_0-E) = \left\{\begin{array}{cr}
	-M(\omega_0) \Theta[n\Omega-\omega_0]\Theta[\omega_0], & n\geq 0 \\
	M(\omega_0) \Theta[|n|\Omega+\omega_0]\Theta[-\omega_0], & n< 0 
\end{array}\right. ,
\end{equation}
where $\Theta(\omega)$ is the Heaviside step function. The correlation function in the frequency domain is then 
\begin{equation}
	\begin{split}
		\hat{G}_l (\omega) = {}&
		\frac{1}{h v_F} \hat{M}_{ee}\left(\omega -l\Omega/2,\omega +l\Omega/2\right) \sum_{n} s_n J_{n} J^*_{n-l} \Theta\left[s_n\left( n\Omega- \omega-l\Omega/2  \right)\right]\Theta\left[s_n\left(\omega +  l\Omega/2\right)\right]
		\\
		 -& 
			\frac{1}{h v_F} \hat{M}_{hh}\left(\omega -l\Omega/2,\omega + l\Omega/2\right) \sum_{n} s_n J^*_{n} J_{n+l} \Theta\left[s_n\left( n\Omega + \omega+ l\Omega/2\right)\right]\Theta\left[-s_n\left( \omega + l\Omega/2 \right)\right] ,
	\end{split} \label{eq:GexcFreq}
\end{equation}
where $s_n=\sgn(n)$ denotes the sign of $n$. 
\end{widetext}

We can repeat the same procedure for the convolution of the excess correlation functions with itself, that is, $\hat{G}\circ\hat{G}\equiv \hat{G}_2$, see \cref{eq:pureFermi}. Namely, 
\begin{equation}\label{eq:G2-time}
	\hat{G}_2 ( t_1, t_2)= \int_{-\infty}^{\infty} \mathrm{d}t \hat{G} ( t_1, t)\hat{G} ( t, t_2) .
\end{equation}
We can again change the time variables and write
\begin{equation}\label{eq:G2-time2}
	\hat{G}_2 (\bar{t}+\tau/2, \bar{t}-\tau/2)= \int_{-\infty}^{\infty} \mathrm{d}t \hat{G} (\bar{t}+\tau/2, t)\hat{G} ( t, \bar{t}-\tau/2).
\end{equation}
Assuming that $\hat{G}_2$ is periodic in $\bar{t}$ but not in $\tau$, we can Fourier transform it as
\begin{equation}\label{eq:G2-freq1}
\begin{split}
    [\hat{G}_2]_l(\omega) ={}& 
	  \int_0^\mathcal{T} \frac{\mathrm{d}\bar{t} }{\mathcal{T}} \e^{\mi l \Omega \bar{t}} \\
 \times & \int_{-\infty}^{\infty} \mathrm{d}\tau \e^{\mi \omega \tau}
	  	\hat{G}_2 (\bar{t}+\tau/2, \bar{t}-\tau/2).
\end{split}
\end{equation}
We now use the inverse Fourier transform, 
\begin{equation}\label{eq:invFT}
	\hat{G}(\bar{t}+\tau/2, \bar{t}-\tau/2)= \sum_l \e^{-\mi l \Omega \bar{t}} \int_{-\infty}^{\infty}  \frac{\mathrm{d} \omega}{2\pi} \e^{-\mi \omega \tau} \hat{G}_l(\omega),
\end{equation}
and then obtain
\begin{align}
     [\hat{G}_2]_l(\omega) = {}&
	\sum_{n_1,n_2}
	\int_{-\infty}^{\infty}  \frac{\mathrm{d} \omega_1}{2\pi} \int_{-\infty}^{\infty}  \frac{\mathrm{d} \omega_2}{2\pi} \hat{G}_{n_1}(\omega_1)  
	 \hat{G}_{n_2}(\omega_2) 
	\notag \\ {}& \times 
	\int_0^\mathcal{T} \frac{\mathrm{d}\bar{t}}{\mathcal{T}} \e^{\mi(\omega_2-\omega_1 - (n_1+n_2-2l)\Omega/2) \bar{t}} 
        \notag \\ {}& \times 
	\int_{-\infty}^{\infty} \mathrm{d}\tau \e^{\mi(\omega - (n_1-n_2)\Omega/4-(\omega_1+\omega_2)/2)\tau} 
        \notag \\ {}& \times 
	 \int_{-\infty}^{\infty} \mathrm{d}t \e^{\mi(\omega_1-\omega_2 - (n_1+n_2)\Omega/2)t} .
    \label{eq:G2-freq2}
\end{align}
Carrying out the integrals, we find after some simplifications the compact expression
\begin{equation}\label{eq:G2-freq}
\begin{split}
    [\hat{G}_2]_l(\omega) = \sum_m & G_{l-m} \left(\omega + (l-m)\Omega/2\right)\\
    \times & G_{m} \left(\omega + (3l-2m)\Omega/4 \right).
\end{split}
\end{equation}

For the analysis of the purity in \cref{sec:purity}, it is important to identify components corresponding to  positive and negative frequencies. To do so, it is useful to begin with the Fourier transform of the correlation function,
\begin{equation}
\begin{split}    
	\hat{G}(\omega_1,\omega_2) = &
	\int_{-\infty}^{\infty}\int_{-\infty}^{\infty}  \mathrm{d}t_1 \mathrm{d}t_2 \e^{\mi \omega_1 t_1}  
	  \e^{-\mi \omega_2 t_2} 
	\hat{G}(t_1,t_2) \\
	=& 	\int_{0}^{\mathcal{T}} \frac{ \mathrm{d}\bar{t} }{\mathcal{T}} \e^{\mi (\omega_1 -\omega_2) \bar{t}}\int_{-\infty}^{\infty} \mathrm{d}\tau \e^{\mi (\omega_1 + \omega_2)\tau/2} \\
 & \times \hat{G}(\bar{t}+\tau/2, \bar{t}-\tau/2),
\end{split}
\end{equation}
which can be further rewritten as
\begin{equation}
\begin{split}
	\hat{G}(\omega_1,\omega_2) = &
	\int_{0}^{\mathcal{T}} \frac{\mathrm{d}\bar{t}}{\mathcal{T}} \e^{\mi (\omega_1 -\omega_2) \bar{t}}  
	\int_{-\infty}^{\infty}  \mathrm{d}\tau \e^{\mi (\omega_1 + \omega_2) \tau/2} \\
 &\times \sum_l \e^{-\mi l \Omega \bar{t}} 
	\int_{-\infty}^{\infty}  \frac{\mathrm{d}\omega}{2\pi}  \e^{-\mi \omega \tau } 
	\hat{G}_l(\omega) 
	\\ = & 
	\sum_l \int_{-\infty}^{\infty}  \frac{\mathrm{d}\omega}{2\pi}  \hat{G}_l(\omega) \\
 &\times 
	\int_{0}^{\mathcal{T}} \frac{\mathrm{d}\bar{t}}{\mathcal{T}} \e^{\mi (\omega_1 - \omega_2 - l\Omega) \bar{t} }\\
 &\times   
	\int_{-\infty}^{\infty} \mathrm{d}\tau \e^{\mi [(\omega_1 + \omega_2)/2 -\omega] \tau} . 
\end{split}
\end{equation}
Carrying out the integrals, we find 
\begin{equation}
\label{eq:2freqFT}
		\hat{G}(\omega_1,\omega_2) = 
		2\pi \sum_l \hat{G}_l[(\omega_1+\omega_2)/2] 
		\delta(\omega_1-\omega_2 - l \Omega) .
\end{equation}
For positive frequencies $\omega_{1,2}\geq0$, we have 
\begin{equation}
\left| \omega_1-\omega_2\right| \leq \omega_1+\omega_2 
 \Rightarrow  \omega_1+\omega_2 \geq |l| \Omega, 
\end{equation}
such that we can write 
\begin{equation}
	[\hat{G}_{++}]_l(\omega) =  
		 \hat{G}_l(\omega)\Theta(\omega - |l| \Omega/2).
\end{equation}

\section{Analysis of purity 
\label{sec:app-purity}}

\subsection{Levitons}

Here, we discuss several properties of levitons which support our conclusions in \cref{sec:purity,sec:results-2}. First, we show that the single-particle states generated by the Lorentzian drive in \cref{eq:volt-lor} are always pure above the Fermi level, before scattering off the superconductor.

Following Ref.~\cite{Roussel_2021}, we describe the generation of exactly one electronic excitation with integer charge $q$ per period of the drive. Each excitation is represented by the state~$\ket{\psi^l_q}$, where the index $l$ labels the period of the emission. In the ideal case, excitations from different periods are orthogonal and the correlation function reads
\begin{equation}\label{eq:app-G-ses}
    \mathbf G = \sum_{l} \ket{\psi^l_q} \!\bra{\psi^l_q}.
\end{equation}
Here, we do not need a description in Nambu space since we are not yet considering superconducting correlations. 

For a periodic train of well-separated levitons, the wavefunction in the frequency domain is~\cite{Grenier_2013,Moskalets_2015,Moskalets_2020,Roussel_2021}
\begin{equation}\label{eq:lev-train-wf-freq}
    \psi_{q}(\omega)=\sqrt{4\pi v_F\tau_0}\mathcal{L}_{q-1}(2\omega\tau_0)\Theta(\omega) \me^{-\omega\tau_0}, 
\end{equation}
where $\tau_0$ the halfwidth of the Lorentzian, and $\mathcal{L}_{q}(x)$ is the $q$'th Laguerre polynomial. 
Thus, for the $l$'th leviton with charge $q=1$, the wave function reads 
\begin{equation}\label{eq:lev-wf-freq}
    \psi^l_{1}(\omega)=\sqrt{4\pi v_F\tau_0} \Theta(\omega) \me^{-\omega\tau_0} \me^{\mi \omega l \mathcal{T}},
\end{equation}
since $\mathcal{L}_{0}(x)=1$. With these expressions we find
\begin{equation}\label{eq:exc-G-lev-train-freq}
    G(\omega,\omega') = 4\pi v_F\tau_0 \Theta(\omega)\Theta(\omega') \me^{-(\omega+\omega')\tau_0} \sum_{l} \me^{\mi(\omega+\omega')l \mathcal{T}} ,
\end{equation}
which is only non-zero for positive frequencies, such that $\mathbf G=\mathbf G_{++}$. We can therefore write
\begin{equation}\label{eq:app-G-lev-train}
    \mathbf G_{++} = \sum_{l} \ket{\Psi^l_{1}} \!\bra{\Psi^l_{1}},
\end{equation}
and we then see that $\mathbf G_{++}^2=\mathbf G_{++}$, if  $\langle \Psi^{l}_{1} | \Psi^{l'}_{1} \rangle=\delta_{ll'}$.

\subsection{Small and large driving frequencies}

We now focus on the scattering of pulses on the superconductor. In particular, we evaluate the purity of states above the Fermi level for small and large excitation energies compared to the superconducting gap. 

From \cref{eq:GexcFreq,eq:G2-freq} we first find that
\begin{equation}
    \mathrm{Tr}\{\mathbf{\hat{G}}_{++}\}= 
    \sum\limits_{n\geq0} |J_n|^2 \int_{0}^{n \Omega} \frac{\mathrm{d}\omega}{2\pi v_F} \sum\limits_{\mu=e,h} |S_{\mu e}(\omega)|^2, 
    \label{eq:G-trace-lev}
\end{equation}
and
\begin{equation}
    \label{eq:G2-trace-lev}
    \begin{split}
    \mathrm{Tr}\{\mathbf{\hat{G}}_{++}^2\}= &
    \sum\limits_{m,n_1,n_2} J^{*}_{n_1} J_{n_1-m} J^{*}_{n_2} J_{n_2-m} 
    \\&
    \times\int_{\mathrm{max}(0,-m) \Omega}^{\mathrm{min}(n_1-m,n_2) \Omega} 
    \frac{\mathrm{d}\omega}{2\pi v_F} \\
    &\times\sum_{\mu,\nu=e,h} |S_{\mu e}(\omega)|^2 |S_{\nu e}(\omega+m\Omega)|^2.    
    \end{split}
\end{equation}
Here, the trace is given as 
\begin{equation}
    \mathrm{Tr}\{\mathbf{\hat{G}}\}= \int \frac{\mathrm{d}\omega}{2\pi} \mathrm{Tr}_\text{N} \{\hat{G}_{l=0}(\omega)\},
\end{equation} 
where $\mathrm{Tr}_\text{N}\{\dots\}$ is the trace in Nambu space. We then evaluate the purity for perfect Andreev conversion ($\alpha=0$) and complete normal transmission ($\alpha=1$).

For complete normal transmission, we have $\mathbf{\hat{G}^\text{out}}=\mathbf{\hat{G}^\text{in}}$, since $|S_{ee}(\omega)|^2=1$ and $|S_{he}(\omega)|^2=0$ for all energies. 
Thus, the state is pure at the single-particle level since there are no Andreev reflections or quasi-particle transfers into the superconductor. 
Moreover, for the Lorentzian drive, the state is also pure above the Fermi level, since all excited particles have positive energies. 
Using the Floquet scattering amplitudes in \cref{table:voltages}, we can also show that the state indeed is pure above the Fermi level using \cref{eq:G-trace-lev,eq:G2-trace-lev}. We then find
\begin{equation}
    \gamma_+\simeq \frac{ \tr \{ (\mathbf{\hat{G}^\text{in}_{++}})^2 \}}{\tr \{\mathbf{\hat{G}^\text{in}_{++}} \}} \simeq \tr \{\mathbf{\hat{G}^\text{in}_{++}} \} 
\end{equation}
for the incoming pulses. In this expression, we only consider particles that are excited above the Fermi level. As a result, \cref{eq:G-trace-lev} reduces to
\begin{equation}
    \gamma_+\simeq \sum\limits_{n>0} n |J_n|^2,\quad \alpha=1.
\end{equation}
We can then show that $\gamma_+=1$ for a sequence of levitons. However, for all other voltage drives, one finds $\gamma_+<1$, since the sum is restricted to non-negative integers. 

\begin{figure*}[t]
    \includegraphics[width=1.0\textwidth]{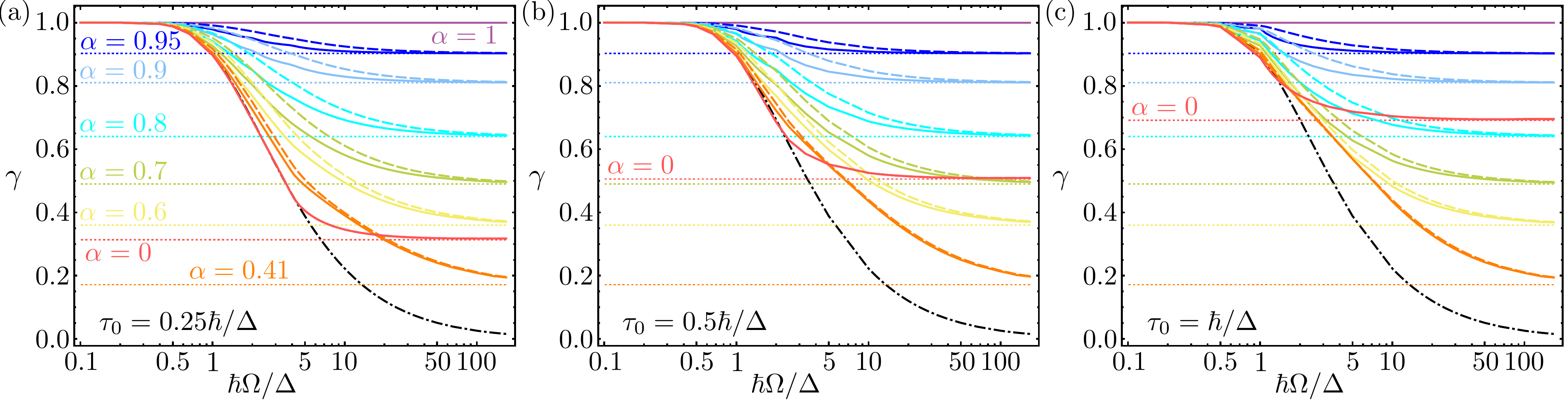}
	\caption{\label{fig:purity-app}
    Purity for Lorentzian pulses. (a) Purity for  pulses with $q=1$ and $\tau_0=0.25\hbar/\Delta$ [same values as in \cref{fig:purity}(a)]. (b)~Purity for pulses with $q=1$ and $\tau_0=0.5\hbar/\Delta$. (c) Purity for pulses with $q=1$ and $\tau_0=\hbar/\Delta$. For each value of $\alpha\neq0$, the dotted lines show the conversion probability $\alpha^2$, which captures the limit of $\hbar\Omega\gg\Delta$. For $\alpha=0$, the red dotted line is given by \cref{eq:app-purity-0}. The dot-dashed black line shows \cref{eq:antilev-prob2}. The purity and the purity above the Fermi level are the same, $\gamma=\gamma_+$. 
	}
\end{figure*}

For perfect Andreev conversion ($\alpha=0$), we find  $|S_{ee}(\omega)|^2=0$, but $|S_{he}(\omega)|^2=1$ only for energies below the gap. For energies above the gap, quasiparticles are transmitted into the superconductor which complicates the integrals in \cref{eq:G-trace-lev,eq:G2-trace-lev}, since the probability for Andreev conversion becomes
\begin{equation}\label{eq:Andreev-prob-over-Delta}
    |S_{he}(\omega)|^2= \left( \hbar\omega/\Delta - \sqrt{ \left( \hbar\omega/\Delta \right)^2-1}\right)^2, \quad \hbar\omega>\Delta,
\end{equation}
which quickly approaches zero as the energy increases. 

If the driving frequency is much larger that the driving amplitude and the superconducting gap, $\hbar\Omega\gg|eV_\text{ac}|,\Delta$, we can assume that the term with $n=1$ in \cref{eq:G-trace-lev} already corresponds to energies that are so large that we have $|S_{he}(\omega + \Omega)|^2\ll 1$ in \cref{eq:Andreev-prob-over-Delta}. 
We then find
\begin{equation}\label{eq:G2-small-delta}
    \begin{split}
    \mathrm{Tr}\{\mathbf{\hat{G}_{++}}^2\}\simeq & 
     \frac{|J_1|^4}{2\pi v_F} \left( \int_{0}^{\Delta/\hbar} \mathrm{d}\omega + \int_{\Delta/\hbar}^{\Omega} \mathrm{d}\omega 
    |S_{he}(\hbar\omega)|^4
    \right) \\
    = & \frac{\Delta |J_1|^4}{h v_F} \frac{1}{15} \left[ 16+ 15\chi-40\chi^3 + 24\chi^5
    \right. \\
    &\left. 
    + \left( \chi^2 - 1 \right)^{3/2}\left(4 - 24\chi^2\right) \right],
        \end{split}
\end{equation}
and
\begin{equation}\label{eq:G-small-delta}
    \begin{split}
    \mathrm{Tr}\{\mathbf{\hat{G}_{++}}\}\simeq &
    \frac{|J_1|^2}{2\pi v_F} \left( \int_{0}^{\Delta/\hbar} \mathrm{d}\omega + \int_{\Delta/\hbar}^{\Omega} \mathrm{d}\omega 
    |S_{he}(\hbar\omega)|^2
    \right) \\    
    = & \frac{\Delta |J_1|^2}{h v_F} \frac{1}{3}\left[4 - 3\chi + 2\chi^3 -2\left(\chi^2 - 1\right)^{3/2} \right],
    \end{split}
\end{equation}
where $\chi=\hbar\Omega/\Delta$. For $\chi\gg1$, all powers of $\chi$ cancel in \cref{eq:G2-small-delta,eq:G-small-delta} and only the constant terms $16/15$ in \cref{eq:G2-small-delta} and $4/3$ in \cref{eq:G-small-delta} survive. 
We then have
\begin{equation}
    \lim\limits_{\chi\to\infty} \mathrm{Tr}\{\mathbf{\hat{G}_{++}}^2\}\simeq 
    16 \Delta |J_1|^4/( 15 h v_F) 
\end{equation}
and
\begin{equation}
    \lim\limits_{\chi\to\infty} \mathrm{Tr}\{\mathbf{\hat{G}_{++}}\}\simeq  
    4 \Delta |J_1|^2/(3h v_F),
\end{equation}
and we thereby find
\begin{equation}
\label{eq:app-purity-gen}
    \gamma_+  \simeq   4 |J_1|^2/5 .
\end{equation}
For levitons with $q=1$, the purity 
reduces to
\begin{equation}\label{eq:app-purity-0}
    \gamma_+ \simeq 16 \sinh^2(\Omega\tau_0)\me^{-2\Omega\tau_0}/5.
\end{equation}
In \cref{fig:purity}, this expression agrees with our results for perfect Andreev reflection ($\alpha=0$) with $\hbar\Omega\gg\Delta$. 

\subsection{Intermediate driving frequencies}

Finally, we consider the driving frequency to be on the order of the superconducting gap. If a single particle is sent towards the superconductor, it can either be converted into a hole with probability~$P_+$, or be absorbed by the superconductor with probability~$1-P_+$. Unlike for small driving frequencies, the wavefunctions of the holes can now be highly deformed. Furthermore, for periodic emission, the scattering events are independent and the outgoing coherence function can be written as $\hat{G}^{\text{out}} = P_+\hat{G}^{\text{train}}$, where $\hat{G}^{\text{train}}$ is a pure train of independent holes. Thus, in this case, we have $\gamma = \gamma_+ = P_+$. For levitons, the outgoing state is a sequence of deformed antilevitons. Furthermore, the probability $P_+$ is given by
\begin{equation}\label{eq:antilev-prob}
    P_+ = \int_{-\infty}^{\infty} \frac{\mathrm{d}\omega}{2\pi v_F}
	\left|\psi^l_{1}(\omega)\right|^2\left|S_{he}(\omega)\right|^2,
\end{equation}
where $\psi_1^l$ is the wavefunction of a single leviton. If we assume that the levitons are well separated, such that $\Omega\tau_0 \ll 1$, we then have $\left|\psi^l_{1}(\omega)\right|^2 \simeq 2\tau_0 \me^{-2\omega\tau_0}$ and
\begin{equation}
\begin{split}    
     P_+ = & 2\tau_0\int_{0}^{\infty} \mathrm{d}\omega \me^{-2\omega\tau_0} \left|S_{he}(\omega)\right|^2 
     \\ 
    = & 
    2\tau_0 \left( \int_{0}^{\Delta/\hbar} \mathrm{d}\omega \me^{-2\omega\tau_0} + \int_{\Delta/\hbar}^{\infty} \mathrm{d}\omega \me^{-2\omega\tau_0} 
    |S_{he}(\omega)|^2 \right)
     \\
    =& 1 + \frac{1+2\tau_0\Delta/\hbar}{\left(\tau_0\Delta/\hbar\right)^2} \me^{-2\tau_0\Delta/\hbar} - 2 \mathcal{K}_2\left( 2\tau_0\Delta/\hbar \right), 
    \label{eq:antilev-prob2}
\end{split}
\end{equation}
where $\mathcal{K}_n(z)$ is the modified Bessel function of the second kind. In \cref{fig:purity,fig:purity-app}, we compare \cref{eq:antilev-prob2} with the purity for periodic levitons and we find good agreement when the excitation energy is much smaller than the gap, $\hbar\Omega\ll\Delta$, since $P_+\simeq1$. The approximation also works well in the region around $\hbar\Omega\simeq \Delta$ as long as $\Omega\tau_0\lesssim1$. However, for $\hbar\Omega\gg\Delta$, the probability vanishes, $P_+\simeq0$, which only corresponds to the behavior of the purity for $\Omega\tau_0\to0$. 
It is thus more appropriate in this regime to consider the asymptotic limit given by \cref{eq:app-purity-0}. 


\begin{thebibliography}{115}%
\makeatletter
\providecommand \@ifxundefined [1]{%
 \@ifx{#1\undefined}
}%
\providecommand \@ifnum [1]{%
 \ifnum #1\expandafter \@firstoftwo
 \else \expandafter \@secondoftwo
 \fi
}%
\providecommand \@ifx [1]{%
 \ifx #1\expandafter \@firstoftwo
 \else \expandafter \@secondoftwo
 \fi
}%
\providecommand \natexlab [1]{#1}%
\providecommand \enquote  [1]{``#1''}%
\providecommand \bibnamefont  [1]{#1}%
\providecommand \bibfnamefont [1]{#1}%
\providecommand \citenamefont [1]{#1}%
\providecommand \href@noop [0]{\@secondoftwo}%
\providecommand \href [0]{\begingroup \@sanitize@url \@href}%
\providecommand \@href[1]{\@@startlink{#1}\@@href}%
\providecommand \@@href[1]{\endgroup#1\@@endlink}%
\providecommand \@sanitize@url [0]{\catcode `\\12\catcode `\$12\catcode `\&12\catcode `\#12\catcode `\^12\catcode `\_12\catcode `\%12\relax}%
\providecommand \@@startlink[1]{}%
\providecommand \@@endlink[0]{}%
\providecommand \url  [0]{\begingroup\@sanitize@url \@url }%
\providecommand \@url [1]{\endgroup\@href {#1}{\urlprefix }}%
\providecommand \urlprefix  [0]{URL }%
\providecommand \Eprint [0]{\href }%
\providecommand \doibase [0]{https://doi.org/}%
\providecommand \selectlanguage [0]{\@gobble}%
\providecommand \bibinfo  [0]{\@secondoftwo}%
\providecommand \bibfield  [0]{\@secondoftwo}%
\providecommand \translation [1]{[#1]}%
\providecommand \BibitemOpen [0]{}%
\providecommand \bibitemStop [0]{}%
\providecommand \bibitemNoStop [0]{.\EOS\space}%
\providecommand \EOS [0]{\spacefactor3000\relax}%
\providecommand \BibitemShut  [1]{\csname bibitem#1\endcsname}%
\let\auto@bib@innerbib\@empty
\bibitem [{\citenamefont {Bocquillon}\ \emph {et~al.}(2014)\citenamefont {Bocquillon}, \citenamefont {Freulon}, \citenamefont {Parmentier}, \citenamefont {Berroir}, \citenamefont {Pla\c{c}ais}, \citenamefont {Wahl}, \citenamefont {Rech}, \citenamefont {Jonckheere}, \citenamefont {Martin}, \citenamefont {Grenier}, \citenamefont {Ferraro}, \citenamefont {Degiovanni},\ and\ \citenamefont {F\'eve}}]{Erwann_AdP}%
  \BibitemOpen
  \bibfield  {author} {\bibinfo {author} {\bibfnamefont {E.}~\bibnamefont {Bocquillon}}, \bibinfo {author} {\bibfnamefont {V.}~\bibnamefont {Freulon}}, \bibinfo {author} {\bibfnamefont {F.~D.}\ \bibnamefont {Parmentier}}, \bibinfo {author} {\bibfnamefont {J.-M.}\ \bibnamefont {Berroir}}, \bibinfo {author} {\bibfnamefont {B.}~\bibnamefont {Pla\c{c}ais}}, \bibinfo {author} {\bibfnamefont {C.}~\bibnamefont {Wahl}}, \bibinfo {author} {\bibfnamefont {J.}~\bibnamefont {Rech}}, \bibinfo {author} {\bibfnamefont {T.}~\bibnamefont {Jonckheere}}, \bibinfo {author} {\bibfnamefont {T.}~\bibnamefont {Martin}}, \bibinfo {author} {\bibfnamefont {C.}~\bibnamefont {Grenier}}, \bibinfo {author} {\bibfnamefont {D.}~\bibnamefont {Ferraro}}, \bibinfo {author} {\bibfnamefont {P.}~\bibnamefont {Degiovanni}},\ and\ \bibinfo {author} {\bibfnamefont {G.}~\bibnamefont {F\'eve}},\ }\bibfield  {title} {\bibinfo {title} {Electron quantum optics in ballistic chiral conductors},\ }\href {https://doi.org/10.1002/andp.201300181} {\bibfield
  {journal} {\bibinfo  {journal} {Ann. Phys.}\ }\textbf {\bibinfo {volume} {526}},\ \bibinfo {pages} {1} (\bibinfo {year} {2014})}\BibitemShut {NoStop}%
\bibitem [{\citenamefont {Splettstoesser}\ and\ \citenamefont {Haug}(2017)}]{Janine_PSS}%
  \BibitemOpen
  \bibfield  {author} {\bibinfo {author} {\bibfnamefont {J.}~\bibnamefont {Splettstoesser}}\ and\ \bibinfo {author} {\bibfnamefont {R.~J.}\ \bibnamefont {Haug}},\ }\bibfield  {title} {\bibinfo {title} {Single-electron control in solid state devices},\ }\href {https://doi.org/10.1002/pssb.201770217} {\bibfield  {journal} {\bibinfo  {journal} {Phys. Status Solidi B}\ }\textbf {\bibinfo {volume} {254}},\ \bibinfo {pages} {1770217} (\bibinfo {year} {2017})}\BibitemShut {NoStop}%
\bibitem [{\citenamefont {B\"auerle}\ \emph {et~al.}(2018)\citenamefont {B\"auerle}, \citenamefont {Glattli}, \citenamefont {Meunier}, \citenamefont {Portier}, \citenamefont {Roche}, \citenamefont {Roulleau}, \citenamefont {Takada},\ and\ \citenamefont {Waintal}}]{Waintal_RPP}%
  \BibitemOpen
  \bibfield  {author} {\bibinfo {author} {\bibfnamefont {C.}~\bibnamefont {B\"auerle}}, \bibinfo {author} {\bibfnamefont {D.~C.}\ \bibnamefont {Glattli}}, \bibinfo {author} {\bibfnamefont {T.}~\bibnamefont {Meunier}}, \bibinfo {author} {\bibfnamefont {F.}~\bibnamefont {Portier}}, \bibinfo {author} {\bibfnamefont {P.}~\bibnamefont {Roche}}, \bibinfo {author} {\bibfnamefont {P.}~\bibnamefont {Roulleau}}, \bibinfo {author} {\bibfnamefont {S.}~\bibnamefont {Takada}},\ and\ \bibinfo {author} {\bibfnamefont {X.}~\bibnamefont {Waintal}},\ }\bibfield  {title} {\bibinfo {title} {Coherent control of single electrons: {A} review of current progress},\ }\href {http://stacks.iop.org/0034-4885/81/i=5/a=056503} {\bibfield  {journal} {\bibinfo  {journal} {Rep. Prog. Phys.}\ }\textbf {\bibinfo {volume} {81}},\ \bibinfo {pages} {056503} (\bibinfo {year} {2018})}\BibitemShut {NoStop}%
\bibitem [{\citenamefont {Weinbub}\ and\ \citenamefont {Kosik}(2022)}]{Weinbub_2022}%
  \BibitemOpen
  \bibfield  {author} {\bibinfo {author} {\bibfnamefont {J.}~\bibnamefont {Weinbub}}\ and\ \bibinfo {author} {\bibfnamefont {R.}~\bibnamefont {Kosik}},\ }\bibfield  {title} {\bibinfo {title} {Computational perspective on recent advances in quantum electronics: {From} electron quantum optics to nanoelectronic devices and systems},\ }\href {https://doi.org/10.1088/1361-648x/ac49c6} {\bibfield  {journal} {\bibinfo  {journal} {J. Phys.: Condens. Matter}\ }\textbf {\bibinfo {volume} {34}},\ \bibinfo {pages} {163001} (\bibinfo {year} {2022})}\BibitemShut {NoStop}%
\bibitem [{\citenamefont {F{\`e}ve}\ \emph {et~al.}(2007)\citenamefont {F{\`e}ve}, \citenamefont {Mah{\'e}}, \citenamefont {Berroir}, \citenamefont {Kontos}, \citenamefont {Pla{\c c}ais}, \citenamefont {Glattli}, \citenamefont {Cavanna}, \citenamefont {Etienne},\ and\ \citenamefont {Jin}}]{Feve_2007}%
  \BibitemOpen
  \bibfield  {author} {\bibinfo {author} {\bibfnamefont {G.}~\bibnamefont {F{\`e}ve}}, \bibinfo {author} {\bibfnamefont {A.}~\bibnamefont {Mah{\'e}}}, \bibinfo {author} {\bibfnamefont {J.-M.}\ \bibnamefont {Berroir}}, \bibinfo {author} {\bibfnamefont {T.}~\bibnamefont {Kontos}}, \bibinfo {author} {\bibfnamefont {B.}~\bibnamefont {Pla{\c c}ais}}, \bibinfo {author} {\bibfnamefont {D.~C.}\ \bibnamefont {Glattli}}, \bibinfo {author} {\bibfnamefont {A.}~\bibnamefont {Cavanna}}, \bibinfo {author} {\bibfnamefont {B.}~\bibnamefont {Etienne}},\ and\ \bibinfo {author} {\bibfnamefont {Y.}~\bibnamefont {Jin}},\ }\bibfield  {title} {\bibinfo {title} {An on-demand coherent single-electron source},\ }\href {https://doi.org/10.1126/science.1141243} {\bibfield  {journal} {\bibinfo  {journal} {Science}\ }\textbf {\bibinfo {volume} {316}},\ \bibinfo {pages} {1169} (\bibinfo {year} {2007})}\BibitemShut {NoStop}%
\bibitem [{\citenamefont {Bocquillon}\ \emph {et~al.}(2013)\citenamefont {Bocquillon}, \citenamefont {Freulon}, \citenamefont {Berroir}, \citenamefont {Degiovanni}, \citenamefont {Pla{\c c}ais}, \citenamefont {Cavanna}, \citenamefont {Jin},\ and\ \citenamefont {F{\`e}ve}}]{Bocquillon_2013}%
  \BibitemOpen
  \bibfield  {author} {\bibinfo {author} {\bibfnamefont {E.}~\bibnamefont {Bocquillon}}, \bibinfo {author} {\bibfnamefont {V.}~\bibnamefont {Freulon}}, \bibinfo {author} {\bibfnamefont {J.-M.}\ \bibnamefont {Berroir}}, \bibinfo {author} {\bibfnamefont {P.}~\bibnamefont {Degiovanni}}, \bibinfo {author} {\bibfnamefont {B.}~\bibnamefont {Pla{\c c}ais}}, \bibinfo {author} {\bibfnamefont {A.}~\bibnamefont {Cavanna}}, \bibinfo {author} {\bibfnamefont {Y.}~\bibnamefont {Jin}},\ and\ \bibinfo {author} {\bibfnamefont {G.}~\bibnamefont {F{\`e}ve}},\ }\bibfield  {title} {\bibinfo {title} {Coherence and indistinguishability of single electrons emitted by independent sources},\ }\href {https://doi.org/10.1126/science.1232572} {\bibfield  {journal} {\bibinfo  {journal} {Science}\ }\textbf {\bibinfo {volume} {339}},\ \bibinfo {pages} {1054} (\bibinfo {year} {2013})}\BibitemShut {NoStop}%
\bibitem [{\citenamefont {Dubois}\ \emph {et~al.}(2013{\natexlab{a}})\citenamefont {Dubois}, \citenamefont {Jullien}, \citenamefont {Portier}, \citenamefont {Roche}, \citenamefont {Cavanna}, \citenamefont {Jin}, \citenamefont {Wegscheider}, \citenamefont {Roulleau},\ and\ \citenamefont {Glattli}}]{Dubois_2013}%
  \BibitemOpen
  \bibfield  {author} {\bibinfo {author} {\bibfnamefont {J.}~\bibnamefont {Dubois}}, \bibinfo {author} {\bibfnamefont {T.}~\bibnamefont {Jullien}}, \bibinfo {author} {\bibfnamefont {F.}~\bibnamefont {Portier}}, \bibinfo {author} {\bibfnamefont {P.}~\bibnamefont {Roche}}, \bibinfo {author} {\bibfnamefont {A.}~\bibnamefont {Cavanna}}, \bibinfo {author} {\bibfnamefont {Y.}~\bibnamefont {Jin}}, \bibinfo {author} {\bibfnamefont {W.}~\bibnamefont {Wegscheider}}, \bibinfo {author} {\bibfnamefont {P.}~\bibnamefont {Roulleau}},\ and\ \bibinfo {author} {\bibfnamefont {D.~C.}\ \bibnamefont {Glattli}},\ }\bibfield  {title} {\bibinfo {title} {{Minimal-excitation states for electron quantum optics using levitons}},\ }\href {https://doi.org/10.1038/nature12713} {\bibfield  {journal} {\bibinfo  {journal} {Nature}\ }\textbf {\bibinfo {volume} {502}},\ \bibinfo {pages} {659} (\bibinfo {year} {2013}{\natexlab{a}})}\BibitemShut {NoStop}%
\bibitem [{\citenamefont {Jullien}\ \emph {et~al.}(2014{\natexlab{a}})\citenamefont {Jullien}, \citenamefont {Roulleau}, \citenamefont {Roche}, \citenamefont {Cavanna}, \citenamefont {Jin},\ and\ \citenamefont {Glattli}}]{Jullien:2014}%
  \BibitemOpen
  \bibfield  {author} {\bibinfo {author} {\bibfnamefont {T.}~\bibnamefont {Jullien}}, \bibinfo {author} {\bibfnamefont {P.}~\bibnamefont {Roulleau}}, \bibinfo {author} {\bibfnamefont {B.}~\bibnamefont {Roche}}, \bibinfo {author} {\bibfnamefont {A.}~\bibnamefont {Cavanna}}, \bibinfo {author} {\bibfnamefont {Y.}~\bibnamefont {Jin}},\ and\ \bibinfo {author} {\bibfnamefont {D.~C.}\ \bibnamefont {Glattli}},\ }\bibfield  {title} {\bibinfo {title} {Quantum tomography of an electron},\ }\href {https://doi.org/10.1038/nature13821} {\bibfield  {journal} {\bibinfo  {journal} {Nature}\ }\textbf {\bibinfo {volume} {514}},\ \bibinfo {pages} {603} (\bibinfo {year} {2014}{\natexlab{a}})}\BibitemShut {NoStop}%
\bibitem [{\citenamefont {Assouline}\ \emph {et~al.}(2023)\citenamefont {Assouline}, \citenamefont {Pugliese}, \citenamefont {Chakraborti}, \citenamefont {Lee}, \citenamefont {Bernabeu}, \citenamefont {Jo}, \citenamefont {Watanabe}, \citenamefont {Taniguchi}, \citenamefont {Glattli}, \citenamefont {Kumada}, \citenamefont {Sim}, \citenamefont {Parmentier},\ and\ \citenamefont {Roulleau}}]{Assouline:2023}%
  \BibitemOpen
  \bibfield  {author} {\bibinfo {author} {\bibfnamefont {A.}~\bibnamefont {Assouline}}, \bibinfo {author} {\bibfnamefont {L.}~\bibnamefont {Pugliese}}, \bibinfo {author} {\bibfnamefont {H.}~\bibnamefont {Chakraborti}}, \bibinfo {author} {\bibfnamefont {S.}~\bibnamefont {Lee}}, \bibinfo {author} {\bibfnamefont {L.}~\bibnamefont {Bernabeu}}, \bibinfo {author} {\bibfnamefont {M.}~\bibnamefont {Jo}}, \bibinfo {author} {\bibfnamefont {K.}~\bibnamefont {Watanabe}}, \bibinfo {author} {\bibfnamefont {T.}~\bibnamefont {Taniguchi}}, \bibinfo {author} {\bibfnamefont {D.~C.}\ \bibnamefont {Glattli}}, \bibinfo {author} {\bibfnamefont {N.}~\bibnamefont {Kumada}}, \bibinfo {author} {\bibfnamefont {H.-S.}\ \bibnamefont {Sim}}, \bibinfo {author} {\bibfnamefont {F.~D.}\ \bibnamefont {Parmentier}},\ and\ \bibinfo {author} {\bibfnamefont {P.}~\bibnamefont {Roulleau}},\ }\bibfield  {title} {\bibinfo {title} {{Emission and coherent control of Levitons in graphene}},\ }\href {https://doi.org/10.1126/science.adf9887} {\bibfield
  {journal} {\bibinfo  {journal} {Science}\ }\textbf {\bibinfo {volume} {382}},\ \bibinfo {pages} {1260} (\bibinfo {year} {2023})}\BibitemShut {NoStop}%
\bibitem [{\citenamefont {Chakraborti}\ \emph {et~al.}(2025)\citenamefont {Chakraborti}, \citenamefont {Pugliese}, \citenamefont {Assouline}, \citenamefont {Watanabe}, \citenamefont {Taniguchi}, \citenamefont {Kumada}, \citenamefont {Glattli}, \citenamefont {Jo}, \citenamefont {Sim},\ and\ \citenamefont {Roulleau}}]{Chakraborti2025}%
  \BibitemOpen
  \bibfield  {author} {\bibinfo {author} {\bibfnamefont {H.}~\bibnamefont {Chakraborti}}, \bibinfo {author} {\bibfnamefont {L.}~\bibnamefont {Pugliese}}, \bibinfo {author} {\bibfnamefont {A.}~\bibnamefont {Assouline}}, \bibinfo {author} {\bibfnamefont {K.}~\bibnamefont {Watanabe}}, \bibinfo {author} {\bibfnamefont {T.}~\bibnamefont {Taniguchi}}, \bibinfo {author} {\bibfnamefont {N.}~\bibnamefont {Kumada}}, \bibinfo {author} {\bibfnamefont {D.~C.}\ \bibnamefont {Glattli}}, \bibinfo {author} {\bibfnamefont {M.}~\bibnamefont {Jo}}, \bibinfo {author} {\bibfnamefont {H.-S.}\ \bibnamefont {Sim}},\ and\ \bibinfo {author} {\bibfnamefont {P.}~\bibnamefont {Roulleau}},\ }\bibfield  {title} {\bibinfo {title} {Electron collision in a two-path graphene interferometer},\ }\href {https://doi.org/10.1126/science.adn4622} {\bibfield  {journal} {\bibinfo  {journal} {Science}\ }\textbf {\bibinfo {volume} {388}},\ \bibinfo {pages} {492} (\bibinfo {year} {2025})}\BibitemShut {NoStop}%
\bibitem [{\citenamefont {Freulon}\ \emph {et~al.}(2015)\citenamefont {Freulon}, \citenamefont {Marguerite}, \citenamefont {Berroir}, \citenamefont {Pla{\c{c}}ais}, \citenamefont {Cavanna}, \citenamefont {Jin},\ and\ \citenamefont {F{\`e}ve}}]{Freulon_2015}%
  \BibitemOpen
  \bibfield  {author} {\bibinfo {author} {\bibfnamefont {V.}~\bibnamefont {Freulon}}, \bibinfo {author} {\bibfnamefont {A.}~\bibnamefont {Marguerite}}, \bibinfo {author} {\bibfnamefont {J.-M.}\ \bibnamefont {Berroir}}, \bibinfo {author} {\bibfnamefont {B.}~\bibnamefont {Pla{\c{c}}ais}}, \bibinfo {author} {\bibfnamefont {A.}~\bibnamefont {Cavanna}}, \bibinfo {author} {\bibfnamefont {Y.}~\bibnamefont {Jin}},\ and\ \bibinfo {author} {\bibfnamefont {G.}~\bibnamefont {F{\`e}ve}},\ }\bibfield  {title} {\bibinfo {title} {{Hong-Ou-Mandel} experiment for temporal investigation of single-electron fractionalization},\ }\href {https://doi.org/10.1038/ncomms7854} {\bibfield  {journal} {\bibinfo  {journal} {Nat. Commun.}\ }\textbf {\bibinfo {volume} {6}},\ \bibinfo {pages} {6854} (\bibinfo {year} {2015})}\BibitemShut {NoStop}%
\bibitem [{\citenamefont {Wang}\ \emph {et~al.}(2023)\citenamefont {Wang}, \citenamefont {Edlbauer}, \citenamefont {Richard}, \citenamefont {Ota}, \citenamefont {Park}, \citenamefont {Shim}, \citenamefont {Ludwig}, \citenamefont {Wieck}, \citenamefont {Sim}, \citenamefont {Urdampilleta}, \citenamefont {Meunier}, \citenamefont {Kodera}, \citenamefont {Kaneko}, \citenamefont {Sellier}, \citenamefont {Waintal}, \citenamefont {Takada},\ and\ \citenamefont {B\"{a}uerle}}]{Wang:2023}%
  \BibitemOpen
  \bibfield  {author} {\bibinfo {author} {\bibfnamefont {J.}~\bibnamefont {Wang}}, \bibinfo {author} {\bibfnamefont {H.}~\bibnamefont {Edlbauer}}, \bibinfo {author} {\bibfnamefont {A.}~\bibnamefont {Richard}}, \bibinfo {author} {\bibfnamefont {S.}~\bibnamefont {Ota}}, \bibinfo {author} {\bibfnamefont {W.}~\bibnamefont {Park}}, \bibinfo {author} {\bibfnamefont {J.}~\bibnamefont {Shim}}, \bibinfo {author} {\bibfnamefont {A.}~\bibnamefont {Ludwig}}, \bibinfo {author} {\bibfnamefont {A.~D.}\ \bibnamefont {Wieck}}, \bibinfo {author} {\bibfnamefont {H.-S.}\ \bibnamefont {Sim}}, \bibinfo {author} {\bibfnamefont {M.}~\bibnamefont {Urdampilleta}}, \bibinfo {author} {\bibfnamefont {T.}~\bibnamefont {Meunier}}, \bibinfo {author} {\bibfnamefont {T.}~\bibnamefont {Kodera}}, \bibinfo {author} {\bibfnamefont {N.-H.}\ \bibnamefont {Kaneko}}, \bibinfo {author} {\bibfnamefont {H.}~\bibnamefont {Sellier}}, \bibinfo {author} {\bibfnamefont {X.}~\bibnamefont {Waintal}}, \bibinfo {author} {\bibfnamefont {S.}~\bibnamefont
  {Takada}},\ and\ \bibinfo {author} {\bibfnamefont {C.}~\bibnamefont {B\"{a}uerle}},\ }\bibfield  {title} {\bibinfo {title} {Coulomb-mediated antibunching of an electron pair surfing on sound},\ }\href {https://doi.org/10.1038/s41565-023-01368-5} {\bibfield  {journal} {\bibinfo  {journal} {Nat. Nanotech.}\ }\textbf {\bibinfo {volume} {18}},\ \bibinfo {pages} {721} (\bibinfo {year} {2023})}\BibitemShut {NoStop}%
\bibitem [{\citenamefont {Fletcher}\ \emph {et~al.}(2023)\citenamefont {Fletcher}, \citenamefont {Park}, \citenamefont {Ryu}, \citenamefont {See}, \citenamefont {Griffiths}, \citenamefont {Jones}, \citenamefont {Farrer}, \citenamefont {Ritchie}, \citenamefont {Sim},\ and\ \citenamefont {Kataoka}}]{Fletcher:2023}%
  \BibitemOpen
  \bibfield  {author} {\bibinfo {author} {\bibfnamefont {J.~D.}\ \bibnamefont {Fletcher}}, \bibinfo {author} {\bibfnamefont {W.}~\bibnamefont {Park}}, \bibinfo {author} {\bibfnamefont {S.}~\bibnamefont {Ryu}}, \bibinfo {author} {\bibfnamefont {P.}~\bibnamefont {See}}, \bibinfo {author} {\bibfnamefont {J.~P.}\ \bibnamefont {Griffiths}}, \bibinfo {author} {\bibfnamefont {G.~A.~C.}\ \bibnamefont {Jones}}, \bibinfo {author} {\bibfnamefont {I.}~\bibnamefont {Farrer}}, \bibinfo {author} {\bibfnamefont {D.~A.}\ \bibnamefont {Ritchie}}, \bibinfo {author} {\bibfnamefont {H.-S.}\ \bibnamefont {Sim}},\ and\ \bibinfo {author} {\bibfnamefont {M.}~\bibnamefont {Kataoka}},\ }\bibfield  {title} {\bibinfo {title} {{Time-resolved Coulomb collision of single electrons}},\ }\href {https://doi.org/10.1038/s41565-023-01369-4} {\bibfield  {journal} {\bibinfo  {journal} {Nat. Nanotech.}\ }\textbf {\bibinfo {volume} {18}},\ \bibinfo {pages} {727} (\bibinfo {year} {2023})}\BibitemShut {NoStop}%
\bibitem [{\citenamefont {Ubbelohde}\ \emph {et~al.}(2023)\citenamefont {Ubbelohde}, \citenamefont {Freise}, \citenamefont {Pavlovska}, \citenamefont {Silvestrov}, \citenamefont {Recher}, \citenamefont {Kokainis}, \citenamefont {Barinovs}, \citenamefont {Hohls}, \citenamefont {Weimann}, \citenamefont {Pierz},\ and\ \citenamefont {Kashcheyevs}}]{Ubbelohde:2023}%
  \BibitemOpen
  \bibfield  {author} {\bibinfo {author} {\bibfnamefont {N.}~\bibnamefont {Ubbelohde}}, \bibinfo {author} {\bibfnamefont {L.}~\bibnamefont {Freise}}, \bibinfo {author} {\bibfnamefont {E.}~\bibnamefont {Pavlovska}}, \bibinfo {author} {\bibfnamefont {P.~G.}\ \bibnamefont {Silvestrov}}, \bibinfo {author} {\bibfnamefont {P.}~\bibnamefont {Recher}}, \bibinfo {author} {\bibfnamefont {M.}~\bibnamefont {Kokainis}}, \bibinfo {author} {\bibfnamefont {G.}~\bibnamefont {Barinovs}}, \bibinfo {author} {\bibfnamefont {F.}~\bibnamefont {Hohls}}, \bibinfo {author} {\bibfnamefont {T.}~\bibnamefont {Weimann}}, \bibinfo {author} {\bibfnamefont {K.}~\bibnamefont {Pierz}},\ and\ \bibinfo {author} {\bibfnamefont {V.}~\bibnamefont {Kashcheyevs}},\ }\bibfield  {title} {\bibinfo {title} {Two electrons interacting at a mesoscopic beam splitter},\ }\href {https://doi.org/10.1038/s41565-023-01370-x} {\bibfield  {journal} {\bibinfo  {journal} {Nat. Nanotech.}\ }\textbf {\bibinfo {volume} {18}},\ \bibinfo {pages} {733} (\bibinfo {year}
  {2023})}\BibitemShut {NoStop}%
\bibitem [{\citenamefont {Wharam}\ \emph {et~al.}(1988)\citenamefont {Wharam}, \citenamefont {Thornton}, \citenamefont {Newbury}, \citenamefont {Pepper}, \citenamefont {Ahmed}, \citenamefont {Frost}, \citenamefont {Hasko}, \citenamefont {Peacock}, \citenamefont {Ritchie},\ and\ \citenamefont {Jones}}]{Jones_1988}%
  \BibitemOpen
  \bibfield  {author} {\bibinfo {author} {\bibfnamefont {D.~A.}\ \bibnamefont {Wharam}}, \bibinfo {author} {\bibfnamefont {T.~J.}\ \bibnamefont {Thornton}}, \bibinfo {author} {\bibfnamefont {R.}~\bibnamefont {Newbury}}, \bibinfo {author} {\bibfnamefont {M.}~\bibnamefont {Pepper}}, \bibinfo {author} {\bibfnamefont {H.}~\bibnamefont {Ahmed}}, \bibinfo {author} {\bibfnamefont {J.~E.~F.}\ \bibnamefont {Frost}}, \bibinfo {author} {\bibfnamefont {D.~G.}\ \bibnamefont {Hasko}}, \bibinfo {author} {\bibfnamefont {D.~C.}\ \bibnamefont {Peacock}}, \bibinfo {author} {\bibfnamefont {D.~A.}\ \bibnamefont {Ritchie}},\ and\ \bibinfo {author} {\bibfnamefont {G.~A.~C.}\ \bibnamefont {Jones}},\ }\bibfield  {title} {\bibinfo {title} {One-dimensional transport and the quantisation of the ballistic resistance},\ }\href {http://stacks.iop.org/0022-3719/21/i=8/a=002} {\bibfield  {journal} {\bibinfo  {journal} {J. Phys. C: Solid State Phys.}\ }\textbf {\bibinfo {volume} {21}},\ \bibinfo {pages} {L209} (\bibinfo {year}
  {1988})}\BibitemShut {NoStop}%
\bibitem [{\citenamefont {van Wees}\ \emph {et~al.}(1988)\citenamefont {van Wees}, \citenamefont {van Houten}, \citenamefont {Beenakker}, \citenamefont {Williamson}, \citenamefont {Kouwenhoven}, \citenamefont {van~der Marel},\ and\ \citenamefont {Foxon}}]{Foxon_1988}%
  \BibitemOpen
  \bibfield  {author} {\bibinfo {author} {\bibfnamefont {B.~J.}\ \bibnamefont {van Wees}}, \bibinfo {author} {\bibfnamefont {H.}~\bibnamefont {van Houten}}, \bibinfo {author} {\bibfnamefont {C.~W.~J.}\ \bibnamefont {Beenakker}}, \bibinfo {author} {\bibfnamefont {J.~G.}\ \bibnamefont {Williamson}}, \bibinfo {author} {\bibfnamefont {L.~P.}\ \bibnamefont {Kouwenhoven}}, \bibinfo {author} {\bibfnamefont {D.}~\bibnamefont {van~der Marel}},\ and\ \bibinfo {author} {\bibfnamefont {C.~T.}\ \bibnamefont {Foxon}},\ }\bibfield  {title} {\bibinfo {title} {Quantized conductance of point contacts in a two-dimensional electron gas},\ }\href {https://doi.org/10.1103/PhysRevLett.60.848} {\bibfield  {journal} {\bibinfo  {journal} {Phys. Rev. Lett.}\ }\textbf {\bibinfo {volume} {60}},\ \bibinfo {pages} {848} (\bibinfo {year} {1988})}\BibitemShut {NoStop}%
\bibitem [{\citenamefont {Ji}\ \emph {et~al.}(2003)\citenamefont {Ji}, \citenamefont {Chung}, \citenamefont {Sprinzak}, \citenamefont {Heiblum}, \citenamefont {Mahalu},\ and\ \citenamefont {Shtrikman}}]{Ji_2003}%
  \BibitemOpen
  \bibfield  {author} {\bibinfo {author} {\bibfnamefont {Y.}~\bibnamefont {Ji}}, \bibinfo {author} {\bibfnamefont {Y.}~\bibnamefont {Chung}}, \bibinfo {author} {\bibfnamefont {D.}~\bibnamefont {Sprinzak}}, \bibinfo {author} {\bibfnamefont {M.}~\bibnamefont {Heiblum}}, \bibinfo {author} {\bibfnamefont {D.}~\bibnamefont {Mahalu}},\ and\ \bibinfo {author} {\bibfnamefont {H.}~\bibnamefont {Shtrikman}},\ }\bibfield  {title} {\bibinfo {title} {An electronic {Mach}-{Zehnder} interferometer},\ }\href {http://dx.doi.org/10.1038/nature01503} {\bibfield  {journal} {\bibinfo  {journal} {Nature}\ }\textbf {\bibinfo {volume} {422}},\ \bibinfo {pages} {415} (\bibinfo {year} {2003})}\BibitemShut {NoStop}%
\bibitem [{\citenamefont {Roulleau}\ \emph {et~al.}(2008)\citenamefont {Roulleau}, \citenamefont {Portier}, \citenamefont {Roche}, \citenamefont {Cavanna}, \citenamefont {Faini}, \citenamefont {Gennser},\ and\ \citenamefont {Mailly}}]{Mailly_2008}%
  \BibitemOpen
  \bibfield  {author} {\bibinfo {author} {\bibfnamefont {P.}~\bibnamefont {Roulleau}}, \bibinfo {author} {\bibfnamefont {F.}~\bibnamefont {Portier}}, \bibinfo {author} {\bibfnamefont {P.}~\bibnamefont {Roche}}, \bibinfo {author} {\bibfnamefont {A.}~\bibnamefont {Cavanna}}, \bibinfo {author} {\bibfnamefont {G.}~\bibnamefont {Faini}}, \bibinfo {author} {\bibfnamefont {U.}~\bibnamefont {Gennser}},\ and\ \bibinfo {author} {\bibfnamefont {D.}~\bibnamefont {Mailly}},\ }\bibfield  {title} {\bibinfo {title} {Direct measurement of the coherence length of edge states in the integer quantum {Hall} regime},\ }\href {https://doi.org/10.1103/PhysRevLett.100.126802} {\bibfield  {journal} {\bibinfo  {journal} {Phys. Rev. Lett.}\ }\textbf {\bibinfo {volume} {100}},\ \bibinfo {pages} {126802} (\bibinfo {year} {2008})}\BibitemShut {NoStop}%
\bibitem [{\citenamefont {Litvin}\ \emph {et~al.}(2008)\citenamefont {Litvin}, \citenamefont {Helzel}, \citenamefont {Tranitz}, \citenamefont {Wegscheider},\ and\ \citenamefont {Strunk}}]{Strunk_2008}%
  \BibitemOpen
  \bibfield  {author} {\bibinfo {author} {\bibfnamefont {L.~V.}\ \bibnamefont {Litvin}}, \bibinfo {author} {\bibfnamefont {A.}~\bibnamefont {Helzel}}, \bibinfo {author} {\bibfnamefont {H.-P.}\ \bibnamefont {Tranitz}}, \bibinfo {author} {\bibfnamefont {W.}~\bibnamefont {Wegscheider}},\ and\ \bibinfo {author} {\bibfnamefont {C.}~\bibnamefont {Strunk}},\ }\bibfield  {title} {\bibinfo {title} {Edge-channel interference controlled by {Landau} level filling},\ }\href {https://doi.org/10.1103/PhysRevB.78.075303} {\bibfield  {journal} {\bibinfo  {journal} {Phys. Rev. B}\ }\textbf {\bibinfo {volume} {78}},\ \bibinfo {pages} {075303} (\bibinfo {year} {2008})}\BibitemShut {NoStop}%
\bibitem [{\citenamefont {Zhang}\ \emph {et~al.}(2009)\citenamefont {Zhang}, \citenamefont {McClure}, \citenamefont {Levenson-Falk}, \citenamefont {Marcus}, \citenamefont {Pfeiffer},\ and\ \citenamefont {West}}]{West_2009}%
  \BibitemOpen
  \bibfield  {author} {\bibinfo {author} {\bibfnamefont {Y.}~\bibnamefont {Zhang}}, \bibinfo {author} {\bibfnamefont {D.~T.}\ \bibnamefont {McClure}}, \bibinfo {author} {\bibfnamefont {E.~M.}\ \bibnamefont {Levenson-Falk}}, \bibinfo {author} {\bibfnamefont {C.~M.}\ \bibnamefont {Marcus}}, \bibinfo {author} {\bibfnamefont {L.~N.}\ \bibnamefont {Pfeiffer}},\ and\ \bibinfo {author} {\bibfnamefont {K.~W.}\ \bibnamefont {West}},\ }\bibfield  {title} {\bibinfo {title} {Distinct signatures for {Coulomb} blockade and {Aharonov}-{Bohm} interference in electronic {Fabry}-{P\'erot} interferometers},\ }\href {https://doi.org/10.1103/PhysRevB.79.241304} {\bibfield  {journal} {\bibinfo  {journal} {Phys. Rev. B}\ }\textbf {\bibinfo {volume} {79}},\ \bibinfo {pages} {241304} (\bibinfo {year} {2009})}\BibitemShut {NoStop}%
\bibitem [{\citenamefont {Tewari}\ \emph {et~al.}(2016)\citenamefont {Tewari}, \citenamefont {Roulleau}, \citenamefont {Grenier}, \citenamefont {Portier}, \citenamefont {Cavanna}, \citenamefont {Gennser}, \citenamefont {Mailly},\ and\ \citenamefont {Roche}}]{Tewari_2016}%
  \BibitemOpen
  \bibfield  {author} {\bibinfo {author} {\bibfnamefont {S.}~\bibnamefont {Tewari}}, \bibinfo {author} {\bibfnamefont {P.}~\bibnamefont {Roulleau}}, \bibinfo {author} {\bibfnamefont {C.}~\bibnamefont {Grenier}}, \bibinfo {author} {\bibfnamefont {F.}~\bibnamefont {Portier}}, \bibinfo {author} {\bibfnamefont {A.}~\bibnamefont {Cavanna}}, \bibinfo {author} {\bibfnamefont {U.}~\bibnamefont {Gennser}}, \bibinfo {author} {\bibfnamefont {D.}~\bibnamefont {Mailly}},\ and\ \bibinfo {author} {\bibfnamefont {P.}~\bibnamefont {Roche}},\ }\bibfield  {title} {\bibinfo {title} {Robust quantum coherence above the {Fermi} sea},\ }\href {https://doi.org/10.1103/PhysRevB.93.035420} {\bibfield  {journal} {\bibinfo  {journal} {Phys. Rev. B}\ }\textbf {\bibinfo {volume} {93}},\ \bibinfo {pages} {035420} (\bibinfo {year} {2016})}\BibitemShut {NoStop}%
\bibitem [{\citenamefont {Jo}\ \emph {et~al.}(2021)\citenamefont {Jo}, \citenamefont {Brasseur}, \citenamefont {Assouline}, \citenamefont {Fleury}, \citenamefont {Sim}, \citenamefont {Watanabe}, \citenamefont {Taniguchi}, \citenamefont {Dumnernpanich}, \citenamefont {Roche}, \citenamefont {Glattli}, \citenamefont {Kumada}, \citenamefont {Parmentier},\ and\ \citenamefont {Roulleau}}]{Roulleau_2021}%
  \BibitemOpen
  \bibfield  {author} {\bibinfo {author} {\bibfnamefont {M.}~\bibnamefont {Jo}}, \bibinfo {author} {\bibfnamefont {P.}~\bibnamefont {Brasseur}}, \bibinfo {author} {\bibfnamefont {A.}~\bibnamefont {Assouline}}, \bibinfo {author} {\bibfnamefont {G.}~\bibnamefont {Fleury}}, \bibinfo {author} {\bibfnamefont {H.-S.}\ \bibnamefont {Sim}}, \bibinfo {author} {\bibfnamefont {K.}~\bibnamefont {Watanabe}}, \bibinfo {author} {\bibfnamefont {T.}~\bibnamefont {Taniguchi}}, \bibinfo {author} {\bibfnamefont {W.}~\bibnamefont {Dumnernpanich}}, \bibinfo {author} {\bibfnamefont {P.}~\bibnamefont {Roche}}, \bibinfo {author} {\bibfnamefont {D.~C.}\ \bibnamefont {Glattli}}, \bibinfo {author} {\bibfnamefont {N.}~\bibnamefont {Kumada}}, \bibinfo {author} {\bibfnamefont {F.~D.}\ \bibnamefont {Parmentier}},\ and\ \bibinfo {author} {\bibfnamefont {P.}~\bibnamefont {Roulleau}},\ }\bibfield  {title} {\bibinfo {title} {Quantum {Hall} valley splitters and a tunable {Mach}-{Zehnder} interferometer in graphene},\ }\href
  {https://doi.org/10.1103/PhysRevLett.126.146803} {\bibfield  {journal} {\bibinfo  {journal} {Phys. Rev. Lett.}\ }\textbf {\bibinfo {volume} {126}},\ \bibinfo {pages} {146803} (\bibinfo {year} {2021})}\BibitemShut {NoStop}%
\bibitem [{\citenamefont {Henny}\ \emph {et~al.}(1999)\citenamefont {Henny}, \citenamefont {Oberholzer}, \citenamefont {Strunk}, \citenamefont {Heinzel}, \citenamefont {Ensslin}, \citenamefont {Holland},\ and\ \citenamefont {Sch\"onenberger}}]{Henny1999}%
  \BibitemOpen
  \bibfield  {author} {\bibinfo {author} {\bibfnamefont {M.}~\bibnamefont {Henny}}, \bibinfo {author} {\bibfnamefont {S.}~\bibnamefont {Oberholzer}}, \bibinfo {author} {\bibfnamefont {C.}~\bibnamefont {Strunk}}, \bibinfo {author} {\bibfnamefont {T.}~\bibnamefont {Heinzel}}, \bibinfo {author} {\bibfnamefont {K.}~\bibnamefont {Ensslin}}, \bibinfo {author} {\bibfnamefont {M.}~\bibnamefont {Holland}},\ and\ \bibinfo {author} {\bibfnamefont {C.}~\bibnamefont {Sch\"onenberger}},\ }\bibfield  {title} {\bibinfo {title} {The fermionic {Hanbury} {Brown} and {Twiss} experiment},\ }\href {https://doi.org/10.1126/science.284.5412.296} {\bibfield  {journal} {\bibinfo  {journal} {Science}\ }\textbf {\bibinfo {volume} {284}},\ \bibinfo {pages} {296} (\bibinfo {year} {1999})}\BibitemShut {NoStop}%
\bibitem [{\citenamefont {Oliver}\ \emph {et~al.}(1999)\citenamefont {Oliver}, \citenamefont {Kim}, \citenamefont {Liu},\ and\ \citenamefont {Yamamoto}}]{Oliver1999}%
  \BibitemOpen
  \bibfield  {author} {\bibinfo {author} {\bibfnamefont {W.~D.}\ \bibnamefont {Oliver}}, \bibinfo {author} {\bibfnamefont {J.}~\bibnamefont {Kim}}, \bibinfo {author} {\bibfnamefont {R.~C.}\ \bibnamefont {Liu}},\ and\ \bibinfo {author} {\bibfnamefont {Y.}~\bibnamefont {Yamamoto}},\ }\bibfield  {title} {\bibinfo {title} {Hanbury {Brown} and {Twiss}-type experiment with electrons},\ }\href {https://doi.org/10.1126/science.284.5412.299} {\bibfield  {journal} {\bibinfo  {journal} {Science}\ }\textbf {\bibinfo {volume} {284}},\ \bibinfo {pages} {299} (\bibinfo {year} {1999})}\BibitemShut {NoStop}%
\bibitem [{\citenamefont {Grenier}\ \emph {et~al.}(2011)\citenamefont {Grenier}, \citenamefont {Herv\'e}, \citenamefont {Parmentier}, \citenamefont {Pla\c{c}ais}, \citenamefont {Berroir}, \citenamefont {F\'eve},\ and\ \citenamefont {Degiovanni}}]{Grenier_2011}%
  \BibitemOpen
  \bibfield  {author} {\bibinfo {author} {\bibfnamefont {C.}~\bibnamefont {Grenier}}, \bibinfo {author} {\bibfnamefont {E.}~\bibnamefont {Herv\'e}, \bibfnamefont {R.~Bocquillon}}, \bibinfo {author} {\bibfnamefont {F.~D.}\ \bibnamefont {Parmentier}}, \bibinfo {author} {\bibfnamefont {B.}~\bibnamefont {Pla\c{c}ais}}, \bibinfo {author} {\bibfnamefont {J.~M.}\ \bibnamefont {Berroir}}, \bibinfo {author} {\bibfnamefont {G.}~\bibnamefont {F\'eve}},\ and\ \bibinfo {author} {\bibfnamefont {P.}~\bibnamefont {Degiovanni}},\ }\bibfield  {title} {\bibinfo {title} {Single-electron quantum tomography in quantum {Hall} edge channels},\ }\href {http://stacks.iop.org/1367-2630/13/i=9/a=093007} {\bibfield  {journal} {\bibinfo  {journal} {New J. Phys.}\ }\textbf {\bibinfo {volume} {13}},\ \bibinfo {pages} {093007} (\bibinfo {year} {2011})}\BibitemShut {NoStop}%
\bibitem [{\citenamefont {Jullien}\ \emph {et~al.}(2014{\natexlab{b}})\citenamefont {Jullien}, \citenamefont {Roulleau}, \citenamefont {Roche}, \citenamefont {Cavanna}, \citenamefont {Jin},\ and\ \citenamefont {Glattli}}]{Jullien_2014}%
  \BibitemOpen
  \bibfield  {author} {\bibinfo {author} {\bibfnamefont {T.}~\bibnamefont {Jullien}}, \bibinfo {author} {\bibfnamefont {P.}~\bibnamefont {Roulleau}}, \bibinfo {author} {\bibfnamefont {B.}~\bibnamefont {Roche}}, \bibinfo {author} {\bibfnamefont {A.}~\bibnamefont {Cavanna}}, \bibinfo {author} {\bibfnamefont {Y.}~\bibnamefont {Jin}},\ and\ \bibinfo {author} {\bibfnamefont {D.~C.}\ \bibnamefont {Glattli}},\ }\bibfield  {title} {\bibinfo {title} {Quantum tomography of an electron},\ }\href {http://dx.doi.org/10.1038/nature13821} {\bibfield  {journal} {\bibinfo  {journal} {Nature}\ }\textbf {\bibinfo {volume} {514}},\ \bibinfo {pages} {603} (\bibinfo {year} {2014}{\natexlab{b}})}\BibitemShut {NoStop}%
\bibitem [{\citenamefont {Fletcher}\ \emph {et~al.}(2019)\citenamefont {Fletcher}, \citenamefont {Johnson}, \citenamefont {Locane}, \citenamefont {See}, \citenamefont {Griffiths}, \citenamefont {Farrer}, \citenamefont {Ritchie}, \citenamefont {Brouwer}, \citenamefont {Kashcheyevs},\ and\ \citenamefont {Kataoka}}]{Fletcher2019}%
  \BibitemOpen
  \bibfield  {author} {\bibinfo {author} {\bibfnamefont {J.~D.}\ \bibnamefont {Fletcher}}, \bibinfo {author} {\bibfnamefont {N.}~\bibnamefont {Johnson}}, \bibinfo {author} {\bibfnamefont {E.}~\bibnamefont {Locane}}, \bibinfo {author} {\bibfnamefont {P.}~\bibnamefont {See}}, \bibinfo {author} {\bibfnamefont {J.~P.}\ \bibnamefont {Griffiths}}, \bibinfo {author} {\bibfnamefont {I.}~\bibnamefont {Farrer}}, \bibinfo {author} {\bibfnamefont {D.~A.}\ \bibnamefont {Ritchie}}, \bibinfo {author} {\bibfnamefont {P.~W.}\ \bibnamefont {Brouwer}}, \bibinfo {author} {\bibfnamefont {V.}~\bibnamefont {Kashcheyevs}},\ and\ \bibinfo {author} {\bibfnamefont {M.}~\bibnamefont {Kataoka}},\ }\bibfield  {title} {\bibinfo {title} {{Continuous-variable tomography of solitary electrons}},\ }\href {https://doi.org/10.1038/s41467-019-13222-1} {\bibfield  {journal} {\bibinfo  {journal} {Nat. Commun.}\ }\textbf {\bibinfo {volume} {10}},\ \bibinfo {pages} {1} (\bibinfo {year} {2019})}\BibitemShut {NoStop}%
\bibitem [{\citenamefont {Bisognin}\ \emph {et~al.}(2019)\citenamefont {Bisognin}, \citenamefont {Marguerite}, \citenamefont {Roussel}, \citenamefont {Kumar}, \citenamefont {Cabart}, \citenamefont {Chapdelaine}, \citenamefont {Mohammad-Djafari}, \citenamefont {Berroir}, \citenamefont {Bocquillon}, \citenamefont {Pla{\c{c}}ais}, \citenamefont {Cavanna}, \citenamefont {Gennser}, \citenamefont {Jin}, \citenamefont {Degiovanni},\ and\ \citenamefont {F{\`e}ve}}]{Bisognin_2019}%
  \BibitemOpen
  \bibfield  {author} {\bibinfo {author} {\bibfnamefont {R.}~\bibnamefont {Bisognin}}, \bibinfo {author} {\bibfnamefont {A.}~\bibnamefont {Marguerite}}, \bibinfo {author} {\bibfnamefont {B.}~\bibnamefont {Roussel}}, \bibinfo {author} {\bibfnamefont {M.}~\bibnamefont {Kumar}}, \bibinfo {author} {\bibfnamefont {C.}~\bibnamefont {Cabart}}, \bibinfo {author} {\bibfnamefont {C.}~\bibnamefont {Chapdelaine}}, \bibinfo {author} {\bibfnamefont {A.}~\bibnamefont {Mohammad-Djafari}}, \bibinfo {author} {\bibfnamefont {J.-M.}\ \bibnamefont {Berroir}}, \bibinfo {author} {\bibfnamefont {E.}~\bibnamefont {Bocquillon}}, \bibinfo {author} {\bibfnamefont {B.}~\bibnamefont {Pla{\c{c}}ais}}, \bibinfo {author} {\bibfnamefont {A.}~\bibnamefont {Cavanna}}, \bibinfo {author} {\bibfnamefont {U.}~\bibnamefont {Gennser}}, \bibinfo {author} {\bibfnamefont {Y.}~\bibnamefont {Jin}}, \bibinfo {author} {\bibfnamefont {P.}~\bibnamefont {Degiovanni}},\ and\ \bibinfo {author} {\bibfnamefont {G.}~\bibnamefont {F{\`e}ve}},\ }\bibfield  {title}
  {\bibinfo {title} {Quantum tomography of electrical currents},\ }\href {https://doi.org/10.1038/s41467-019-11369-5} {\bibfield  {journal} {\bibinfo  {journal} {Nat. Commun.}\ }\textbf {\bibinfo {volume} {10}},\ \bibinfo {pages} {3379} (\bibinfo {year} {2019})}\BibitemShut {NoStop}%
\bibitem [{\citenamefont {Fletcher}\ \emph {et~al.}(2013)\citenamefont {Fletcher}, \citenamefont {See}, \citenamefont {Howe}, \citenamefont {Pepper}, \citenamefont {Giblin}, \citenamefont {Griffiths}, \citenamefont {Jones}, \citenamefont {Farrer}, \citenamefont {Ritchie}, \citenamefont {Janssen},\ and\ \citenamefont {Kataoka}}]{Kataoka_2013}%
  \BibitemOpen
  \bibfield  {author} {\bibinfo {author} {\bibfnamefont {J.~D.}\ \bibnamefont {Fletcher}}, \bibinfo {author} {\bibfnamefont {P.}~\bibnamefont {See}}, \bibinfo {author} {\bibfnamefont {H.}~\bibnamefont {Howe}}, \bibinfo {author} {\bibfnamefont {M.}~\bibnamefont {Pepper}}, \bibinfo {author} {\bibfnamefont {S.~P.}\ \bibnamefont {Giblin}}, \bibinfo {author} {\bibfnamefont {J.~P.}\ \bibnamefont {Griffiths}}, \bibinfo {author} {\bibfnamefont {G.~A.~C.}\ \bibnamefont {Jones}}, \bibinfo {author} {\bibfnamefont {I.}~\bibnamefont {Farrer}}, \bibinfo {author} {\bibfnamefont {D.~A.}\ \bibnamefont {Ritchie}}, \bibinfo {author} {\bibfnamefont {T.~J. B.~M.}\ \bibnamefont {Janssen}},\ and\ \bibinfo {author} {\bibfnamefont {M.}~\bibnamefont {Kataoka}},\ }\bibfield  {title} {\bibinfo {title} {Clock-controlled emission of single-electron wave packets in a solid-state circuit},\ }\href {https://doi.org/10.1103/PhysRevLett.111.216807} {\bibfield  {journal} {\bibinfo  {journal} {Phys. Rev. Lett.}\ }\textbf {\bibinfo {volume}
  {111}},\ \bibinfo {pages} {216807} (\bibinfo {year} {2013})}\BibitemShut {NoStop}%
\bibitem [{\citenamefont {Waldie}\ \emph {et~al.}(2015)\citenamefont {Waldie}, \citenamefont {See}, \citenamefont {Kashcheyevs}, \citenamefont {Griffiths}, \citenamefont {Farrer}, \citenamefont {Jones}, \citenamefont {Ritchie}, \citenamefont {Janssen},\ and\ \citenamefont {Kataoka}}]{Kataoka_2015}%
  \BibitemOpen
  \bibfield  {author} {\bibinfo {author} {\bibfnamefont {J.}~\bibnamefont {Waldie}}, \bibinfo {author} {\bibfnamefont {P.}~\bibnamefont {See}}, \bibinfo {author} {\bibfnamefont {V.}~\bibnamefont {Kashcheyevs}}, \bibinfo {author} {\bibfnamefont {J.~P.}\ \bibnamefont {Griffiths}}, \bibinfo {author} {\bibfnamefont {I.}~\bibnamefont {Farrer}}, \bibinfo {author} {\bibfnamefont {G.~A.~C.}\ \bibnamefont {Jones}}, \bibinfo {author} {\bibfnamefont {D.~A.}\ \bibnamefont {Ritchie}}, \bibinfo {author} {\bibfnamefont {T.~J. B.~M.}\ \bibnamefont {Janssen}},\ and\ \bibinfo {author} {\bibfnamefont {M.}~\bibnamefont {Kataoka}},\ }\bibfield  {title} {\bibinfo {title} {Measurement and control of electron wave packets from a single-electron source},\ }\href {https://doi.org/10.1103/PhysRevB.92.125305} {\bibfield  {journal} {\bibinfo  {journal} {Phys. Rev. B}\ }\textbf {\bibinfo {volume} {92}},\ \bibinfo {pages} {125305} (\bibinfo {year} {2015})}\BibitemShut {NoStop}%
\bibitem [{\citenamefont {Kataoka}\ \emph {et~al.}(2016)\citenamefont {Kataoka}, \citenamefont {Johnson}, \citenamefont {Emary}, \citenamefont {See}, \citenamefont {Griffiths}, \citenamefont {Jones}, \citenamefont {Farrer}, \citenamefont {Ritchie}, \citenamefont {Pepper},\ and\ \citenamefont {Janssen}}]{Kataoka_2016}%
  \BibitemOpen
  \bibfield  {author} {\bibinfo {author} {\bibfnamefont {M.}~\bibnamefont {Kataoka}}, \bibinfo {author} {\bibfnamefont {N.}~\bibnamefont {Johnson}}, \bibinfo {author} {\bibfnamefont {C.}~\bibnamefont {Emary}}, \bibinfo {author} {\bibfnamefont {P.}~\bibnamefont {See}}, \bibinfo {author} {\bibfnamefont {J.~P.}\ \bibnamefont {Griffiths}}, \bibinfo {author} {\bibfnamefont {G.~A.~C.}\ \bibnamefont {Jones}}, \bibinfo {author} {\bibfnamefont {I.}~\bibnamefont {Farrer}}, \bibinfo {author} {\bibfnamefont {D.~A.}\ \bibnamefont {Ritchie}}, \bibinfo {author} {\bibfnamefont {M.}~\bibnamefont {Pepper}},\ and\ \bibinfo {author} {\bibfnamefont {T.~J. B.~M.}\ \bibnamefont {Janssen}},\ }\bibfield  {title} {\bibinfo {title} {Time-of-flight measurements of single-electron wave packets in quantum {Hall} edge states},\ }\href {https://doi.org/10.1103/PhysRevLett.116.126803} {\bibfield  {journal} {\bibinfo  {journal} {Phys. Rev. Lett.}\ }\textbf {\bibinfo {volume} {116}},\ \bibinfo {pages} {126803} (\bibinfo {year}
  {2016})}\BibitemShut {NoStop}%
\bibitem [{\citenamefont {Roussely}\ \emph {et~al.}(2018)\citenamefont {Roussely}, \citenamefont {Arrighi}, \citenamefont {Georgiou}, \citenamefont {Takada}, \citenamefont {Schalk}, \citenamefont {Urdampilleta}, \citenamefont {Ludwig}, \citenamefont {Wieck}, \citenamefont {Armagnat}, \citenamefont {Kloss}, \citenamefont {Waintal}, \citenamefont {Meunier},\ and\ \citenamefont {B{\"a}uerle}}]{Bauerle_2018}%
  \BibitemOpen
  \bibfield  {author} {\bibinfo {author} {\bibfnamefont {G.}~\bibnamefont {Roussely}}, \bibinfo {author} {\bibfnamefont {E.}~\bibnamefont {Arrighi}}, \bibinfo {author} {\bibfnamefont {G.}~\bibnamefont {Georgiou}}, \bibinfo {author} {\bibfnamefont {S.}~\bibnamefont {Takada}}, \bibinfo {author} {\bibfnamefont {M.}~\bibnamefont {Schalk}}, \bibinfo {author} {\bibfnamefont {M.}~\bibnamefont {Urdampilleta}}, \bibinfo {author} {\bibfnamefont {A.}~\bibnamefont {Ludwig}}, \bibinfo {author} {\bibfnamefont {A.~D.}\ \bibnamefont {Wieck}}, \bibinfo {author} {\bibfnamefont {P.}~\bibnamefont {Armagnat}}, \bibinfo {author} {\bibfnamefont {T.}~\bibnamefont {Kloss}}, \bibinfo {author} {\bibfnamefont {X.}~\bibnamefont {Waintal}}, \bibinfo {author} {\bibfnamefont {T.}~\bibnamefont {Meunier}},\ and\ \bibinfo {author} {\bibfnamefont {C.}~\bibnamefont {B{\"a}uerle}},\ }\bibfield  {title} {\bibinfo {title} {Unveiling the bosonic nature of an ultrashort few-electron pulse},\ }\href {https://doi.org/10.1038/s41467-018-05203-7}
  {\bibfield  {journal} {\bibinfo  {journal} {Nat, Commun.}\ }\textbf {\bibinfo {volume} {9}},\ \bibinfo {pages} {2811} (\bibinfo {year} {2018})}\BibitemShut {NoStop}%
\bibitem [{\citenamefont {Beenakker}\ \emph {et~al.}(2003)\citenamefont {Beenakker}, \citenamefont {Emary}, \citenamefont {Kindermann},\ and\ \citenamefont {van Velsen}}]{Beenakker_2003}%
  \BibitemOpen
  \bibfield  {author} {\bibinfo {author} {\bibfnamefont {C.~W.~J.}\ \bibnamefont {Beenakker}}, \bibinfo {author} {\bibfnamefont {C.}~\bibnamefont {Emary}}, \bibinfo {author} {\bibfnamefont {M.}~\bibnamefont {Kindermann}},\ and\ \bibinfo {author} {\bibfnamefont {J.~L.}\ \bibnamefont {van Velsen}},\ }\bibfield  {title} {\bibinfo {title} {Proposal for production and detection of entangled electron-hole pairs in a degenerate electron gas},\ }\href {https://doi.org/10.1103/PhysRevLett.91.147901} {\bibfield  {journal} {\bibinfo  {journal} {Phys. Rev. Lett.}\ }\textbf {\bibinfo {volume} {91}},\ \bibinfo {pages} {147901} (\bibinfo {year} {2003})}\BibitemShut {NoStop}%
\bibitem [{\citenamefont {Beenakker}\ \emph {et~al.}(2005)\citenamefont {Beenakker}, \citenamefont {Titov},\ and\ \citenamefont {Trauzettel}}]{Trauzettel_2005}%
  \BibitemOpen
  \bibfield  {author} {\bibinfo {author} {\bibfnamefont {C.~W.~J.}\ \bibnamefont {Beenakker}}, \bibinfo {author} {\bibfnamefont {M.}~\bibnamefont {Titov}},\ and\ \bibinfo {author} {\bibfnamefont {B.}~\bibnamefont {Trauzettel}},\ }\bibfield  {title} {\bibinfo {title} {Optimal spin-entangled electron-hole pair pump},\ }\href {https://doi.org/10.1103/PhysRevLett.94.186804} {\bibfield  {journal} {\bibinfo  {journal} {Phys. Rev. Lett.}\ }\textbf {\bibinfo {volume} {94}},\ \bibinfo {pages} {186804} (\bibinfo {year} {2005})}\BibitemShut {NoStop}%
\bibitem [{\citenamefont {Dasenbrook}\ and\ \citenamefont {Flindt}(2015)}]{Dasenbrook_2015}%
  \BibitemOpen
  \bibfield  {author} {\bibinfo {author} {\bibfnamefont {D.}~\bibnamefont {Dasenbrook}}\ and\ \bibinfo {author} {\bibfnamefont {C.}~\bibnamefont {Flindt}},\ }\bibfield  {title} {\bibinfo {title} {Dynamical generation and detection of entanglement in neutral leviton pairs},\ }\href {https://doi.org/10.1103/PhysRevB.92.161412} {\bibfield  {journal} {\bibinfo  {journal} {Phys. Rev. B}\ }\textbf {\bibinfo {volume} {92}},\ \bibinfo {pages} {161412} (\bibinfo {year} {2015})}\BibitemShut {NoStop}%
\bibitem [{\citenamefont {Luo}\ \emph {et~al.}(2022)\citenamefont {Luo}, \citenamefont {Geng}, \citenamefont {Xing}, \citenamefont {Blatter},\ and\ \citenamefont {Chen}}]{Chen2022}%
  \BibitemOpen
  \bibfield  {author} {\bibinfo {author} {\bibfnamefont {W.}~\bibnamefont {Luo}}, \bibinfo {author} {\bibfnamefont {H.}~\bibnamefont {Geng}}, \bibinfo {author} {\bibfnamefont {D.~Y.}\ \bibnamefont {Xing}}, \bibinfo {author} {\bibfnamefont {G.}~\bibnamefont {Blatter}},\ and\ \bibinfo {author} {\bibfnamefont {W.}~\bibnamefont {Chen}},\ }\bibfield  {title} {\bibinfo {title} {Entanglement of {Nambu} spinors and {Bell} inequality test without beam splitters},\ }\href {https://doi.org/10.1103/PhysRevLett.129.120507} {\bibfield  {journal} {\bibinfo  {journal} {Phys. Rev. Lett.}\ }\textbf {\bibinfo {volume} {129}},\ \bibinfo {pages} {120507} (\bibinfo {year} {2022})}\BibitemShut {NoStop}%
\bibitem [{\citenamefont {Bertin-Johannet}\ \emph {et~al.}(2023)\citenamefont {Bertin-Johannet}, \citenamefont {Raymond}, \citenamefont {Ronetti}, \citenamefont {Rech}, \citenamefont {Jonckheere}, \citenamefont {Gr{\ifmmode\acute{e}\else\'{e}\fi}maud},\ and\ \citenamefont {Martin}}]{Martin_2023}%
  \BibitemOpen
  \bibfield  {author} {\bibinfo {author} {\bibfnamefont {B.}~\bibnamefont {Bertin-Johannet}}, \bibinfo {author} {\bibfnamefont {L.}~\bibnamefont {Raymond}}, \bibinfo {author} {\bibfnamefont {F.}~\bibnamefont {Ronetti}}, \bibinfo {author} {\bibfnamefont {J.}~\bibnamefont {Rech}}, \bibinfo {author} {\bibfnamefont {T.}~\bibnamefont {Jonckheere}}, \bibinfo {author} {\bibfnamefont {B.}~\bibnamefont {Gr{\ifmmode\acute{e}\else\'{e}\fi}maud}},\ and\ \bibinfo {author} {\bibfnamefont {T.}~\bibnamefont {Martin}},\ }\bibfield  {title} {\bibinfo {title} {An on-demand source of energy-entangled electrons using levitons},\ }\href {https://doi.org/10.1063/5.0148041} {\bibfield  {journal} {\bibinfo  {journal} {Appl. Phys. Lett.}\ }\textbf {\bibinfo {volume} {122}},\ \bibinfo {pages} {202601} (\bibinfo {year} {2023})}\BibitemShut {NoStop}%
\bibitem [{\citenamefont {Li}\ and\ \citenamefont {Chen}(2025)}]{Chen2025}%
  \BibitemOpen
  \bibfield  {author} {\bibinfo {author} {\bibfnamefont {H.-D.}\ \bibnamefont {Li}}\ and\ \bibinfo {author} {\bibfnamefont {W.}~\bibnamefont {Chen}},\ }\bibfield  {title} {\bibinfo {title} {Anomalous exchange correlation of quasiparticles with entangled nambu spinors},\ }\href {https://doi.org/10.1103/PhysRevB.111.195306} {\bibfield  {journal} {\bibinfo  {journal} {Phys. Rev. B}\ }\textbf {\bibinfo {volume} {111}},\ \bibinfo {pages} {195306} (\bibinfo {year} {2025})}\BibitemShut {NoStop}%
\bibitem [{\citenamefont {Eroms}\ \emph {et~al.}(2005)\citenamefont {Eroms}, \citenamefont {Weiss}, \citenamefont {Boeck}, \citenamefont {Borghs},\ and\ \citenamefont {Z\"ulicke}}]{Zulicke_2005}%
  \BibitemOpen
  \bibfield  {author} {\bibinfo {author} {\bibfnamefont {J.}~\bibnamefont {Eroms}}, \bibinfo {author} {\bibfnamefont {D.}~\bibnamefont {Weiss}}, \bibinfo {author} {\bibfnamefont {J.~D.}\ \bibnamefont {Boeck}}, \bibinfo {author} {\bibfnamefont {G.}~\bibnamefont {Borghs}},\ and\ \bibinfo {author} {\bibfnamefont {U.}~\bibnamefont {Z\"ulicke}},\ }\bibfield  {title} {\bibinfo {title} {Andreev reflection at high magnetic fields: {Evidence} for electron and hole transport in edge states},\ }\href {https://doi.org/10.1103/PhysRevLett.95.107001} {\bibfield  {journal} {\bibinfo  {journal} {Phys. Rev. Lett.}\ }\textbf {\bibinfo {volume} {95}},\ \bibinfo {pages} {107001} (\bibinfo {year} {2005})}\BibitemShut {NoStop}%
\bibitem [{\citenamefont {Batov}\ \emph {et~al.}(2007)\citenamefont {Batov}, \citenamefont {Sch\"apers}, \citenamefont {Chtchelkatchev}, \citenamefont {Hardtdegen},\ and\ \citenamefont {Ustinov}}]{Ustinov_2007}%
  \BibitemOpen
  \bibfield  {author} {\bibinfo {author} {\bibfnamefont {I.~E.}\ \bibnamefont {Batov}}, \bibinfo {author} {\bibfnamefont {T.}~\bibnamefont {Sch\"apers}}, \bibinfo {author} {\bibfnamefont {N.~M.}\ \bibnamefont {Chtchelkatchev}}, \bibinfo {author} {\bibfnamefont {H.}~\bibnamefont {Hardtdegen}},\ and\ \bibinfo {author} {\bibfnamefont {A.~V.}\ \bibnamefont {Ustinov}},\ }\bibfield  {title} {\bibinfo {title} {Andreev reflection and strongly enhanced magnetoresistance oscillations in {Ga}$_x${In}$_{1-x}${As/InP} heterostructures with superconducting contacts},\ }\href {https://doi.org/10.1103/PhysRevB.76.115313} {\bibfield  {journal} {\bibinfo  {journal} {Phys. Rev. B}\ }\textbf {\bibinfo {volume} {76}},\ \bibinfo {pages} {115313} (\bibinfo {year} {2007})}\BibitemShut {NoStop}%
\bibitem [{\citenamefont {Rickhaus}\ \emph {et~al.}(2012)\citenamefont {Rickhaus}, \citenamefont {Weiss}, \citenamefont {Marot},\ and\ \citenamefont {Sch\"onenberger}}]{Schonenberger_2012}%
  \BibitemOpen
  \bibfield  {author} {\bibinfo {author} {\bibfnamefont {P.}~\bibnamefont {Rickhaus}}, \bibinfo {author} {\bibfnamefont {M.}~\bibnamefont {Weiss}}, \bibinfo {author} {\bibfnamefont {L.}~\bibnamefont {Marot}},\ and\ \bibinfo {author} {\bibfnamefont {C.}~\bibnamefont {Sch\"onenberger}},\ }\bibfield  {title} {\bibinfo {title} {Quantum {Hall} effect in graphene with superconducting electrodes},\ }\href {https://doi.org/10.1021/nl204415s} {\bibfield  {journal} {\bibinfo  {journal} {Nano Lett.}\ }\textbf {\bibinfo {volume} {12}},\ \bibinfo {pages} {1942} (\bibinfo {year} {2012})}\BibitemShut {NoStop}%
\bibitem [{\citenamefont {Wan}\ \emph {et~al.}(2015)\citenamefont {Wan}, \citenamefont {Kazakov}, \citenamefont {Manfra}, \citenamefont {Pfeiffer}, \citenamefont {West},\ and\ \citenamefont {Rokhinson}}]{Rokhinson_2015}%
  \BibitemOpen
  \bibfield  {author} {\bibinfo {author} {\bibfnamefont {Z.}~\bibnamefont {Wan}}, \bibinfo {author} {\bibfnamefont {A.}~\bibnamefont {Kazakov}}, \bibinfo {author} {\bibfnamefont {M.~J.}\ \bibnamefont {Manfra}}, \bibinfo {author} {\bibfnamefont {L.~N.}\ \bibnamefont {Pfeiffer}}, \bibinfo {author} {\bibfnamefont {K.~W.}\ \bibnamefont {West}},\ and\ \bibinfo {author} {\bibfnamefont {L.~P.}\ \bibnamefont {Rokhinson}},\ }\bibfield  {title} {\bibinfo {title} {{Induced superconductivity in high-mobility two-dimensional electron gas in gallium arsenide heterostructures}},\ }\href {https://doi.org/10.1038/ncomms8426} {\bibfield  {journal} {\bibinfo  {journal} {Nat. Commun.}\ }\textbf {\bibinfo {volume} {6}},\ \bibinfo {pages} {1} (\bibinfo {year} {2015})}\BibitemShut {NoStop}%
\bibitem [{\citenamefont {Calado}\ \emph {et~al.}(2015)\citenamefont {Calado}, \citenamefont {Goswami}, \citenamefont {Nanda}, \citenamefont {Diez}, \citenamefont {Akhmerov}, \citenamefont {Watanabe}, \citenamefont {Taniguchi}, \citenamefont {Klapwijk},\ and\ \citenamefont {Vandersypen}}]{Calado2015}%
  \BibitemOpen
  \bibfield  {author} {\bibinfo {author} {\bibfnamefont {V.~E.}\ \bibnamefont {Calado}}, \bibinfo {author} {\bibfnamefont {S.}~\bibnamefont {Goswami}}, \bibinfo {author} {\bibfnamefont {G.}~\bibnamefont {Nanda}}, \bibinfo {author} {\bibfnamefont {M.}~\bibnamefont {Diez}}, \bibinfo {author} {\bibfnamefont {A.~R.}\ \bibnamefont {Akhmerov}}, \bibinfo {author} {\bibfnamefont {K.}~\bibnamefont {Watanabe}}, \bibinfo {author} {\bibfnamefont {T.}~\bibnamefont {Taniguchi}}, \bibinfo {author} {\bibfnamefont {T.~M.}\ \bibnamefont {Klapwijk}},\ and\ \bibinfo {author} {\bibfnamefont {L.~M.~K.}\ \bibnamefont {Vandersypen}},\ }\bibfield  {title} {\bibinfo {title} {{Ballistic Josephson junctions in edge-contacted graphene}},\ }\href {https://doi.org/10.1038/nnano.2015.156} {\bibfield  {journal} {\bibinfo  {journal} {Nat. Nanotech.}\ }\textbf {\bibinfo {volume} {10}},\ \bibinfo {pages} {761} (\bibinfo {year} {2015})}\BibitemShut {NoStop}%
\bibitem [{\citenamefont {Ben~Shalom}\ \emph {et~al.}(2016)\citenamefont {Ben~Shalom}, \citenamefont {Zhu}, \citenamefont {Fal{'}ko}, \citenamefont {Mishchenko}, \citenamefont {Kretinin}, \citenamefont {Novoselov}, \citenamefont {Woods}, \citenamefont {Watanabe}, \citenamefont {Taniguchi}, \citenamefont {Geim},\ and\ \citenamefont {Prance}}]{BenShalom2016}%
  \BibitemOpen
  \bibfield  {author} {\bibinfo {author} {\bibfnamefont {M.}~\bibnamefont {Ben~Shalom}}, \bibinfo {author} {\bibfnamefont {M.~J.}\ \bibnamefont {Zhu}}, \bibinfo {author} {\bibfnamefont {V.~I.}\ \bibnamefont {Fal{'}ko}}, \bibinfo {author} {\bibfnamefont {A.}~\bibnamefont {Mishchenko}}, \bibinfo {author} {\bibfnamefont {A.~V.}\ \bibnamefont {Kretinin}}, \bibinfo {author} {\bibfnamefont {K.~S.}\ \bibnamefont {Novoselov}}, \bibinfo {author} {\bibfnamefont {C.~R.}\ \bibnamefont {Woods}}, \bibinfo {author} {\bibfnamefont {K.}~\bibnamefont {Watanabe}}, \bibinfo {author} {\bibfnamefont {T.}~\bibnamefont {Taniguchi}}, \bibinfo {author} {\bibfnamefont {A.~K.}\ \bibnamefont {Geim}},\ and\ \bibinfo {author} {\bibfnamefont {J.~R.}\ \bibnamefont {Prance}},\ }\bibfield  {title} {\bibinfo {title} {{Quantum oscillations of the critical current and high-field superconducting proximity in ballistic graphene}},\ }\href {https://doi.org/10.1038/nphys3592} {\bibfield  {journal} {\bibinfo  {journal} {Nat. Phys.}\ }\textbf {\bibinfo
  {volume} {12}},\ \bibinfo {pages} {318} (\bibinfo {year} {2016})}\BibitemShut {NoStop}%
\bibitem [{\citenamefont {Amet}\ \emph {et~al.}(2016)\citenamefont {Amet}, \citenamefont {Ke}, \citenamefont {Borzenets}, \citenamefont {Wang}, \citenamefont {Watanabe}, \citenamefont {Taniguchi}, \citenamefont {Deacon}, \citenamefont {Yamamoto}, \citenamefont {Bomze}, \citenamefont {Tarucha},\ and\ \citenamefont {Finkelstein}}]{Amet_2016}%
  \BibitemOpen
  \bibfield  {author} {\bibinfo {author} {\bibfnamefont {F.}~\bibnamefont {Amet}}, \bibinfo {author} {\bibfnamefont {C.~T.}\ \bibnamefont {Ke}}, \bibinfo {author} {\bibfnamefont {I.~V.}\ \bibnamefont {Borzenets}}, \bibinfo {author} {\bibfnamefont {J.}~\bibnamefont {Wang}}, \bibinfo {author} {\bibfnamefont {K.}~\bibnamefont {Watanabe}}, \bibinfo {author} {\bibfnamefont {T.}~\bibnamefont {Taniguchi}}, \bibinfo {author} {\bibfnamefont {R.~S.}\ \bibnamefont {Deacon}}, \bibinfo {author} {\bibfnamefont {M.}~\bibnamefont {Yamamoto}}, \bibinfo {author} {\bibfnamefont {Y.}~\bibnamefont {Bomze}}, \bibinfo {author} {\bibfnamefont {S.}~\bibnamefont {Tarucha}},\ and\ \bibinfo {author} {\bibfnamefont {G.}~\bibnamefont {Finkelstein}},\ }\bibfield  {title} {\bibinfo {title} {Supercurrent in the quantum {H}all regime},\ }\href {https://doi.org/10.1126/science.aad6203} {\bibfield  {journal} {\bibinfo  {journal} {Science}\ }\textbf {\bibinfo {volume} {352}},\ \bibinfo {pages} {966} (\bibinfo {year} {2016})}\BibitemShut
  {NoStop}%
\bibitem [{\citenamefont {Lee}\ \emph {et~al.}(2017)\citenamefont {Lee}, \citenamefont {Huang}, \citenamefont {Efetov}, \citenamefont {Wei}, \citenamefont {Hart}, \citenamefont {Taniguchi}, \citenamefont {Watanabe}, \citenamefont {Yacoby},\ and\ \citenamefont {Kim}}]{Lee_2017}%
  \BibitemOpen
  \bibfield  {author} {\bibinfo {author} {\bibfnamefont {G.-H.}\ \bibnamefont {Lee}}, \bibinfo {author} {\bibfnamefont {K.-F.}\ \bibnamefont {Huang}}, \bibinfo {author} {\bibfnamefont {D.~K.}\ \bibnamefont {Efetov}}, \bibinfo {author} {\bibfnamefont {D.~S.}\ \bibnamefont {Wei}}, \bibinfo {author} {\bibfnamefont {S.}~\bibnamefont {Hart}}, \bibinfo {author} {\bibfnamefont {T.}~\bibnamefont {Taniguchi}}, \bibinfo {author} {\bibfnamefont {K.}~\bibnamefont {Watanabe}}, \bibinfo {author} {\bibfnamefont {A.}~\bibnamefont {Yacoby}},\ and\ \bibinfo {author} {\bibfnamefont {P.}~\bibnamefont {Kim}},\ }\bibfield  {title} {\bibinfo {title} {Inducing superconducting correlation in quantum {H}all edge states},\ }\href {https://doi.org/10.1038/nphys4084} {\bibfield  {journal} {\bibinfo  {journal} {Nat. Phys.}\ }\textbf {\bibinfo {volume} {13}},\ \bibinfo {pages} {693} (\bibinfo {year} {2017})}\BibitemShut {NoStop}%
\bibitem [{\citenamefont {Xu}\ \emph {et~al.}(2017)\citenamefont {Xu}, \citenamefont {Song}, \citenamefont {Liu}, \citenamefont {Chen}, \citenamefont {Wang}, \citenamefont {Fan}, \citenamefont {Kang}, \citenamefont {Ma}, \citenamefont {Cheng},\ and\ \citenamefont {Ren}}]{Ren_2017}%
  \BibitemOpen
  \bibfield  {author} {\bibinfo {author} {\bibfnamefont {C.}~\bibnamefont {Xu}}, \bibinfo {author} {\bibfnamefont {S.}~\bibnamefont {Song}}, \bibinfo {author} {\bibfnamefont {Z.}~\bibnamefont {Liu}}, \bibinfo {author} {\bibfnamefont {L.}~\bibnamefont {Chen}}, \bibinfo {author} {\bibfnamefont {L.}~\bibnamefont {Wang}}, \bibinfo {author} {\bibfnamefont {D.}~\bibnamefont {Fan}}, \bibinfo {author} {\bibfnamefont {N.}~\bibnamefont {Kang}}, \bibinfo {author} {\bibfnamefont {X.}~\bibnamefont {Ma}}, \bibinfo {author} {\bibfnamefont {H.-M.}\ \bibnamefont {Cheng}},\ and\ \bibinfo {author} {\bibfnamefont {W.}~\bibnamefont {Ren}},\ }\bibfield  {title} {\bibinfo {title} {Strongly coupled high-quality graphene/2d superconducting {Mo$_2$C} vertical heterostructures with aligned orientation},\ }\href {https://doi.org/10.1021/acsnano.7b01638} {\bibfield  {journal} {\bibinfo  {journal} {ACS Nano}\ }\textbf {\bibinfo {volume} {11}},\ \bibinfo {pages} {5906} (\bibinfo {year} {2017})}\BibitemShut {NoStop}%
\bibitem [{\citenamefont {Sahu}\ \emph {et~al.}(2018)\citenamefont {Sahu}, \citenamefont {Liu}, \citenamefont {Paul}, \citenamefont {Das}, \citenamefont {Raychaudhuri}, \citenamefont {Jain},\ and\ \citenamefont {Das}}]{Das_2018}%
  \BibitemOpen
  \bibfield  {author} {\bibinfo {author} {\bibfnamefont {M.~R.}\ \bibnamefont {Sahu}}, \bibinfo {author} {\bibfnamefont {X.}~\bibnamefont {Liu}}, \bibinfo {author} {\bibfnamefont {A.~K.}\ \bibnamefont {Paul}}, \bibinfo {author} {\bibfnamefont {S.}~\bibnamefont {Das}}, \bibinfo {author} {\bibfnamefont {P.}~\bibnamefont {Raychaudhuri}}, \bibinfo {author} {\bibfnamefont {J.~K.}\ \bibnamefont {Jain}},\ and\ \bibinfo {author} {\bibfnamefont {A.}~\bibnamefont {Das}},\ }\bibfield  {title} {\bibinfo {title} {Inter-{Landau}-level {Andreev} reflection at the {Dirac} point in a graphene quantum {Hall} state coupled to a {NbSe}$_{2}$ superconductor},\ }\href {https://doi.org/10.1103/PhysRevLett.121.086809} {\bibfield  {journal} {\bibinfo  {journal} {Phys. Rev. Lett.}\ }\textbf {\bibinfo {volume} {121}},\ \bibinfo {pages} {086809} (\bibinfo {year} {2018})}\BibitemShut {NoStop}%
\bibitem [{\citenamefont {Zhi}\ \emph {et~al.}(2019)\citenamefont {Zhi}, \citenamefont {Kang}, \citenamefont {Su}, \citenamefont {Fan}, \citenamefont {Li}, \citenamefont {Pan}, \citenamefont {Zhao}, \citenamefont {Zhao},\ and\ \citenamefont {Xu}}]{Xu_2019}%
  \BibitemOpen
  \bibfield  {author} {\bibinfo {author} {\bibfnamefont {J.}~\bibnamefont {Zhi}}, \bibinfo {author} {\bibfnamefont {N.}~\bibnamefont {Kang}}, \bibinfo {author} {\bibfnamefont {F.}~\bibnamefont {Su}}, \bibinfo {author} {\bibfnamefont {D.}~\bibnamefont {Fan}}, \bibinfo {author} {\bibfnamefont {S.}~\bibnamefont {Li}}, \bibinfo {author} {\bibfnamefont {D.}~\bibnamefont {Pan}}, \bibinfo {author} {\bibfnamefont {S.~P.}\ \bibnamefont {Zhao}}, \bibinfo {author} {\bibfnamefont {J.}~\bibnamefont {Zhao}},\ and\ \bibinfo {author} {\bibfnamefont {H.~Q.}\ \bibnamefont {Xu}},\ }\bibfield  {title} {\bibinfo {title} {Coexistence of induced superconductivity and quantum {Hall} states in {InSb} nanosheets},\ }\href {https://doi.org/10.1103/PhysRevB.99.245302} {\bibfield  {journal} {\bibinfo  {journal} {Phys. Rev. B}\ }\textbf {\bibinfo {volume} {99}},\ \bibinfo {pages} {245302} (\bibinfo {year} {2019})}\BibitemShut {NoStop}%
\bibitem [{\citenamefont {Zhao}\ \emph {et~al.}(2020)\citenamefont {Zhao}, \citenamefont {Arnault}, \citenamefont {Bondarev}, \citenamefont {Seredinski}, \citenamefont {Larson}, \citenamefont {Draelos}, \citenamefont {Li}, \citenamefont {Watanabe}, \citenamefont {Taniguchi}, \citenamefont {Amet}, \citenamefont {Baranger},\ and\ \citenamefont {Finkelstein}}]{Finkelstein_2020}%
  \BibitemOpen
  \bibfield  {author} {\bibinfo {author} {\bibfnamefont {L.}~\bibnamefont {Zhao}}, \bibinfo {author} {\bibfnamefont {E.~G.}\ \bibnamefont {Arnault}}, \bibinfo {author} {\bibfnamefont {A.}~\bibnamefont {Bondarev}}, \bibinfo {author} {\bibfnamefont {A.}~\bibnamefont {Seredinski}}, \bibinfo {author} {\bibfnamefont {T.~F.~Q.}\ \bibnamefont {Larson}}, \bibinfo {author} {\bibfnamefont {A.~W.}\ \bibnamefont {Draelos}}, \bibinfo {author} {\bibfnamefont {H.}~\bibnamefont {Li}}, \bibinfo {author} {\bibfnamefont {K.}~\bibnamefont {Watanabe}}, \bibinfo {author} {\bibfnamefont {T.}~\bibnamefont {Taniguchi}}, \bibinfo {author} {\bibfnamefont {F.}~\bibnamefont {Amet}}, \bibinfo {author} {\bibfnamefont {H.~U.}\ \bibnamefont {Baranger}},\ and\ \bibinfo {author} {\bibfnamefont {G.}~\bibnamefont {Finkelstein}},\ }\bibfield  {title} {\bibinfo {title} {Interference of chiral {Andreev} edge states},\ }\href {https://doi.org/10.1038/s41567-020-0898-5} {\bibfield  {journal} {\bibinfo  {journal} {Nat. Phys.}\ }\textbf {\bibinfo
  {volume} {16}},\ \bibinfo {pages} {862} (\bibinfo {year} {2020})}\BibitemShut {NoStop}%
\bibitem [{\citenamefont {Sahu}\ \emph {et~al.}(2021)\citenamefont {Sahu}, \citenamefont {Paul}, \citenamefont {Sutradhar}, \citenamefont {Watanabe}, \citenamefont {Taniguchi}, \citenamefont {Singh}, \citenamefont {Mukerjee}, \citenamefont {Banerjee},\ and\ \citenamefont {Das}}]{Das_2021}%
  \BibitemOpen
  \bibfield  {author} {\bibinfo {author} {\bibfnamefont {M.~R.}\ \bibnamefont {Sahu}}, \bibinfo {author} {\bibfnamefont {A.~K.}\ \bibnamefont {Paul}}, \bibinfo {author} {\bibfnamefont {J.}~\bibnamefont {Sutradhar}}, \bibinfo {author} {\bibfnamefont {K.}~\bibnamefont {Watanabe}}, \bibinfo {author} {\bibfnamefont {T.}~\bibnamefont {Taniguchi}}, \bibinfo {author} {\bibfnamefont {V.}~\bibnamefont {Singh}}, \bibinfo {author} {\bibfnamefont {S.}~\bibnamefont {Mukerjee}}, \bibinfo {author} {\bibfnamefont {S.}~\bibnamefont {Banerjee}},\ and\ \bibinfo {author} {\bibfnamefont {A.}~\bibnamefont {Das}},\ }\bibfield  {title} {\bibinfo {title} {Quantized conductance with nonzero shot noise as a signature of {Andreev} edge state},\ }\href {https://doi.org/10.1103/PhysRevB.104.L081404} {\bibfield  {journal} {\bibinfo  {journal} {Phys. Rev. B}\ }\textbf {\bibinfo {volume} {104}},\ \bibinfo {pages} {L081404} (\bibinfo {year} {2021})}\BibitemShut {NoStop}%
\bibitem [{\citenamefont {G\"ul}\ \emph {et~al.}(2022)\citenamefont {G\"ul}, \citenamefont {Ronen}, \citenamefont {Lee}, \citenamefont {Shapourian}, \citenamefont {Zauberman}, \citenamefont {Lee}, \citenamefont {Watanabe}, \citenamefont {Taniguchi}, \citenamefont {Vishwanath}, \citenamefont {Yacoby},\ and\ \citenamefont {Kim}}]{Kim2022}%
  \BibitemOpen
  \bibfield  {author} {\bibinfo {author} {\bibfnamefont {O.}~\bibnamefont {G\"ul}}, \bibinfo {author} {\bibfnamefont {Y.}~\bibnamefont {Ronen}}, \bibinfo {author} {\bibfnamefont {S.~Y.}\ \bibnamefont {Lee}}, \bibinfo {author} {\bibfnamefont {H.}~\bibnamefont {Shapourian}}, \bibinfo {author} {\bibfnamefont {J.}~\bibnamefont {Zauberman}}, \bibinfo {author} {\bibfnamefont {Y.~H.}\ \bibnamefont {Lee}}, \bibinfo {author} {\bibfnamefont {K.}~\bibnamefont {Watanabe}}, \bibinfo {author} {\bibfnamefont {T.}~\bibnamefont {Taniguchi}}, \bibinfo {author} {\bibfnamefont {A.}~\bibnamefont {Vishwanath}}, \bibinfo {author} {\bibfnamefont {A.}~\bibnamefont {Yacoby}},\ and\ \bibinfo {author} {\bibfnamefont {P.}~\bibnamefont {Kim}},\ }\bibfield  {title} {\bibinfo {title} {Andreev reflection in the fractional quantum {Hall} state},\ }\href {https://doi.org/10.1103/PhysRevX.12.021057} {\bibfield  {journal} {\bibinfo  {journal} {Phys. Rev. X}\ }\textbf {\bibinfo {volume} {12}},\ \bibinfo {pages} {021057} (\bibinfo {year}
  {2022})}\BibitemShut {NoStop}%
\bibitem [{\citenamefont {Hatefipour}\ \emph {et~al.}(2022)\citenamefont {Hatefipour}, \citenamefont {Cuozzo}, \citenamefont {Kanter}, \citenamefont {Strickland}, \citenamefont {Allemang}, \citenamefont {Lu}, \citenamefont {Rossi},\ and\ \citenamefont {Shabani}}]{Shabani_2022}%
  \BibitemOpen
  \bibfield  {author} {\bibinfo {author} {\bibfnamefont {M.}~\bibnamefont {Hatefipour}}, \bibinfo {author} {\bibfnamefont {J.~J.}\ \bibnamefont {Cuozzo}}, \bibinfo {author} {\bibfnamefont {J.}~\bibnamefont {Kanter}}, \bibinfo {author} {\bibfnamefont {W.~M.}\ \bibnamefont {Strickland}}, \bibinfo {author} {\bibfnamefont {C.~R.}\ \bibnamefont {Allemang}}, \bibinfo {author} {\bibfnamefont {T.-M.}\ \bibnamefont {Lu}}, \bibinfo {author} {\bibfnamefont {E.}~\bibnamefont {Rossi}},\ and\ \bibinfo {author} {\bibfnamefont {J.}~\bibnamefont {Shabani}},\ }\bibfield  {title} {\bibinfo {title} {Induced superconducting pairing in integer quantum {Hall} edge states},\ }\href {https://doi.org/10.1021/acs.nanolett.2c01413} {\bibfield  {journal} {\bibinfo  {journal} {Nano Lett.}\ }\textbf {\bibinfo {volume} {22}},\ \bibinfo {pages} {6173} (\bibinfo {year} {2022})}\BibitemShut {NoStop}%
\bibitem [{\citenamefont {Zhao}\ \emph {et~al.}(2023)\citenamefont {Zhao}, \citenamefont {Iftikhar}, \citenamefont {Larson}, \citenamefont {Arnault}, \citenamefont {Watanabe}, \citenamefont {Taniguchi}, \citenamefont {Amet},\ and\ \citenamefont {Finkelstein}}]{Finkelstein_2023}%
  \BibitemOpen
  \bibfield  {author} {\bibinfo {author} {\bibfnamefont {L.}~\bibnamefont {Zhao}}, \bibinfo {author} {\bibfnamefont {Z.}~\bibnamefont {Iftikhar}}, \bibinfo {author} {\bibfnamefont {T.~F.~Q.}\ \bibnamefont {Larson}}, \bibinfo {author} {\bibfnamefont {E.~G.}\ \bibnamefont {Arnault}}, \bibinfo {author} {\bibfnamefont {K.}~\bibnamefont {Watanabe}}, \bibinfo {author} {\bibfnamefont {T.}~\bibnamefont {Taniguchi}}, \bibinfo {author} {\bibfnamefont {F.}~\bibnamefont {Amet}},\ and\ \bibinfo {author} {\bibfnamefont {G.}~\bibnamefont {Finkelstein}},\ }\bibfield  {title} {\bibinfo {title} {Loss and decoherence at the quantum {Hall}-superconductor interface},\ }\href {https://doi.org/10.1103/PhysRevLett.131.176604} {\bibfield  {journal} {\bibinfo  {journal} {Phys. Rev. Lett.}\ }\textbf {\bibinfo {volume} {131}},\ \bibinfo {pages} {176604} (\bibinfo {year} {2023})}\BibitemShut {NoStop}%
\bibitem [{\citenamefont {Vignaud}\ \emph {et~al.}(2023)\citenamefont {Vignaud}, \citenamefont {Perconte}, \citenamefont {Yang}, \citenamefont {Kousar}, \citenamefont {Wagner}, \citenamefont {Gay}, \citenamefont {Watanabe}, \citenamefont {Taniguchi}, \citenamefont {Courtois}, \citenamefont {Han}, \citenamefont {Sellier},\ and\ \citenamefont {Sac{\ifmmode\acute{e}\else\'{e}\fi}p{\ifmmode\acute{e}\else\'{e}\fi}}}]{Vignaud2023}%
  \BibitemOpen
  \bibfield  {author} {\bibinfo {author} {\bibfnamefont {H.}~\bibnamefont {Vignaud}}, \bibinfo {author} {\bibfnamefont {D.}~\bibnamefont {Perconte}}, \bibinfo {author} {\bibfnamefont {W.}~\bibnamefont {Yang}}, \bibinfo {author} {\bibfnamefont {B.}~\bibnamefont {Kousar}}, \bibinfo {author} {\bibfnamefont {E.}~\bibnamefont {Wagner}}, \bibinfo {author} {\bibfnamefont {F.}~\bibnamefont {Gay}}, \bibinfo {author} {\bibfnamefont {K.}~\bibnamefont {Watanabe}}, \bibinfo {author} {\bibfnamefont {T.}~\bibnamefont {Taniguchi}}, \bibinfo {author} {\bibfnamefont {H.}~\bibnamefont {Courtois}}, \bibinfo {author} {\bibfnamefont {Z.}~\bibnamefont {Han}}, \bibinfo {author} {\bibfnamefont {H.}~\bibnamefont {Sellier}},\ and\ \bibinfo {author} {\bibfnamefont {B.}~\bibnamefont {Sac{\ifmmode\acute{e}\else\'{e}\fi}p{\ifmmode\acute{e}\else\'{e}\fi}}},\ }\bibfield  {title} {\bibinfo {title} {{Evidence for chiral supercurrent in quantum Hall Josephson junctions}},\ }\href {https://doi.org/10.1038/s41586-023-06764-4} {\bibfield
  {journal} {\bibinfo  {journal} {Nature}\ }\textbf {\bibinfo {volume} {624}},\ \bibinfo {pages} {545} (\bibinfo {year} {2023})}\BibitemShut {NoStop}%
\bibitem [{\citenamefont {Uday}\ \emph {et~al.}(2024)\citenamefont {Uday}, \citenamefont {Lippertz}, \citenamefont {Moors}, \citenamefont {Legg}, \citenamefont {Joris}, \citenamefont {Bliesener}, \citenamefont {Pereira}, \citenamefont {Taskin},\ and\ \citenamefont {Ando}}]{Uday2023}%
  \BibitemOpen
  \bibfield  {author} {\bibinfo {author} {\bibfnamefont {A.}~\bibnamefont {Uday}}, \bibinfo {author} {\bibfnamefont {G.}~\bibnamefont {Lippertz}}, \bibinfo {author} {\bibfnamefont {K.}~\bibnamefont {Moors}}, \bibinfo {author} {\bibfnamefont {H.~F.}\ \bibnamefont {Legg}}, \bibinfo {author} {\bibfnamefont {R.}~\bibnamefont {Joris}}, \bibinfo {author} {\bibfnamefont {A.}~\bibnamefont {Bliesener}}, \bibinfo {author} {\bibfnamefont {L.~M.~C.}\ \bibnamefont {Pereira}}, \bibinfo {author} {\bibfnamefont {A.~A.}\ \bibnamefont {Taskin}},\ and\ \bibinfo {author} {\bibfnamefont {Y.}~\bibnamefont {Ando}},\ }\bibfield  {title} {\bibinfo {title} {{Induced superconducting correlations in a quantum anomalous Hall insulator}},\ }\href {https://doi.org/10.1038/s41567-024-02574-1} {\bibfield  {journal} {\bibinfo  {journal} {Nat. Phys.}\ }\textbf {\bibinfo {volume} {20}},\ \bibinfo {pages} {1589} (\bibinfo {year} {2024})}\BibitemShut {NoStop}%
\bibitem [{\citenamefont {van Ostaay}\ \emph {et~al.}(2011)\citenamefont {van Ostaay}, \citenamefont {Akhmerov},\ and\ \citenamefont {Beenakker}}]{Beenakker_2011}%
  \BibitemOpen
  \bibfield  {author} {\bibinfo {author} {\bibfnamefont {J.~A.~M.}\ \bibnamefont {van Ostaay}}, \bibinfo {author} {\bibfnamefont {A.~R.}\ \bibnamefont {Akhmerov}},\ and\ \bibinfo {author} {\bibfnamefont {C.~W.~J.}\ \bibnamefont {Beenakker}},\ }\bibfield  {title} {\bibinfo {title} {Spin-triplet supercurrent carried by quantum {H}all edge states through a {Josephson} junction},\ }\href {https://doi.org/10.1103/PhysRevB.83.195441} {\bibfield  {journal} {\bibinfo  {journal} {Phys. Rev. B}\ }\textbf {\bibinfo {volume} {83}},\ \bibinfo {pages} {195441} (\bibinfo {year} {2011})}\BibitemShut {NoStop}%
\bibitem [{\citenamefont {Beenakker}(2014)}]{Beenakker_2014}%
  \BibitemOpen
  \bibfield  {author} {\bibinfo {author} {\bibfnamefont {C.~W.~J.}\ \bibnamefont {Beenakker}},\ }\bibfield  {title} {\bibinfo {title} {Annihilation of colliding {Bogoliubov} quasiparticles reveals their {Majorana} nature},\ }\href {https://doi.org/10.1103/PhysRevLett.112.070604} {\bibfield  {journal} {\bibinfo  {journal} {Phys. Rev. Lett.}\ }\textbf {\bibinfo {volume} {112}},\ \bibinfo {pages} {070604} (\bibinfo {year} {2014})}\BibitemShut {NoStop}%
\bibitem [{\citenamefont {Clarke}\ \emph {et~al.}(2014)\citenamefont {Clarke}, \citenamefont {Alicea},\ and\ \citenamefont {Shtengel}}]{Clarke_2014}%
  \BibitemOpen
  \bibfield  {author} {\bibinfo {author} {\bibfnamefont {D.~J.}\ \bibnamefont {Clarke}}, \bibinfo {author} {\bibfnamefont {J.}~\bibnamefont {Alicea}},\ and\ \bibinfo {author} {\bibfnamefont {K.}~\bibnamefont {Shtengel}},\ }\bibfield  {title} {\bibinfo {title} {Exotic circuit elements from zero-modes in hybrid superconductor-quantum-{Hall} systems},\ }\href {https://doi.org/10.1038/nphys3114} {\bibfield  {journal} {\bibinfo  {journal} {Nat. Phys.}\ }\textbf {\bibinfo {volume} {10}},\ \bibinfo {pages} {877} (\bibinfo {year} {2014})}\BibitemShut {NoStop}%
\bibitem [{\citenamefont {Huang}\ and\ \citenamefont {Nazarov}(2017)}]{Nazarov_2017}%
  \BibitemOpen
  \bibfield  {author} {\bibinfo {author} {\bibfnamefont {X.-L.}\ \bibnamefont {Huang}}\ and\ \bibinfo {author} {\bibfnamefont {Y.~V.}\ \bibnamefont {Nazarov}},\ }\bibfield  {title} {\bibinfo {title} {Supercurrents in unidirectional channels originate from information transfer in the opposite direction: {A} theoretical prediction},\ }\href {https://doi.org/10.1103/PhysRevLett.118.177001} {\bibfield  {journal} {\bibinfo  {journal} {Phys. Rev. Lett.}\ }\textbf {\bibinfo {volume} {118}},\ \bibinfo {pages} {177001} (\bibinfo {year} {2017})}\BibitemShut {NoStop}%
\bibitem [{\citenamefont {Huang}\ and\ \citenamefont {Nazarov}(2019)}]{Nazarov_2019}%
  \BibitemOpen
  \bibfield  {author} {\bibinfo {author} {\bibfnamefont {X.-L.}\ \bibnamefont {Huang}}\ and\ \bibinfo {author} {\bibfnamefont {Y.~V.}\ \bibnamefont {Nazarov}},\ }\bibfield  {title} {\bibinfo {title} {Interaction-induced supercurrent in quantum {Hall} setups},\ }\href {https://doi.org/10.1103/PhysRevB.100.155411} {\bibfield  {journal} {\bibinfo  {journal} {Phys. Rev. B}\ }\textbf {\bibinfo {volume} {100}},\ \bibinfo {pages} {155411} (\bibinfo {year} {2019})}\BibitemShut {NoStop}%
\bibitem [{\citenamefont {Zhang}\ and\ \citenamefont {Trauzettel}(2019)}]{Trauzettel_2019}%
  \BibitemOpen
  \bibfield  {author} {\bibinfo {author} {\bibfnamefont {S.-B.}\ \bibnamefont {Zhang}}\ and\ \bibinfo {author} {\bibfnamefont {B.}~\bibnamefont {Trauzettel}},\ }\bibfield  {title} {\bibinfo {title} {Perfect crossed {Andreev} reflection in {Dirac} hybrid junctions in the quantum {Hall} regime},\ }\href {https://doi.org/10.1103/PhysRevLett.122.257701} {\bibfield  {journal} {\bibinfo  {journal} {Phys. Rev. Lett.}\ }\textbf {\bibinfo {volume} {122}},\ \bibinfo {pages} {257701} (\bibinfo {year} {2019})}\BibitemShut {NoStop}%
\bibitem [{\citenamefont {Michelsen}\ \emph {et~al.}(2020)\citenamefont {Michelsen}, \citenamefont {Schmidt},\ and\ \citenamefont {Idrisov}}]{Idrisov_2020}%
  \BibitemOpen
  \bibfield  {author} {\bibinfo {author} {\bibfnamefont {A.~B.}\ \bibnamefont {Michelsen}}, \bibinfo {author} {\bibfnamefont {T.~L.}\ \bibnamefont {Schmidt}},\ and\ \bibinfo {author} {\bibfnamefont {E.~G.}\ \bibnamefont {Idrisov}},\ }\bibfield  {title} {\bibinfo {title} {Current correlations of {Cooper}-pair tunneling into a quantum {Hall} system},\ }\href {https://doi.org/10.1103/PhysRevB.102.125402} {\bibfield  {journal} {\bibinfo  {journal} {Phys. Rev. B}\ }\textbf {\bibinfo {volume} {102}},\ \bibinfo {pages} {125402} (\bibinfo {year} {2020})}\BibitemShut {NoStop}%
\bibitem [{\citenamefont {Manesco}\ \emph {et~al.}(2022)\citenamefont {Manesco}, \citenamefont {Flór}, \citenamefont {Liu},\ and\ \citenamefont {Akhmerov}}]{Akhmerov_2022}%
  \BibitemOpen
  \bibfield  {author} {\bibinfo {author} {\bibfnamefont {A.~L.~R.}\ \bibnamefont {Manesco}}, \bibinfo {author} {\bibfnamefont {I.~M.}\ \bibnamefont {Flór}}, \bibinfo {author} {\bibfnamefont {C.-X.}\ \bibnamefont {Liu}},\ and\ \bibinfo {author} {\bibfnamefont {A.~R.}\ \bibnamefont {Akhmerov}},\ }\bibfield  {title} {\bibinfo {title} {{Mechanisms of Andreev reflection in quantum Hall graphene}},\ }\href {https://doi.org/10.21468/SciPostPhysCore.5.3.045} {\bibfield  {journal} {\bibinfo  {journal} {SciPost Phys. Core}\ }\textbf {\bibinfo {volume} {5}},\ \bibinfo {pages} {045} (\bibinfo {year} {2022})}\BibitemShut {NoStop}%
\bibitem [{\citenamefont {Galambos}\ \emph {et~al.}(2022)\citenamefont {Galambos}, \citenamefont {Ronetti}, \citenamefont {Het\'enyi}, \citenamefont {Loss},\ and\ \citenamefont {Klinovaja}}]{Klinovaja_2022}%
  \BibitemOpen
  \bibfield  {author} {\bibinfo {author} {\bibfnamefont {T.~H.}\ \bibnamefont {Galambos}}, \bibinfo {author} {\bibfnamefont {F.}~\bibnamefont {Ronetti}}, \bibinfo {author} {\bibfnamefont {B.}~\bibnamefont {Het\'enyi}}, \bibinfo {author} {\bibfnamefont {D.}~\bibnamefont {Loss}},\ and\ \bibinfo {author} {\bibfnamefont {J.}~\bibnamefont {Klinovaja}},\ }\bibfield  {title} {\bibinfo {title} {Crossed {Andreev} reflection in spin-polarized chiral edge states due to the {Meissner} effect},\ }\href {https://doi.org/10.1103/PhysRevB.106.075410} {\bibfield  {journal} {\bibinfo  {journal} {Phys. Rev. B}\ }\textbf {\bibinfo {volume} {106}},\ \bibinfo {pages} {075410} (\bibinfo {year} {2022})}\BibitemShut {NoStop}%
\bibitem [{\citenamefont {Kurilovich}\ \emph {et~al.}(2023)\citenamefont {Kurilovich}, \citenamefont {Raines},\ and\ \citenamefont {Glazman}}]{Kurilovich2023}%
  \BibitemOpen
  \bibfield  {author} {\bibinfo {author} {\bibfnamefont {V.~D.}\ \bibnamefont {Kurilovich}}, \bibinfo {author} {\bibfnamefont {Z.~M.}\ \bibnamefont {Raines}},\ and\ \bibinfo {author} {\bibfnamefont {L.~I.}\ \bibnamefont {Glazman}},\ }\bibfield  {title} {\bibinfo {title} {Disorder-enabled {Andreev} reflection of a quantum {Hall} edge},\ }\href {https://doi.org/10.1038/s41467-023-37794-1} {\bibfield  {journal} {\bibinfo  {journal} {Nat. Commun.}\ }\textbf {\bibinfo {volume} {14}},\ \bibinfo {pages} {1} (\bibinfo {year} {2023})}\BibitemShut {NoStop}%
\bibitem [{\citenamefont {Schiller}\ \emph {et~al.}(2023)\citenamefont {Schiller}, \citenamefont {Katzir}, \citenamefont {Stern}, \citenamefont {Berg}, \citenamefont {Lindner},\ and\ \citenamefont {Oreg}}]{Oreg_2023}%
  \BibitemOpen
  \bibfield  {author} {\bibinfo {author} {\bibfnamefont {N.}~\bibnamefont {Schiller}}, \bibinfo {author} {\bibfnamefont {B.~A.}\ \bibnamefont {Katzir}}, \bibinfo {author} {\bibfnamefont {A.}~\bibnamefont {Stern}}, \bibinfo {author} {\bibfnamefont {E.}~\bibnamefont {Berg}}, \bibinfo {author} {\bibfnamefont {N.~H.}\ \bibnamefont {Lindner}},\ and\ \bibinfo {author} {\bibfnamefont {Y.}~\bibnamefont {Oreg}},\ }\bibfield  {title} {\bibinfo {title} {Superconductivity and fermionic dissipation in quantum {Hall} edges},\ }\href {https://doi.org/10.1103/PhysRevB.107.L161105} {\bibfield  {journal} {\bibinfo  {journal} {Phys. Rev. B}\ }\textbf {\bibinfo {volume} {107}},\ \bibinfo {pages} {L161105} (\bibinfo {year} {2023})}\BibitemShut {NoStop}%
\bibitem [{\citenamefont {Michelsen}\ \emph {et~al.}(2023)\citenamefont {Michelsen}, \citenamefont {Recher}, \citenamefont {Braunecker},\ and\ \citenamefont {Schmidt}}]{Schmidt_2023}%
  \BibitemOpen
  \bibfield  {author} {\bibinfo {author} {\bibfnamefont {A.~B.}\ \bibnamefont {Michelsen}}, \bibinfo {author} {\bibfnamefont {P.}~\bibnamefont {Recher}}, \bibinfo {author} {\bibfnamefont {B.}~\bibnamefont {Braunecker}},\ and\ \bibinfo {author} {\bibfnamefont {T.~L.}\ \bibnamefont {Schmidt}},\ }\bibfield  {title} {\bibinfo {title} {Supercurrent-enabled {Andreev} reflection in a chiral quantum {Hall} edge state},\ }\href {https://doi.org/10.1103/PhysRevResearch.5.013066} {\bibfield  {journal} {\bibinfo  {journal} {Phys. Rev. Res.}\ }\textbf {\bibinfo {volume} {5}},\ \bibinfo {pages} {013066} (\bibinfo {year} {2023})}\BibitemShut {NoStop}%
\bibitem [{\citenamefont {David}\ \emph {et~al.}(2023)\citenamefont {David}, \citenamefont {Meyer},\ and\ \citenamefont {Houzet}}]{Houzet_2023}%
  \BibitemOpen
  \bibfield  {author} {\bibinfo {author} {\bibfnamefont {A.}~\bibnamefont {David}}, \bibinfo {author} {\bibfnamefont {J.~S.}\ \bibnamefont {Meyer}},\ and\ \bibinfo {author} {\bibfnamefont {M.}~\bibnamefont {Houzet}},\ }\bibfield  {title} {\bibinfo {title} {Geometrical effects on the downstream conductance in quantum-{Hall}--superconductor hybrid systems},\ }\href {https://doi.org/10.1103/PhysRevB.107.125416} {\bibfield  {journal} {\bibinfo  {journal} {Phys. Rev. B}\ }\textbf {\bibinfo {volume} {107}},\ \bibinfo {pages} {125416} (\bibinfo {year} {2023})}\BibitemShut {NoStop}%
\bibitem [{\citenamefont {Arrachea}\ \emph {et~al.}(2024)\citenamefont {Arrachea}, \citenamefont {Yeyati},\ and\ \citenamefont {Balseiro}}]{Balseiro_2023}%
  \BibitemOpen
  \bibfield  {author} {\bibinfo {author} {\bibfnamefont {L.}~\bibnamefont {Arrachea}}, \bibinfo {author} {\bibfnamefont {A.~L.}\ \bibnamefont {Yeyati}},\ and\ \bibinfo {author} {\bibfnamefont {C.~A.}\ \bibnamefont {Balseiro}},\ }\bibfield  {title} {\bibinfo {title} {Signatures of triplet superconductivity in $\ensuremath{\nu}=2$ chiral {Andreev} states},\ }\href {https://doi.org/10.1103/PhysRevB.109.064519} {\bibfield  {journal} {\bibinfo  {journal} {Phys. Rev. B}\ }\textbf {\bibinfo {volume} {109}},\ \bibinfo {pages} {064519} (\bibinfo {year} {2024})}\BibitemShut {NoStop}%
\bibitem [{\citenamefont {Baba}\ \emph {et~al.}(2025)\citenamefont {Baba}, \citenamefont {Levy~Yeyati},\ and\ \citenamefont {Burset}}]{Baba2025}%
  \BibitemOpen
  \bibfield  {author} {\bibinfo {author} {\bibfnamefont {Y.}~\bibnamefont {Baba}}, \bibinfo {author} {\bibfnamefont {A.}~\bibnamefont {Levy~Yeyati}},\ and\ \bibinfo {author} {\bibfnamefont {P.}~\bibnamefont {Burset}},\ }\bibfield  {title} {\bibinfo {title} {Emergent topology by {Landau} level mixing in quantum {Hall}-superconductor nanostructures},\ }\Eprint {https://arxiv.org/abs/2507.14074} {arXiv:2507.14074}  (\bibinfo {year} {2025})\BibitemShut {NoStop}%
\bibitem [{\citenamefont {Tien}\ and\ \citenamefont {Gordon}(1963)}]{Tien_1963}%
  \BibitemOpen
  \bibfield  {author} {\bibinfo {author} {\bibfnamefont {P.~K.}\ \bibnamefont {Tien}}\ and\ \bibinfo {author} {\bibfnamefont {J.~P.}\ \bibnamefont {Gordon}},\ }\bibfield  {title} {\bibinfo {title} {Multiphoton process observed in the interaction of microwave fields with the tunneling between superconductor films},\ }\href {https://doi.org/10.1103/PhysRev.129.647} {\bibfield  {journal} {\bibinfo  {journal} {Phys. Rev.}\ }\textbf {\bibinfo {volume} {129}},\ \bibinfo {pages} {647} (\bibinfo {year} {1963})}\BibitemShut {NoStop}%
\bibitem [{\citenamefont {Belzig}\ and\ \citenamefont {Vanevic}(2016)}]{Vanevic_2015}%
  \BibitemOpen
  \bibfield  {author} {\bibinfo {author} {\bibfnamefont {W.}~\bibnamefont {Belzig}}\ and\ \bibinfo {author} {\bibfnamefont {M.}~\bibnamefont {Vanevic}},\ }\bibfield  {title} {\bibinfo {title} {Elementary {Andreev} processes in a driven superconductor–normal metal contact},\ }\href {https://doi.org/https://doi.org/10.1016/j.physe.2015.08.036} {\bibfield  {journal} {\bibinfo  {journal} {Physica E}\ }\textbf {\bibinfo {volume} {75}},\ \bibinfo {pages} {22 } (\bibinfo {year} {2016})}\BibitemShut {NoStop}%
\bibitem [{\citenamefont {Acciai}\ \emph {et~al.}(2019)\citenamefont {Acciai}, \citenamefont {Ronetti}, \citenamefont {Ferraro}, \citenamefont {Rech}, \citenamefont {Jonckheere}, \citenamefont {Sassetti},\ and\ \citenamefont {Martin}}]{Martin_2019}%
  \BibitemOpen
  \bibfield  {author} {\bibinfo {author} {\bibfnamefont {M.}~\bibnamefont {Acciai}}, \bibinfo {author} {\bibfnamefont {F.}~\bibnamefont {Ronetti}}, \bibinfo {author} {\bibfnamefont {D.}~\bibnamefont {Ferraro}}, \bibinfo {author} {\bibfnamefont {J.}~\bibnamefont {Rech}}, \bibinfo {author} {\bibfnamefont {T.}~\bibnamefont {Jonckheere}}, \bibinfo {author} {\bibfnamefont {M.}~\bibnamefont {Sassetti}},\ and\ \bibinfo {author} {\bibfnamefont {T.}~\bibnamefont {Martin}},\ }\bibfield  {title} {\bibinfo {title} {Levitons in superconducting point contacts},\ }\href {https://doi.org/10.1103/PhysRevB.100.085418} {\bibfield  {journal} {\bibinfo  {journal} {Phys. Rev. B}\ }\textbf {\bibinfo {volume} {100}},\ \bibinfo {pages} {085418} (\bibinfo {year} {2019})}\BibitemShut {NoStop}%
\bibitem [{\citenamefont {Ronetti}\ \emph {et~al.}(2020)\citenamefont {Ronetti}, \citenamefont {Carrega},\ and\ \citenamefont {Sassetti}}]{Sassetti_2020}%
  \BibitemOpen
  \bibfield  {author} {\bibinfo {author} {\bibfnamefont {F.}~\bibnamefont {Ronetti}}, \bibinfo {author} {\bibfnamefont {M.}~\bibnamefont {Carrega}},\ and\ \bibinfo {author} {\bibfnamefont {M.}~\bibnamefont {Sassetti}},\ }\bibfield  {title} {\bibinfo {title} {Levitons in helical liquids with {Rashba} spin-orbit coupling probed by a superconducting contact},\ }\href {https://doi.org/10.1103/PhysRevResearch.2.013203} {\bibfield  {journal} {\bibinfo  {journal} {Phys. Rev. Res.}\ }\textbf {\bibinfo {volume} {2}},\ \bibinfo {pages} {013203} (\bibinfo {year} {2020})}\BibitemShut {NoStop}%
\bibitem [{\citenamefont {Averin}\ \emph {et~al.}(2020)\citenamefont {Averin}, \citenamefont {Wang},\ and\ \citenamefont {Vasenko}}]{Vasenko_2020}%
  \BibitemOpen
  \bibfield  {author} {\bibinfo {author} {\bibfnamefont {D.~V.}\ \bibnamefont {Averin}}, \bibinfo {author} {\bibfnamefont {G.}~\bibnamefont {Wang}},\ and\ \bibinfo {author} {\bibfnamefont {A.~S.}\ \bibnamefont {Vasenko}},\ }\bibfield  {title} {\bibinfo {title} {Time-dependent {Andreev} reflection},\ }\href {https://doi.org/10.1103/PhysRevB.102.144516} {\bibfield  {journal} {\bibinfo  {journal} {Phys. Rev. B}\ }\textbf {\bibinfo {volume} {102}},\ \bibinfo {pages} {144516} (\bibinfo {year} {2020})}\BibitemShut {NoStop}%
\bibitem [{\citenamefont {Bertin-Johannet}\ \emph {et~al.}(2024)\citenamefont {Bertin-Johannet}, \citenamefont {Gr\'emaud}, \citenamefont {Ronetti}, \citenamefont {Raymond}, \citenamefont {Rech}, \citenamefont {Jonckheere},\ and\ \citenamefont {Martin}}]{Martin_2023c}%
  \BibitemOpen
  \bibfield  {author} {\bibinfo {author} {\bibfnamefont {B.}~\bibnamefont {Bertin-Johannet}}, \bibinfo {author} {\bibfnamefont {B.}~\bibnamefont {Gr\'emaud}}, \bibinfo {author} {\bibfnamefont {F.}~\bibnamefont {Ronetti}}, \bibinfo {author} {\bibfnamefont {L.}~\bibnamefont {Raymond}}, \bibinfo {author} {\bibfnamefont {J.}~\bibnamefont {Rech}}, \bibinfo {author} {\bibfnamefont {T.}~\bibnamefont {Jonckheere}},\ and\ \bibinfo {author} {\bibfnamefont {T.}~\bibnamefont {Martin}},\ }\bibfield  {title} {\bibinfo {title} {Current and shot noise in a normal metal--superconductor junction driven by spin-dependent periodic pulse sequence},\ }\href {https://doi.org/10.1103/PhysRevB.109.174514} {\bibfield  {journal} {\bibinfo  {journal} {Phys. Rev. B}\ }\textbf {\bibinfo {volume} {109}},\ \bibinfo {pages} {174514} (\bibinfo {year} {2024})}\BibitemShut {NoStop}%
\bibitem [{\citenamefont {Ronetti}\ \emph {et~al.}(2024)\citenamefont {Ronetti}, \citenamefont {Bertin-Johannet}, \citenamefont {Popoff}, \citenamefont {Rech}, \citenamefont {Jonckheere}, \citenamefont {Gr{\ifmmode\acute{e}\else\'{e}\fi}maud}, \citenamefont {Raymond},\ and\ \citenamefont {Martin}}]{Ronetti2024}%
  \BibitemOpen
  \bibfield  {author} {\bibinfo {author} {\bibfnamefont {F.}~\bibnamefont {Ronetti}}, \bibinfo {author} {\bibfnamefont {B.}~\bibnamefont {Bertin-Johannet}}, \bibinfo {author} {\bibfnamefont {A.}~\bibnamefont {Popoff}}, \bibinfo {author} {\bibfnamefont {J.}~\bibnamefont {Rech}}, \bibinfo {author} {\bibfnamefont {T.}~\bibnamefont {Jonckheere}}, \bibinfo {author} {\bibfnamefont {B.}~\bibnamefont {Gr{\ifmmode\acute{e}\else\'{e}\fi}maud}}, \bibinfo {author} {\bibfnamefont {L.}~\bibnamefont {Raymond}},\ and\ \bibinfo {author} {\bibfnamefont {T.}~\bibnamefont {Martin}},\ }\bibfield  {title} {\bibinfo {title} {{Levitons in correlated nano-scale systems}},\ }\href {https://doi.org/10.1063/5.0199567} {\bibfield  {journal} {\bibinfo  {journal} {Chaos}\ }\textbf {\bibinfo {volume} {34}},\ \bibinfo {pages} {042103} (\bibinfo {year} {2024})}\BibitemShut {NoStop}%
\bibitem [{\citenamefont {Burset}\ \emph {et~al.}(2023)\citenamefont {Burset}, \citenamefont {Roussel}, \citenamefont {Moskalets},\ and\ \citenamefont {Flindt}}]{Burset_short}%
  \BibitemOpen
  \bibfield  {author} {\bibinfo {author} {\bibfnamefont {P.}~\bibnamefont {Burset}}, \bibinfo {author} {\bibfnamefont {B.}~\bibnamefont {Roussel}}, \bibinfo {author} {\bibfnamefont {M.}~\bibnamefont {Moskalets}},\ and\ \bibinfo {author} {\bibfnamefont {C.}~\bibnamefont {Flindt}},\ }\bibfield  {title} {\bibinfo {title} {{Tunable {Andreev}-conversion of single-electron charge pulses}},\ }\Eprint {https://arxiv.org/abs/2312.13145} {arXiv:2312.13145}  (\bibinfo {year} {2023})\BibitemShut {NoStop}%
\bibitem [{\citenamefont {Keeling}\ \emph {et~al.}(2006)\citenamefont {Keeling}, \citenamefont {Klich},\ and\ \citenamefont {Levitov}}]{Levitov_2006}%
  \BibitemOpen
  \bibfield  {author} {\bibinfo {author} {\bibfnamefont {J.}~\bibnamefont {Keeling}}, \bibinfo {author} {\bibfnamefont {I.}~\bibnamefont {Klich}},\ and\ \bibinfo {author} {\bibfnamefont {L.~S.}\ \bibnamefont {Levitov}},\ }\bibfield  {title} {\bibinfo {title} {Minimal excitation states of electrons in one-dimensional wires},\ }\href {https://doi.org/10.1103/PhysRevLett.97.116403} {\bibfield  {journal} {\bibinfo  {journal} {Phys. Rev. Lett.}\ }\textbf {\bibinfo {volume} {97}},\ \bibinfo {pages} {116403} (\bibinfo {year} {2006})}\BibitemShut {NoStop}%
\bibitem [{\citenamefont {Anantram}\ and\ \citenamefont {Datta}(1996)}]{Anantram-Datta}%
  \BibitemOpen
  \bibfield  {author} {\bibinfo {author} {\bibfnamefont {M.~P.}\ \bibnamefont {Anantram}}\ and\ \bibinfo {author} {\bibfnamefont {S.}~\bibnamefont {Datta}},\ }\bibfield  {title} {\bibinfo {title} {Current fluctuations in mesoscopic systems with {Andreev} scattering},\ }\href {https://doi.org/10.1103/PhysRevB.53.16390} {\bibfield  {journal} {\bibinfo  {journal} {Phys. Rev. B}\ }\textbf {\bibinfo {volume} {53}},\ \bibinfo {pages} {16390} (\bibinfo {year} {1996})}\BibitemShut {NoStop}%
\bibitem [{\citenamefont {Blonder}\ \emph {et~al.}(1982)\citenamefont {Blonder}, \citenamefont {Tinkham},\ and\ \citenamefont {Klapwijk}}]{BTK}%
  \BibitemOpen
  \bibfield  {author} {\bibinfo {author} {\bibfnamefont {G.~E.}\ \bibnamefont {Blonder}}, \bibinfo {author} {\bibfnamefont {M.}~\bibnamefont {Tinkham}},\ and\ \bibinfo {author} {\bibfnamefont {T.~M.}\ \bibnamefont {Klapwijk}},\ }\bibfield  {title} {\bibinfo {title} {Transition from metallic to tunneling regimes in superconducting microconstrictions: {Excess} current, charge imbalance, and supercurrent conversion},\ }\href {https://doi.org/10.1103/PhysRevB.25.4515} {\bibfield  {journal} {\bibinfo  {journal} {Phys. Rev. B}\ }\textbf {\bibinfo {volume} {25}},\ \bibinfo {pages} {4515} (\bibinfo {year} {1982})}\BibitemShut {NoStop}%
\bibitem [{\citenamefont {Cayao}\ \emph {et~al.}(2021)\citenamefont {Cayao}, \citenamefont {Triola},\ and\ \citenamefont {Black-Schaffer}}]{Cayao2021Mar}%
  \BibitemOpen
  \bibfield  {author} {\bibinfo {author} {\bibfnamefont {J.}~\bibnamefont {Cayao}}, \bibinfo {author} {\bibfnamefont {C.}~\bibnamefont {Triola}},\ and\ \bibinfo {author} {\bibfnamefont {A.~M.}\ \bibnamefont {Black-Schaffer}},\ }\bibfield  {title} {\bibinfo {title} {Floquet engineering bulk odd-frequency superconducting pairs},\ }\href {https://doi.org/10.1103/PhysRevB.103.104505} {\bibfield  {journal} {\bibinfo  {journal} {Phys. Rev. B}\ }\textbf {\bibinfo {volume} {103}},\ \bibinfo {pages} {104505} (\bibinfo {year} {2021})}\BibitemShut {NoStop}%
\bibitem [{\citenamefont {Ahmed}\ \emph {et~al.}(2025)\citenamefont {Ahmed}, \citenamefont {Tamura}, \citenamefont {Tanaka},\ and\ \citenamefont {Cayao}}]{Ahmed2025Jan}%
  \BibitemOpen
  \bibfield  {author} {\bibinfo {author} {\bibfnamefont {E.}~\bibnamefont {Ahmed}}, \bibinfo {author} {\bibfnamefont {S.}~\bibnamefont {Tamura}}, \bibinfo {author} {\bibfnamefont {Y.}~\bibnamefont {Tanaka}},\ and\ \bibinfo {author} {\bibfnamefont {J.}~\bibnamefont {Cayao}},\ }\bibfield  {title} {\bibinfo {title} {Odd-frequency superconducting pairing due to multiple majorana edge modes in driven topological superconductors},\ }\href {https://doi.org/10.1103/PhysRevB.111.024507} {\bibfield  {journal} {\bibinfo  {journal} {Phys. Rev. B}\ }\textbf {\bibinfo {volume} {111}},\ \bibinfo {pages} {024507} (\bibinfo {year} {2025})}\BibitemShut {NoStop}%
\bibitem [{\citenamefont {Moskalets}\ \emph {et~al.}(2020)\citenamefont {Moskalets}, \citenamefont {Kotilahti}, \citenamefont {Burset},\ and\ \citenamefont {Flindt}}]{Moskalets_2020}%
  \BibitemOpen
  \bibfield  {author} {\bibinfo {author} {\bibfnamefont {M.}~\bibnamefont {Moskalets}}, \bibinfo {author} {\bibfnamefont {J.}~\bibnamefont {Kotilahti}}, \bibinfo {author} {\bibfnamefont {P.}~\bibnamefont {Burset}},\ and\ \bibinfo {author} {\bibfnamefont {C.}~\bibnamefont {Flindt}},\ }\bibfield  {title} {\bibinfo {title} {Composite two-particle sources},\ }\href {https://doi.org/10.1140/epjst/e2019-900121-x} {\bibfield  {journal} {\bibinfo  {journal} {Eur. Phys. J. Spec. Top.}\ }\textbf {\bibinfo {volume} {229}},\ \bibinfo {pages} {647} (\bibinfo {year} {2020})}\BibitemShut {NoStop}%
\bibitem [{\citenamefont {Kotilahti}\ \emph {et~al.}(2021)\citenamefont {Kotilahti}, \citenamefont {Burset}, \citenamefont {Moskalets},\ and\ \citenamefont {Flindt}}]{Kotilahti_2021}%
  \BibitemOpen
  \bibfield  {author} {\bibinfo {author} {\bibfnamefont {J.}~\bibnamefont {Kotilahti}}, \bibinfo {author} {\bibfnamefont {P.}~\bibnamefont {Burset}}, \bibinfo {author} {\bibfnamefont {M.}~\bibnamefont {Moskalets}},\ and\ \bibinfo {author} {\bibfnamefont {C.}~\bibnamefont {Flindt}},\ }\bibfield  {title} {\bibinfo {title} {Multi-particle interference in an electronic {Mach–Zehnder} interferometer},\ }\href {https://doi.org/10.3390/e23060736} {\bibfield  {journal} {\bibinfo  {journal} {Entropy}\ }\textbf {\bibinfo {volume} {23}},\ \bibinfo {pages} {736} (\bibinfo {year} {2021})}\BibitemShut {NoStop}%
\bibitem [{\citenamefont {Moskalets}(2011)}]{Moskalets_book}%
  \BibitemOpen
  \bibfield  {author} {\bibinfo {author} {\bibfnamefont {M.~V.}\ \bibnamefont {Moskalets}},\ }\href {https://doi.org/10.1142/p822} {\emph {\bibinfo {title} {Scattering {M}atrix {A}pproach to {N}on-{S}tationary {Q}uantum {T}ransport}}}\ (\bibinfo  {publisher} {Imperial College Press},\ \bibinfo {year} {2011})\BibitemShut {NoStop}%
\bibitem [{\citenamefont {Hofer}\ and\ \citenamefont {Flindt}(2014)}]{hofer:2014}%
  \BibitemOpen
  \bibfield  {author} {\bibinfo {author} {\bibfnamefont {P.~P.}\ \bibnamefont {Hofer}}\ and\ \bibinfo {author} {\bibfnamefont {C.}~\bibnamefont {Flindt}},\ }\bibfield  {title} {\bibinfo {title} {{Mach-Zehnder interferometry with periodic voltage pulses}},\ }\href {https://doi.org/10.1103/PhysRevB.90.235416} {\bibfield  {journal} {\bibinfo  {journal} {Phys. Rev. B}\ }\textbf {\bibinfo {volume} {90}},\ \bibinfo {pages} {235416} (\bibinfo {year} {2014})}\BibitemShut {NoStop}%
\bibitem [{\citenamefont {Dubois}\ \emph {et~al.}(2013{\natexlab{b}})\citenamefont {Dubois}, \citenamefont {Jullien}, \citenamefont {Grenier}, \citenamefont {Degiovanni}, \citenamefont {Roulleau},\ and\ \citenamefont {Glattli}}]{Dubois_2013b}%
  \BibitemOpen
  \bibfield  {author} {\bibinfo {author} {\bibfnamefont {J.}~\bibnamefont {Dubois}}, \bibinfo {author} {\bibfnamefont {T.}~\bibnamefont {Jullien}}, \bibinfo {author} {\bibfnamefont {C.}~\bibnamefont {Grenier}}, \bibinfo {author} {\bibfnamefont {P.}~\bibnamefont {Degiovanni}}, \bibinfo {author} {\bibfnamefont {P.}~\bibnamefont {Roulleau}},\ and\ \bibinfo {author} {\bibfnamefont {D.~C.}\ \bibnamefont {Glattli}},\ }\bibfield  {title} {\bibinfo {title} {Integer and fractional charge {Lorentzian} voltage pulses analyzed in the framework of photon-assisted shot noise},\ }\href {https://doi.org/10.1103/PhysRevB.88.085301} {\bibfield  {journal} {\bibinfo  {journal} {Phys. Rev. B}\ }\textbf {\bibinfo {volume} {88}},\ \bibinfo {pages} {085301} (\bibinfo {year} {2013}{\natexlab{b}})}\BibitemShut {NoStop}%
\bibitem [{\citenamefont {Aluffi}\ \emph {et~al.}(2023)\citenamefont {Aluffi}, \citenamefont {Vasselon}, \citenamefont {Ouacel}, \citenamefont {Edlbauer}, \citenamefont {Geffroy}, \citenamefont {Roulleau}, \citenamefont {Glattli}, \citenamefont {Georgiou},\ and\ \citenamefont {B\"auerle}}]{Aluffi2023}%
  \BibitemOpen
  \bibfield  {author} {\bibinfo {author} {\bibfnamefont {M.}~\bibnamefont {Aluffi}}, \bibinfo {author} {\bibfnamefont {T.}~\bibnamefont {Vasselon}}, \bibinfo {author} {\bibfnamefont {S.}~\bibnamefont {Ouacel}}, \bibinfo {author} {\bibfnamefont {H.}~\bibnamefont {Edlbauer}}, \bibinfo {author} {\bibfnamefont {C.}~\bibnamefont {Geffroy}}, \bibinfo {author} {\bibfnamefont {P.}~\bibnamefont {Roulleau}}, \bibinfo {author} {\bibfnamefont {D.~C.}\ \bibnamefont {Glattli}}, \bibinfo {author} {\bibfnamefont {G.}~\bibnamefont {Georgiou}},\ and\ \bibinfo {author} {\bibfnamefont {C.}~\bibnamefont {B\"auerle}},\ }\bibfield  {title} {\bibinfo {title} {Ultrashort electron wave packets via frequency-comb synthesis},\ }\href {https://doi.org/10.1103/PhysRevApplied.20.034005} {\bibfield  {journal} {\bibinfo  {journal} {Phys. Rev. Appl.}\ }\textbf {\bibinfo {volume} {20}},\ \bibinfo {pages} {034005} (\bibinfo {year} {2023})}\BibitemShut {NoStop}%
\bibitem [{\citenamefont {Moskalets}\ and\ \citenamefont {B\"uttiker}(2002)}]{Moskalets_2002}%
  \BibitemOpen
  \bibfield  {author} {\bibinfo {author} {\bibfnamefont {M.}~\bibnamefont {Moskalets}}\ and\ \bibinfo {author} {\bibfnamefont {M.}~\bibnamefont {B\"uttiker}},\ }\bibfield  {title} {\bibinfo {title} {{Floquet} scattering theory of quantum pumps},\ }\href {https://doi.org/10.1103/PhysRevB.66.205320} {\bibfield  {journal} {\bibinfo  {journal} {Phys. Rev. B}\ }\textbf {\bibinfo {volume} {66}},\ \bibinfo {pages} {205320} (\bibinfo {year} {2002})}\BibitemShut {NoStop}%
\bibitem [{\citenamefont {Lambert}\ and\ \citenamefont {Raimondi}(1998)}]{Lambert-Raimondi}%
  \BibitemOpen
  \bibfield  {author} {\bibinfo {author} {\bibfnamefont {C.~J.}\ \bibnamefont {Lambert}}\ and\ \bibinfo {author} {\bibfnamefont {R.}~\bibnamefont {Raimondi}},\ }\bibfield  {title} {\bibinfo {title} {{Phase-coherent transport in hybrid superconducting nanostructures}},\ }\href {https://doi.org/10.1088/0953-8984/10/5/003} {\bibfield  {journal} {\bibinfo  {journal} {J. Phys.: Condens. Matter}\ }\textbf {\bibinfo {volume} {10}},\ \bibinfo {pages} {901} (\bibinfo {year} {1998})}\BibitemShut {NoStop}%
\bibitem [{\citenamefont {Moskalets}\ and\ \citenamefont {Haack}(2017)}]{Moskalets_2016b}%
  \BibitemOpen
  \bibfield  {author} {\bibinfo {author} {\bibfnamefont {M.}~\bibnamefont {Moskalets}}\ and\ \bibinfo {author} {\bibfnamefont {G.}~\bibnamefont {Haack}},\ }\bibfield  {title} {\bibinfo {title} {Heat and charge transport measurements to access single-electron quantum characteristics},\ }\href {https://doi.org/10.1002/pssb.201600616} {\bibfield  {journal} {\bibinfo  {journal} {Phys. Status Solidi B}\ }\textbf {\bibinfo {volume} {254}},\ \bibinfo {pages} {1600616} (\bibinfo {year} {2017})}\BibitemShut {NoStop}%
\bibitem [{\citenamefont {Roussel}\ \emph {et~al.}(2025)\citenamefont {Roussel}, \citenamefont {Burset},\ and\ \citenamefont {Flindt}}]{Roussel_2023}%
  \BibitemOpen
  \bibfield  {author} {\bibinfo {author} {\bibfnamefont {B.}~\bibnamefont {Roussel}}, \bibinfo {author} {\bibfnamefont {P.}~\bibnamefont {Burset}},\ and\ \bibinfo {author} {\bibfnamefont {C.}~\bibnamefont {Flindt}},\ }\bibfield  {title} {\bibinfo {title} {Wigner representation of andreev-reflected charge pulses},\ }\href {https://doi.org/10.1103/g8z1-f2fk} {\bibfield  {journal} {\bibinfo  {journal} {Phys. Rev. B}\ }\textbf {\bibinfo {volume} {112}},\ \bibinfo {pages} {014506} (\bibinfo {year} {2025})}\BibitemShut {NoStop}%
\bibitem [{\citenamefont {Roussel}\ \emph {et~al.}(2021)\citenamefont {Roussel}, \citenamefont {Cabart}, \citenamefont {F\`eve},\ and\ \citenamefont {Degiovanni}}]{Roussel_2021}%
  \BibitemOpen
  \bibfield  {author} {\bibinfo {author} {\bibfnamefont {B.}~\bibnamefont {Roussel}}, \bibinfo {author} {\bibfnamefont {C.}~\bibnamefont {Cabart}}, \bibinfo {author} {\bibfnamefont {G.}~\bibnamefont {F\`eve}},\ and\ \bibinfo {author} {\bibfnamefont {P.}~\bibnamefont {Degiovanni}},\ }\bibfield  {title} {\bibinfo {title} {Processing quantum signals carried by electrical currents},\ }\href {https://doi.org/10.1103/PRXQuantum.2.020314} {\bibfield  {journal} {\bibinfo  {journal} {PRX Quantum}\ }\textbf {\bibinfo {volume} {2}},\ \bibinfo {pages} {020314} (\bibinfo {year} {2021})}\BibitemShut {NoStop}%
\bibitem [{\citenamefont {Shimizu}\ \emph {et~al.}(2025)\citenamefont {Shimizu}, \citenamefont {Iyoda}, \citenamefont {Sasaki}, \citenamefont {Endo}, \citenamefont {Katsumoto}, \citenamefont {Kumada},\ and\ \citenamefont {Hashisaka}}]{Shimizu2024}%
  \BibitemOpen
  \bibfield  {author} {\bibinfo {author} {\bibfnamefont {T.}~\bibnamefont {Shimizu}}, \bibinfo {author} {\bibfnamefont {E.}~\bibnamefont {Iyoda}}, \bibinfo {author} {\bibfnamefont {S.}~\bibnamefont {Sasaki}}, \bibinfo {author} {\bibfnamefont {A.}~\bibnamefont {Endo}}, \bibinfo {author} {\bibfnamefont {S.}~\bibnamefont {Katsumoto}}, \bibinfo {author} {\bibfnamefont {N.}~\bibnamefont {Kumada}},\ and\ \bibinfo {author} {\bibfnamefont {M.}~\bibnamefont {Hashisaka}},\ }\bibfield  {title} {\bibinfo {title} {Mach-zehnder interference of fractionalized electron-spin excitations},\ }\href {https://doi.org/10.1103/PhysRevB.111.L161406} {\bibfield  {journal} {\bibinfo  {journal} {Phys. Rev. B}\ }\textbf {\bibinfo {volume} {111}},\ \bibinfo {pages} {L161406} (\bibinfo {year} {2025})}\BibitemShut {NoStop}%
\bibitem [{\citenamefont {Iyoda}\ \emph {et~al.}(2024)\citenamefont {Iyoda}, \citenamefont {Shimizu},\ and\ \citenamefont {Hashisaka}}]{Iyoda2024}%
  \BibitemOpen
  \bibfield  {author} {\bibinfo {author} {\bibfnamefont {E.}~\bibnamefont {Iyoda}}, \bibinfo {author} {\bibfnamefont {T.}~\bibnamefont {Shimizu}},\ and\ \bibinfo {author} {\bibfnamefont {M.}~\bibnamefont {Hashisaka}},\ }\bibfield  {title} {\bibinfo {title} {{Coherent electron splitting in interacting chiral edge channels}},\ }\Eprint {https://arxiv.org/abs/2407.11491} {arXiv:2407.11491}  (\bibinfo {year} {2024})\BibitemShut {NoStop}%
\bibitem [{\citenamefont {Fukuzawa}\ \emph {et~al.}(2023)\citenamefont {Fukuzawa}, \citenamefont {Kato}, \citenamefont {Jonckheere}, \citenamefont {Rech},\ and\ \citenamefont {Martin}}]{Martin_2023b}%
  \BibitemOpen
  \bibfield  {author} {\bibinfo {author} {\bibfnamefont {K.}~\bibnamefont {Fukuzawa}}, \bibinfo {author} {\bibfnamefont {T.}~\bibnamefont {Kato}}, \bibinfo {author} {\bibfnamefont {T.}~\bibnamefont {Jonckheere}}, \bibinfo {author} {\bibfnamefont {J.}~\bibnamefont {Rech}},\ and\ \bibinfo {author} {\bibfnamefont {T.}~\bibnamefont {Martin}},\ }\bibfield  {title} {\bibinfo {title} {Minimal alternating current injection into carbon nanotubes},\ }\href {https://doi.org/10.1103/PhysRevB.108.125307} {\bibfield  {journal} {\bibinfo  {journal} {Phys. Rev. B}\ }\textbf {\bibinfo {volume} {108}},\ \bibinfo {pages} {125307} (\bibinfo {year} {2023})}\BibitemShut {NoStop}%
\bibitem [{\citenamefont {Moskalets}(2015)}]{Moskalets_2015}%
  \BibitemOpen
  \bibfield  {author} {\bibinfo {author} {\bibfnamefont {M.}~\bibnamefont {Moskalets}},\ }\bibfield  {title} {\bibinfo {title} {First-order correlation function of a stream of single-electron wave packets},\ }\href {https://doi.org/10.1103/PhysRevB.91.195431} {\bibfield  {journal} {\bibinfo  {journal} {Phys. Rev. B}\ }\textbf {\bibinfo {volume} {91}},\ \bibinfo {pages} {195431} (\bibinfo {year} {2015})}\BibitemShut {NoStop}%
\bibitem [{\citenamefont {Tanaka}\ \emph {et~al.}(2011)\citenamefont {Tanaka}, \citenamefont {Sato},\ and\ \citenamefont {Nagaosa}}]{Tanaka2011}%
  \BibitemOpen
  \bibfield  {author} {\bibinfo {author} {\bibfnamefont {Y.}~\bibnamefont {Tanaka}}, \bibinfo {author} {\bibfnamefont {M.}~\bibnamefont {Sato}},\ and\ \bibinfo {author} {\bibfnamefont {N.}~\bibnamefont {Nagaosa}},\ }\bibfield  {title} {\bibinfo {title} {Symmetry and topology in superconductors {\textendash}odd-frequency pairing and edge states{\textendash}},\ }\href {https://doi.org/10.1143/JPSJ.81.011013} {\bibfield  {journal} {\bibinfo  {journal} {J. Phys. Soc. Jpn.}\ }\textbf {\bibinfo {volume} {81}},\ \bibinfo {pages} {011013} (\bibinfo {year} {2011})}\BibitemShut {NoStop}%
\bibitem [{\citenamefont {Roussel}\ \emph {et~al.}(2017)\citenamefont {Roussel}, \citenamefont {Cabart}, \citenamefont {F{\ifmmode\grave{e}\else\`{e}\fi}ve}, \citenamefont {Thibierge},\ and\ \citenamefont {Degiovanni}}]{Roussel_2017}%
  \BibitemOpen
  \bibfield  {author} {\bibinfo {author} {\bibfnamefont {B.}~\bibnamefont {Roussel}}, \bibinfo {author} {\bibfnamefont {C.}~\bibnamefont {Cabart}}, \bibinfo {author} {\bibfnamefont {G.}~\bibnamefont {F{\ifmmode\grave{e}\else\`{e}\fi}ve}}, \bibinfo {author} {\bibfnamefont {E.}~\bibnamefont {Thibierge}},\ and\ \bibinfo {author} {\bibfnamefont {P.}~\bibnamefont {Degiovanni}},\ }\bibfield  {title} {\bibinfo {title} {Electron quantum optics as quantum signal processing},\ }\href {https://doi.org/10.1002/pssb.201600621} {\bibfield  {journal} {\bibinfo  {journal} {Phys. Status Solidi B}\ }\textbf {\bibinfo {volume} {254}},\ \bibinfo {pages} {1600621} (\bibinfo {year} {2017})}\BibitemShut {NoStop}%
\bibitem [{\citenamefont {Marguerite}\ \emph {et~al.}(2017)\citenamefont {Marguerite}, \citenamefont {Bocquillon}, \citenamefont {Berroir}, \citenamefont {Pla{\ifmmode\mbox{\c{c}}\else\c{c}\fi}ais}, \citenamefont {Cavanna}, \citenamefont {Jin}, \citenamefont {Degiovanni},\ and\ \citenamefont {F{\ifmmode\grave{e}\else\`{e}\fi}ve}}]{Marguerite2017}%
  \BibitemOpen
  \bibfield  {author} {\bibinfo {author} {\bibfnamefont {A.}~\bibnamefont {Marguerite}}, \bibinfo {author} {\bibfnamefont {E.}~\bibnamefont {Bocquillon}}, \bibinfo {author} {\bibfnamefont {J.-M.}\ \bibnamefont {Berroir}}, \bibinfo {author} {\bibfnamefont {B.}~\bibnamefont {Pla{\ifmmode\mbox{\c{c}}\else\c{c}\fi}ais}}, \bibinfo {author} {\bibfnamefont {A.}~\bibnamefont {Cavanna}}, \bibinfo {author} {\bibfnamefont {Y.}~\bibnamefont {Jin}}, \bibinfo {author} {\bibfnamefont {P.}~\bibnamefont {Degiovanni}},\ and\ \bibinfo {author} {\bibfnamefont {G.}~\bibnamefont {F{\ifmmode\grave{e}\else\`{e}\fi}ve}},\ }\bibfield  {title} {\bibinfo {title} {{Two-particle interferometry in quantum Hall edge channels}},\ }\href {https://doi.org/10.1002/pssb.201600618} {\bibfield  {journal} {\bibinfo  {journal} {Phys. Status Solidi B}\ }\textbf {\bibinfo {volume} {254}},\ \bibinfo {pages} {1600618} (\bibinfo {year} {2017})}\BibitemShut {NoStop}%
\bibitem [{\citenamefont {Kashcheyevs}\ \emph {et~al.}(2022)\citenamefont {Kashcheyevs}, \citenamefont {Degiovanni}, \citenamefont {Roussel}, \citenamefont {Kataoka}, \citenamefont {Fletcher}, \citenamefont {Freise}, \citenamefont {Ubbelohde}, \citenamefont {F{\ifmmode\grave{e}\else\`{e}\fi}ve}, \citenamefont {Bartolomei}, \citenamefont {Cou{\ifmmode\ddot{e}\else\"{e}\fi}do}, \citenamefont {Kadykov}, \citenamefont {Poirier}, \citenamefont {Roulleau}, \citenamefont {Parmentier},\ and\ \citenamefont {Hohls}}]{sequoia2022}%
  \BibitemOpen
  \bibfield  {author} {\bibinfo {author} {\bibfnamefont {V.}~\bibnamefont {Kashcheyevs}}, \bibinfo {author} {\bibfnamefont {P.}~\bibnamefont {Degiovanni}}, \bibinfo {author} {\bibfnamefont {B.}~\bibnamefont {Roussel}}, \bibinfo {author} {\bibfnamefont {M.}~\bibnamefont {Kataoka}}, \bibinfo {author} {\bibfnamefont {J.~D.}\ \bibnamefont {Fletcher}}, \bibinfo {author} {\bibfnamefont {L.}~\bibnamefont {Freise}}, \bibinfo {author} {\bibfnamefont {N.}~\bibnamefont {Ubbelohde}}, \bibinfo {author} {\bibfnamefont {G.}~\bibnamefont {F{\ifmmode\grave{e}\else\`{e}\fi}ve}}, \bibinfo {author} {\bibfnamefont {H.}~\bibnamefont {Bartolomei}}, \bibinfo {author} {\bibfnamefont {F.}~\bibnamefont {Cou{\ifmmode\ddot{e}\else\"{e}\fi}do}}, \bibinfo {author} {\bibfnamefont {A.}~\bibnamefont {Kadykov}}, \bibinfo {author} {\bibfnamefont {W.}~\bibnamefont {Poirier}}, \bibinfo {author} {\bibfnamefont {P.}~\bibnamefont {Roulleau}}, \bibinfo {author} {\bibfnamefont {F.~D.}\ \bibnamefont {Parmentier}},\ and\ \bibinfo {author} {\bibfnamefont
  {F.}~\bibnamefont {Hohls}},\ }\bibfield  {title} {\bibinfo {title} {{Single-electron wave packets for quantum metrology: concepts, implementations, and applications}},\ }\bibfield  {journal} {\bibinfo  {journal} {PTB-OAR}\ }\href {https://doi.org/10.7795/EMPIR.17FUN04.RE.20220228} {10.7795/EMPIR.17FUN04.RE.20220228} (\bibinfo {year} {2022})\BibitemShut {NoStop}%
\bibitem [{\citenamefont {Edlbauer}\ \emph {et~al.}(2022)\citenamefont {Edlbauer}, \citenamefont {Wang}, \citenamefont {Crozes}, \citenamefont {Perrier}, \citenamefont {Ouacel}, \citenamefont {Geffroy}, \citenamefont {Georgiou}, \citenamefont {Chatzikyriakou}, \citenamefont {Lacerda-Santos}, \citenamefont {Waintal}, \citenamefont {Glattli}, \citenamefont {Roulleau}, \citenamefont {Nath}, \citenamefont {Kataoka}, \citenamefont {Splettstoesser}, \citenamefont {Acciai}, \citenamefont {da~Silva~Figueira}, \citenamefont {{\ifmmode\ddot{O}\else\"{O}\fi}ztas}, \citenamefont {Trellakis}, \citenamefont {Grange}, \citenamefont {Yevtushenko}, \citenamefont {Birner},\ and\ \citenamefont {B{\ifmmode\ddot{a}\else\"{a}\fi}uerle}}]{Edlbauer2022}%
  \BibitemOpen
  \bibfield  {author} {\bibinfo {author} {\bibfnamefont {H.}~\bibnamefont {Edlbauer}}, \bibinfo {author} {\bibfnamefont {J.}~\bibnamefont {Wang}}, \bibinfo {author} {\bibfnamefont {T.}~\bibnamefont {Crozes}}, \bibinfo {author} {\bibfnamefont {P.}~\bibnamefont {Perrier}}, \bibinfo {author} {\bibfnamefont {S.}~\bibnamefont {Ouacel}}, \bibinfo {author} {\bibfnamefont {C.}~\bibnamefont {Geffroy}}, \bibinfo {author} {\bibfnamefont {G.}~\bibnamefont {Georgiou}}, \bibinfo {author} {\bibfnamefont {E.}~\bibnamefont {Chatzikyriakou}}, \bibinfo {author} {\bibfnamefont {A.}~\bibnamefont {Lacerda-Santos}}, \bibinfo {author} {\bibfnamefont {X.}~\bibnamefont {Waintal}}, \bibinfo {author} {\bibfnamefont {D.~C.}\ \bibnamefont {Glattli}}, \bibinfo {author} {\bibfnamefont {P.}~\bibnamefont {Roulleau}}, \bibinfo {author} {\bibfnamefont {J.}~\bibnamefont {Nath}}, \bibinfo {author} {\bibfnamefont {M.}~\bibnamefont {Kataoka}}, \bibinfo {author} {\bibfnamefont {J.}~\bibnamefont {Splettstoesser}}, \bibinfo {author} {\bibfnamefont
  {M.}~\bibnamefont {Acciai}}, \bibinfo {author} {\bibfnamefont {M.~C.}\ \bibnamefont {da~Silva~Figueira}}, \bibinfo {author} {\bibfnamefont {K.}~\bibnamefont {{\ifmmode\ddot{O}\else\"{O}\fi}ztas}}, \bibinfo {author} {\bibfnamefont {A.}~\bibnamefont {Trellakis}}, \bibinfo {author} {\bibfnamefont {T.}~\bibnamefont {Grange}}, \bibinfo {author} {\bibfnamefont {O.~M.}\ \bibnamefont {Yevtushenko}}, \bibinfo {author} {\bibfnamefont {S.}~\bibnamefont {Birner}},\ and\ \bibinfo {author} {\bibfnamefont {C.}~\bibnamefont {B{\ifmmode\ddot{a}\else\"{a}\fi}uerle}},\ }\bibfield  {title} {\bibinfo {title} {{Semiconductor-based electron flying qubits: review on recent progress accelerated by numerical modelling}},\ }\href {https://doi.org/10.1140/epjqt/s40507-022-00139-w} {\bibfield  {journal} {\bibinfo  {journal} {EPJ Quantum Technol.}\ }\textbf {\bibinfo {volume} {9}},\ \bibinfo {pages} {1} (\bibinfo {year} {2022})}\BibitemShut {NoStop}%
\bibitem [{\citenamefont {Ferraro}\ \emph {et~al.}(2014)\citenamefont {Ferraro}, \citenamefont {Roussel}, \citenamefont {Cabart}, \citenamefont {Thibierge}, \citenamefont {F\`eve}, \citenamefont {Grenier},\ and\ \citenamefont {Degiovanni}}]{Ferraro_2014}%
  \BibitemOpen
  \bibfield  {author} {\bibinfo {author} {\bibfnamefont {D.}~\bibnamefont {Ferraro}}, \bibinfo {author} {\bibfnamefont {B.}~\bibnamefont {Roussel}}, \bibinfo {author} {\bibfnamefont {C.}~\bibnamefont {Cabart}}, \bibinfo {author} {\bibfnamefont {E.}~\bibnamefont {Thibierge}}, \bibinfo {author} {\bibfnamefont {G.}~\bibnamefont {F\`eve}}, \bibinfo {author} {\bibfnamefont {C.}~\bibnamefont {Grenier}},\ and\ \bibinfo {author} {\bibfnamefont {P.}~\bibnamefont {Degiovanni}},\ }\bibfield  {title} {\bibinfo {title} {Real-time decoherence of {Landau} and {Levitov} quasiparticles in quantum {Hall} edge channels},\ }\href {https://doi.org/10.1103/PhysRevLett.113.166403} {\bibfield  {journal} {\bibinfo  {journal} {Phys. Rev. Lett.}\ }\textbf {\bibinfo {volume} {113}},\ \bibinfo {pages} {166403} (\bibinfo {year} {2014})}\BibitemShut {NoStop}%
\bibitem [{\citenamefont {Cabart}\ \emph {et~al.}(2018)\citenamefont {Cabart}, \citenamefont {Roussel}, \citenamefont {F\`eve},\ and\ \citenamefont {Degiovanni}}]{Cabart_2018}%
  \BibitemOpen
  \bibfield  {author} {\bibinfo {author} {\bibfnamefont {C.}~\bibnamefont {Cabart}}, \bibinfo {author} {\bibfnamefont {B.}~\bibnamefont {Roussel}}, \bibinfo {author} {\bibfnamefont {G.}~\bibnamefont {F\`eve}},\ and\ \bibinfo {author} {\bibfnamefont {P.}~\bibnamefont {Degiovanni}},\ }\bibfield  {title} {\bibinfo {title} {Taming electronic decoherence in one-dimensional chiral ballistic quantum conductors},\ }\href {https://doi.org/10.1103/PhysRevB.98.155302} {\bibfield  {journal} {\bibinfo  {journal} {Phys. Rev. B}\ }\textbf {\bibinfo {volume} {98}},\ \bibinfo {pages} {155302} (\bibinfo {year} {2018})}\BibitemShut {NoStop}%
\bibitem [{\citenamefont {Marguerite}\ \emph {et~al.}(2016)\citenamefont {Marguerite}, \citenamefont {Cabart}, \citenamefont {Wahl}, \citenamefont {Roussel}, \citenamefont {Freulon}, \citenamefont {Ferraro}, \citenamefont {Grenier}, \citenamefont {Berroir}, \citenamefont {Pla\ifmmode~\mbox{\c{c}}\else \c{c}\fi{}ais}, \citenamefont {Jonckheere}, \citenamefont {Rech}, \citenamefont {Martin}, \citenamefont {Degiovanni}, \citenamefont {Cavanna}, \citenamefont {Jin},\ and\ \citenamefont {F\`eve}}]{Marguerite_2016}%
  \BibitemOpen
  \bibfield  {author} {\bibinfo {author} {\bibfnamefont {A.}~\bibnamefont {Marguerite}}, \bibinfo {author} {\bibfnamefont {C.}~\bibnamefont {Cabart}}, \bibinfo {author} {\bibfnamefont {C.}~\bibnamefont {Wahl}}, \bibinfo {author} {\bibfnamefont {B.}~\bibnamefont {Roussel}}, \bibinfo {author} {\bibfnamefont {V.}~\bibnamefont {Freulon}}, \bibinfo {author} {\bibfnamefont {D.}~\bibnamefont {Ferraro}}, \bibinfo {author} {\bibfnamefont {C.}~\bibnamefont {Grenier}}, \bibinfo {author} {\bibfnamefont {J.-M.}\ \bibnamefont {Berroir}}, \bibinfo {author} {\bibfnamefont {B.}~\bibnamefont {Pla\ifmmode~\mbox{\c{c}}\else \c{c}\fi{}ais}}, \bibinfo {author} {\bibfnamefont {T.}~\bibnamefont {Jonckheere}}, \bibinfo {author} {\bibfnamefont {J.}~\bibnamefont {Rech}}, \bibinfo {author} {\bibfnamefont {T.}~\bibnamefont {Martin}}, \bibinfo {author} {\bibfnamefont {P.}~\bibnamefont {Degiovanni}}, \bibinfo {author} {\bibfnamefont {A.}~\bibnamefont {Cavanna}}, \bibinfo {author} {\bibfnamefont {Y.}~\bibnamefont {Jin}},\ and\ \bibinfo
  {author} {\bibfnamefont {G.}~\bibnamefont {F\`eve}},\ }\bibfield  {title} {\bibinfo {title} {Decoherence and relaxation of a single electron in a one-dimensional conductor},\ }\href {https://doi.org/10.1103/PhysRevB.94.115311} {\bibfield  {journal} {\bibinfo  {journal} {Phys. Rev. B}\ }\textbf {\bibinfo {volume} {94}},\ \bibinfo {pages} {115311} (\bibinfo {year} {2016})}\BibitemShut {NoStop}%
\bibitem [{\citenamefont {Altimiras}\ \emph {et~al.}(2010)\citenamefont {Altimiras}, \citenamefont {le~Sueur}, \citenamefont {Gennser}, \citenamefont {Cavanna}, \citenamefont {Mailly},\ and\ \citenamefont {Pierre}}]{Altimiras_2010}%
  \BibitemOpen
  \bibfield  {author} {\bibinfo {author} {\bibfnamefont {C.}~\bibnamefont {Altimiras}}, \bibinfo {author} {\bibfnamefont {H.}~\bibnamefont {le~Sueur}}, \bibinfo {author} {\bibfnamefont {U.}~\bibnamefont {Gennser}}, \bibinfo {author} {\bibfnamefont {A.}~\bibnamefont {Cavanna}}, \bibinfo {author} {\bibfnamefont {D.}~\bibnamefont {Mailly}},\ and\ \bibinfo {author} {\bibfnamefont {F.}~\bibnamefont {Pierre}},\ }\bibfield  {title} {\bibinfo {title} {Tuning energy relaxation along quantum {Hall} channels},\ }\href {https://doi.org/10.1103/PhysRevLett.105.226804} {\bibfield  {journal} {\bibinfo  {journal} {Phys. Rev. Lett.}\ }\textbf {\bibinfo {volume} {105}},\ \bibinfo {pages} {226804} (\bibinfo {year} {2010})}\BibitemShut {NoStop}%
\bibitem [{\citenamefont {Huynh}\ \emph {et~al.}(2012)\citenamefont {Huynh}, \citenamefont {Portier}, \citenamefont {le~Sueur}, \citenamefont {Faini}, \citenamefont {Gennser}, \citenamefont {Mailly}, \citenamefont {Pierre}, \citenamefont {Wegscheider},\ and\ \citenamefont {Roche}}]{Huynh_2012}%
  \BibitemOpen
  \bibfield  {author} {\bibinfo {author} {\bibfnamefont {P.-A.}\ \bibnamefont {Huynh}}, \bibinfo {author} {\bibfnamefont {F.}~\bibnamefont {Portier}}, \bibinfo {author} {\bibfnamefont {H.}~\bibnamefont {le~Sueur}}, \bibinfo {author} {\bibfnamefont {G.}~\bibnamefont {Faini}}, \bibinfo {author} {\bibfnamefont {U.}~\bibnamefont {Gennser}}, \bibinfo {author} {\bibfnamefont {D.}~\bibnamefont {Mailly}}, \bibinfo {author} {\bibfnamefont {F.}~\bibnamefont {Pierre}}, \bibinfo {author} {\bibfnamefont {W.}~\bibnamefont {Wegscheider}},\ and\ \bibinfo {author} {\bibfnamefont {P.}~\bibnamefont {Roche}},\ }\bibfield  {title} {\bibinfo {title} {Quantum coherence engineering in the integer quantum {Hall} regime},\ }\href {https://doi.org/10.1103/PhysRevLett.108.256802} {\bibfield  {journal} {\bibinfo  {journal} {Phys. Rev. Lett.}\ }\textbf {\bibinfo {volume} {108}},\ \bibinfo {pages} {256802} (\bibinfo {year} {2012})}\BibitemShut {NoStop}%
\bibitem [{\citenamefont {K{\"o}nig}\ \emph {et~al.}(2007)\citenamefont {K{\"o}nig}, \citenamefont {Wiedmann}, \citenamefont {Br{\"u}ne}, \citenamefont {Roth}, \citenamefont {Buhmann}, \citenamefont {Molenkamp}, \citenamefont {Qi},\ and\ \citenamefont {Zhang}}]{QSHI_2007}%
  \BibitemOpen
  \bibfield  {author} {\bibinfo {author} {\bibfnamefont {M.}~\bibnamefont {K{\"o}nig}}, \bibinfo {author} {\bibfnamefont {S.}~\bibnamefont {Wiedmann}}, \bibinfo {author} {\bibfnamefont {C.}~\bibnamefont {Br{\"u}ne}}, \bibinfo {author} {\bibfnamefont {A.}~\bibnamefont {Roth}}, \bibinfo {author} {\bibfnamefont {H.}~\bibnamefont {Buhmann}}, \bibinfo {author} {\bibfnamefont {L.~W.}\ \bibnamefont {Molenkamp}}, \bibinfo {author} {\bibfnamefont {X.-L.}\ \bibnamefont {Qi}},\ and\ \bibinfo {author} {\bibfnamefont {S.-C.}\ \bibnamefont {Zhang}},\ }\bibfield  {title} {\bibinfo {title} {Quantum spin {Hall} insulator state in {HgTe} quantum wells},\ }\href {https://doi.org/10.1126/science.1148047} {\bibfield  {journal} {\bibinfo  {journal} {Science}\ }\textbf {\bibinfo {volume} {318}},\ \bibinfo {pages} {766} (\bibinfo {year} {2007})}\BibitemShut {NoStop}%
\bibitem [{\citenamefont {Wu}\ \emph {et~al.}(2018)\citenamefont {Wu}, \citenamefont {Fatemi}, \citenamefont {Gibson}, \citenamefont {Watanabe}, \citenamefont {Taniguchi}, \citenamefont {Cava},\ and\ \citenamefont {Jarillo-Herrero}}]{Wu2018}%
  \BibitemOpen
  \bibfield  {author} {\bibinfo {author} {\bibfnamefont {S.}~\bibnamefont {Wu}}, \bibinfo {author} {\bibfnamefont {V.}~\bibnamefont {Fatemi}}, \bibinfo {author} {\bibfnamefont {Q.~D.}\ \bibnamefont {Gibson}}, \bibinfo {author} {\bibfnamefont {K.}~\bibnamefont {Watanabe}}, \bibinfo {author} {\bibfnamefont {T.}~\bibnamefont {Taniguchi}}, \bibinfo {author} {\bibfnamefont {R.~J.}\ \bibnamefont {Cava}},\ and\ \bibinfo {author} {\bibfnamefont {P.}~\bibnamefont {Jarillo-Herrero}},\ }\bibfield  {title} {\bibinfo {title} {{Observation of the quantum spin Hall effect up to 100 kelvin in a monolayer crystal}},\ }\href {https://doi.org/10.1126/science.aan6003} {\bibfield  {journal} {\bibinfo  {journal} {Science}\ }\textbf {\bibinfo {volume} {359}},\ \bibinfo {pages} {76} (\bibinfo {year} {2018})}\BibitemShut {NoStop}%
\bibitem [{\citenamefont {Takeda}\ and\ \citenamefont {Furusawa}(2017)}]{Furusawa2017}%
  \BibitemOpen
  \bibfield  {author} {\bibinfo {author} {\bibfnamefont {S.}~\bibnamefont {Takeda}}\ and\ \bibinfo {author} {\bibfnamefont {A.}~\bibnamefont {Furusawa}},\ }\bibfield  {title} {\bibinfo {title} {Universal quantum computing with measurement-induced continuous-variable gate sequence in a loop-based architecture},\ }\href {https://doi.org/10.1103/PhysRevLett.119.120504} {\bibfield  {journal} {\bibinfo  {journal} {Phys. Rev. Lett.}\ }\textbf {\bibinfo {volume} {119}},\ \bibinfo {pages} {120504} (\bibinfo {year} {2017})}\BibitemShut {NoStop}%
\bibitem [{\citenamefont {Burkard}\ \emph {et~al.}(2000)\citenamefont {Burkard}, \citenamefont {Loss},\ and\ \citenamefont {Sukhorukov}}]{Burkard2000}%
  \BibitemOpen
  \bibfield  {author} {\bibinfo {author} {\bibfnamefont {G.}~\bibnamefont {Burkard}}, \bibinfo {author} {\bibfnamefont {D.}~\bibnamefont {Loss}},\ and\ \bibinfo {author} {\bibfnamefont {E.~V.}\ \bibnamefont {Sukhorukov}},\ }\bibfield  {title} {\bibinfo {title} {Noise of entangled electrons: Bunching and antibunching},\ }\href {https://doi.org/10.1103/PhysRevB.61.R16303} {\bibfield  {journal} {\bibinfo  {journal} {Phys. Rev. B}\ }\textbf {\bibinfo {volume} {61}},\ \bibinfo {pages} {R16303} (\bibinfo {year} {2000})}\BibitemShut {NoStop}%
\bibitem [{\citenamefont {Mazza}\ \emph {et~al.}(2013)\citenamefont {Mazza}, \citenamefont {Braunecker}, \citenamefont {Recher},\ and\ \citenamefont {Levy~Yeyati}}]{Mazza2013}%
  \BibitemOpen
  \bibfield  {author} {\bibinfo {author} {\bibfnamefont {F.}~\bibnamefont {Mazza}}, \bibinfo {author} {\bibfnamefont {B.}~\bibnamefont {Braunecker}}, \bibinfo {author} {\bibfnamefont {P.}~\bibnamefont {Recher}},\ and\ \bibinfo {author} {\bibfnamefont {A.}~\bibnamefont {Levy~Yeyati}},\ }\bibfield  {title} {\bibinfo {title} {Spin filtering and entanglement detection due to spin-orbit interaction in carbon nanotube cross-junctions},\ }\href {https://doi.org/10.1103/PhysRevB.88.195403} {\bibfield  {journal} {\bibinfo  {journal} {Phys. Rev. B}\ }\textbf {\bibinfo {volume} {88}},\ \bibinfo {pages} {195403} (\bibinfo {year} {2013})}\BibitemShut {NoStop}%
\bibitem [{\citenamefont {Grenier}\ \emph {et~al.}(2013)\citenamefont {Grenier}, \citenamefont {Dubois}, \citenamefont {Jullien}, \citenamefont {Roulleau}, \citenamefont {Glattli},\ and\ \citenamefont {Degiovanni}}]{Grenier_2013}%
  \BibitemOpen
  \bibfield  {author} {\bibinfo {author} {\bibfnamefont {C.}~\bibnamefont {Grenier}}, \bibinfo {author} {\bibfnamefont {J.}~\bibnamefont {Dubois}}, \bibinfo {author} {\bibfnamefont {T.}~\bibnamefont {Jullien}}, \bibinfo {author} {\bibfnamefont {P.}~\bibnamefont {Roulleau}}, \bibinfo {author} {\bibfnamefont {D.~C.}\ \bibnamefont {Glattli}},\ and\ \bibinfo {author} {\bibfnamefont {P.}~\bibnamefont {Degiovanni}},\ }\bibfield  {title} {\bibinfo {title} {Fractionalization of minimal excitations in integer quantum {Hall} edge channels},\ }\href {https://doi.org/10.1103/PhysRevB.88.085302} {\bibfield  {journal} {\bibinfo  {journal} {Phys. Rev. B}\ }\textbf {\bibinfo {volume} {88}},\ \bibinfo {pages} {085302} (\bibinfo {year} {2013})}\BibitemShut {NoStop}%
\end{thebibliography}

%

\end{document}